\documentclass[preprint,5p,times,twocolumn]{elsarticle}
 
\usepackage{hyperref}
\hypersetup{
  colorlinks=true, 
}
\usepackage{amssymb}
\usepackage{amsmath}
\usepackage{bbm}
\usepackage{nicefrac}

\DeclareMathAlphabet{\mathdutchcal}{U}{dutchcal}{m}{n}
\newcommand{\ii}{\ensuremath\mathbbm{i}}

\newcommand{\kk}{\ensuremath\mathdutchcal{k}}

\newcommand{\sss}{\ensuremath\mathdutchcal{s}}

\newcommand{\pp}{\ensuremath\mathdutchcal{p}}
\newcommand{\qq}{\ensuremath\mathdutchcal{q}}
\newcommand{\mm}{\ensuremath\mathdutchcal{m}}

\newcommand{\op}[1]{\ensuremath\hat{{\mathcal{#1}}} }

\providecommand{\footref}[1]{\textsuperscript{\ref{#1}}}
\newcommand{\mum}{\ensuremath\mu\mathrm{m}}

\usepackage{tikz}

\journal{Computer Physics Communications}

\begin{document}

\begin{frontmatter}


\title{High-order exponential solver method for particle-in-cell simulations in cylindrical geometry }

\author[label1]{Szil\'ard Majorosi\corref{cor1} }
\ead{szilard.majorosi@eli-alps.hu}
\cortext[cor1]{Corresponding author.}

\author[label1,label2]{Nasr A.M. Hafz}

\author[label1]{Zsolt L\'ecz}

\affiliation[label1]{organization={ELI ALPS, The Extreme Light Infrastructure ERIC},
             addressline={Wolfgang Sandner utca 3.},
             city={Szeged},
             postcode={H-6728},
             country={Hungary}}


\affiliation[label2]{organization={Doctoral School of Physics, Faculty of Science and Informatics, University of Szeged},
             addressline={Dóm tér 9.},
             city={Szeged},
            postcode={H-6720},
             country={Hungary}}

\begin{abstract}

Recent developments in high peak-power table-top laser systems reaching highly relativistic light intensities have led to significant advances in laser-driven particle acceleration schemes (mainly the laser wakefield acceleration, LWFA) that heavily rely on particle-in-cell (PIC) simulations for the microscopic understanding of the acceleration process. Efficient algorithms have been developed by taking advantage of the cylindrical geometry of the laser-plasma acceleration interaction, which reduces the computational and memory costs of these simulations, but with the trade-off of reduced accuracy compared to the 3D simulations. The most successful solution solves the Maxwell equations on a Fourier-Bessel spectral basis in this geometry, as used by the well-known FBPIC code. In this work, we present a solution that is a real-space equivalent of the latter using the finite difference exponential time-domain method. Spatially, we represent the derivatives with high-order staggered finite differences locally and address issues of the near-axis particle representation. Additionally, we also develop an exponential solution to propagate the laser envelope potential with high accuracy in the cylindrically symmetric PIC model.
We show that this method provides a very high accuracy without relying on a transformation to special basis functions. We verified the accuracy and the convergence of these methods in various benchmarks involving laser propagation in vacuum and in underdense plasma. Electron injection in the non-linear laser wakefield regime has also been simulated and the results are compared with 3D simulations, and to the cylindrical spectral solution of FBPIC. We found good agreement between these methods; however, the spectral solution resulted in less energetic electrons and a smoother spatial distribution near the cylindrical axis.

\end{abstract}

\end{frontmatter}


\section{Introduction} \label{sec:introduction}

The invention of the chirped-pulse amplification technique (CPA) \cite{Strickland85} opened the era of relativistic laser-plasma interactions where highly relativistic intensities are used for novel charged-particle acceleration schemes \cite{Tajima79,esarey2009laser_plasma_accelerators, lu2007electron_laser_wakefield, albert2016applications_lwfa}.  The most successful laser-driven acceleration mechanism so far is called the laser wakefield electron acceleration (LWFA), which has advanced significantly in  the past 20 years due to the continuous laser and target developments. In this scheme an intense laser pulse propagates in an underdense gaseous plasma, where the relativistic quiver motion of the electrons, driven by the radiation pressure, lead to the formation of ion cavities or wakefields. After self-injection (or other injection physical mechanism), electrons enter these cavities where they can be accelerated to GeV energies over few centimeters. Such a LWFA process can be accurately modeled with particle-in-cell simulations \cite{BOOK_PLASMA_SIMULATION, fonseca2002OSIRIS, pritchett2003pic_tutorial, arber2015pic_epoch, belayev2015PICsar, derouillat2018pic_smilei}. This require accurate simulation tools since strong nonlinear effects, like shock formation and laser self-phase modulation, occur. 

By taking advantage of the near cylindrical symmetry of the LWFA acceleration scheme, Maxwell-PIC algorithms have been developed within cylindrical geometry, initially using finite difference Yee-methods with azimuthal modes decomposition \cite{lifschitz2009pic_fourier_decomposition, davidson2015OSIRIS_RZ}. These methods reduce the computational and memory cost of the simulations by orders of magnitude compared to the full three dimensional (3D) simulations, allowing to perform long distance propagation feasibly, but with the trade-offs of reduced sampling in momentum space and artificial representation of the particles, flawed especially near the cylindrical axis. The most successful solution solves the Maxwell-equations on Fourier-Bessel spectral basis in this geometry, implemented in the well-known FBPIC code \cite{lehe2016fbpic}. It enables artifact free propagation of the Maxwell-fields using the Fourier basis in the forward direction, and Bessel-Hankel transforms in the radial direction. The computational cost of these transformations has raised the questions whether it could be done without the need of special basis functions, but reaching the same high accuracy. However, we found that such one-to-one approximation of the Fourier-Bessel method was not possible this way.

The presented work here relies on our exponential PIC method \cite{majorosi2026exponentialpic}, where we introduced exponential Maxwell-solver solution in Cartesian geometry and developed PIC method using that formalism.  Here we apply that solution in cylindrical geometry and using real azimuthal modes decomposition for the angular variable. The spatial differences are represented with high order finite differences (6th-32nd) and the time stepping with exponential operators \cite{leforestier1991tdsecomparisom, castro2004kohnshampropagators, bandrauk2013splitting} – in practice a lot of matrix multiplications. This approach is the real space approximation of spectral solution of the respective partial differential equations. The cylindrical geometry has a Cartesian axis, $z$, and along this axis all numerical methods are the same as those described in our Cartesian paper, including their accuracy characteristics. This also makes it possible to directly compare the 3D and cylindrical simulations with the same numerical parameters.

Additionally, in this work we introduce this kind of exponential solution for scalar wave equation, particularly to propagate the potential of a laser pulse envelope highly accurately. This can be used in laser envelope PIC method \cite{esarey2009laser_plasma_accelerators, cowan2011envelope_model, benedetti2018envelope_solver, terzani2019envelope}, in which the particles only interact with cycle averaged fields of the laser potential forming the so-called ponderomotive guiding center approximation. This simplified model can be utilized for even faster and efficient simulations of the laser pulse propagation in underdense plasma, making this type of computations feasible without HPC resources, if the laser pulse is much longer than few cycles and the intensity envelope is cylindrical symmetric.

This paper is organized as follows. In Section \ref{sec:MaxwellAM}, we discuss how the Maxwell-equations can be solved in cylindrical geometry with exponential operators and discuss how the required matrices are calculated in the finite difference spatial representation, and we also present the actual numerical algorithms. Then, in Section \ref{sec:particlesAM} we develop our cylindrical particle-in-cell routine compatible with our field solutions. In Section \ref{sec:laserA} we apply these methods to the propagation of the cylindrically symmetric laser envelope PIC model. Finally, in  Section \ref{sec:BenchmarkAM} we benchmark our method in various cases mostly focusing on laser propagation in vacuum, in an underdense plasma, producing accurate wakefields and nonlinear laser wakefield electron acceleration (LWFA) and we show that this method could provide very high accuracy in the spectral range of physical interest.

In the following, we write all quantities in the normalized units commonly used in plasma physics unless stated otherwise. In these units we set the speed of light ($c$), the elementary charge ($e$), the  electron mass ($\mm_e$) equal to unity. For the precise definitions see \cite{derouillat2018pic_smilei, majorosi2026exponentialpic}.  These will simplify the physical and matrix equations related to the electromagnetic fields and the relativistic particles.

\section{Electromagnetic fields in cylindrical coordinates}  \label{sec:MaxwellAM}

\subsection{Maxwell-equations in cylindrical coordinates} \label{subsec:MaxwellAM}

In the following, let us summarize the Maxwell equations in cylindrical coordinates $(\rho, \theta, z)$. We first write out the Amp\'ere-equation, using the cylindrical form of $\nabla\times{\bf B}$:
\begin{align}
\partial_t E_\rho     &= -\partial_z B_\theta +\rho^{-1}\partial_\theta B_z - J_\rho, \label{eq:maxwell3DC_Er}\\
\partial_t E_\theta  &=  \partial_z B_\rho -\partial_\rho B_z - J_\phi,\label{eq:maxwell3DC_Ephi}\\
\partial_t E_z        &=  -\rho^{-1}\partial_\theta B_\rho +\rho^{-1}\partial_\rho (\rho B_\theta) - J_z, \label{eq:maxwell3DC_Ez}
\end{align}
and the cylindrical form of the Faraday-equation reads
\begin{align}
\partial_t B_\rho     &= \partial_z E_\theta -\rho^{-1}\partial_\theta E_z, \label{eq:maxwell3DC_Br} \\
\partial_t B_\theta  &=  -\partial_z E_\rho +\partial_\rho E_z ,\label{eq:maxwell3DC_Bphi}\\
\partial_t B_z        &= \rho^{-1}\partial_\theta E_\rho -\rho^{-1}\partial_\rho (\rho E_\theta). \label{eq:maxwell3DC_Bz}
\end{align}

Let us also write out the Gauss's law and Poisson's equation for a scalar potential $\phi(\rho, \theta, z)$ to list the cylindrical form of gradient, divergence and the Laplace operator: 
\begin{align}
\nabla \cdot {\bf E}  &=  \rho^{-1}\partial_\rho (\rho E_\rho) + \rho^{-1}\partial_\theta E_\theta + \partial_z E_z = \varrho,  \label{eq:maxwell3DC_divE} \\
{\bf E}   &= -\nabla \phi = -\left( \partial_\rho \phi \right) {\bf \hat{e}_\rho} - \left( \rho^{-1} \partial_\theta \phi \right) {\bf \hat{e}_\theta} - \left( \partial_z \phi \right) {\bf \hat{e}_z}, \label{eq:maxwell3DC_gradphi}\\
\nabla^2 \phi  &= \rho^{-1}\left( \partial_\rho \rho  \partial_\rho\phi \right) + \rho^{-2} \partial_\theta^2 \phi + \partial_z^2 \phi = - \varrho.\label{eq:maxwell3DC_Poisson}
\end{align}

A real Fourier-series for a periodic scalar field $\phi=\phi(\rho, \theta, z) $ in variable  $\theta$ is given by \cite{BOOK_NUMERICAL_RECIPIES, constantinescu2002flow_cylindrical}:
\begin{equation} \label{eq:maxwellAM_fourier}
 \phi = \phi_0 +\phi_{1} \cos \left( M \theta \right) + \sum_{m = 1}^{M-1} \left[ \cos \left(m\theta  \right) \phi_{2m} + \sin \left(m \theta  \right) \phi_{2m+1}   \right],
\end{equation}
where $\phi_{2m}(\rho,z)$, $\phi_{2m+1}(\rho,z)$ are the expansion coefficients. If $\theta$ is discretized on a grid that has even number of points, then its discrete Fourier-transform must also have an even number of basis functions as in Eq. (\ref{eq:maxwellAM_fourier}). The formula Eq. (\ref{eq:maxwellAM_fourier}) is equivalent to the usual complex form of the azimuthal modes decomposition with $M$ complex modes \cite{lifschitz2009pic_fourier_decomposition}, except that the former has an additional component $\phi_1(\rho,z)$.

In the complex spectral basis $\partial_\theta$ is replaced by $\ii m$, in the real formalism, however, it is replaced by a 2x2 matrix acting on pairs of coefficients:
\begin{equation} \label{eq:maxwellAM_dtheta}
\partial_\theta \rightarrow
\left( \begin{array}{cc}
0  & -m \\
m  &  0
\end{array} \right)
\left( \begin{array}{c}
\phi_{2m} \\
\phi_{2m+1}
\end{array} \right),
\end{equation}
Because the expansion does not have $\sin (M \theta)$ basis function, $\phi_1$ can be thought as the "imaginary" part of the complex Fourier-transform coefficient with $m = 0$. From physics perspective the inclusion of this modal coefficient can yield nonphysical results. For example, the continuity equation $\partial_t\varrho = -\nabla\cdot{\bf J}$ is not valid for these modal coefficients, because for a true $m = 0$ mode the charge must be conserved, but it is not the case for these because of the $\cos(M\theta)$ angular dependence. Since the validity of the former is important in PIC description, we regard $\phi_1=0$ always.

Since the Maxwell-fields are real quantities we choose to keep their form real after the expansion in the following. For each $m$ number we can formally form pairs of coefficients forming index subsets $2m$ and $2m+1$ for the expansion coefficients, which in practice is equivalent to the complex azimuthal modes. Combining these we form 6-component real vectors for each $m$ mode from the expansion coefficients of ${\bf E}$, ${\bf B}$ vector fields as:
\begin{align} \label{eq:maxwellAM_vecE}
{\bf E}_{2m} &=
\left( \begin{array}{ccccccc}
E_{x,2m} & E_{x,2m+1} & E_{y,2m} & E_{y,2m+1} & E_{z,2m} & E_{z,2m+1}
\end{array} \right)  ^T , \\ \label{eq:maxwellAM_vecB}
{\bf B}_{2m} &=
\left( \begin{array}{ccccccc}
B_{x,2m} & B_{x,2m+1} & B_{y,2m} & B_{y,2m+1} & B_{z,2m} & B_{z,2m+1}
\end{array} \right)  ^T .
\end{align}
 Here the even indices of $2m$ can be thought as the real parts, the odd indices of $2m+1$ the imaginary parts of the equivalent complex Fourier-transforms of index $|m|$. 

Unfortunately, in this real formalism the amount of Maxwell-equations are doubled (for the $2m$ and $2m+1$ index subsets), however, we write out only half of them for simplicity - the rest follow straightforwardly.  In the following presentation we select the half which forms a closed set of equations, and is complete for a reflectional symmetric system in $y$. This half of the Ampére equations are written as:
\begin{align}
\partial_t E_{\rho_,2m}     &= -\partial_z B_{\theta,2m} - m\rho^{-1}B_{z,2m+1} - J_{\rho,2m}, \label{eq:maxwellAM_Er}\\
\partial_t E_{\theta,2m+1}  &=  \partial_z B_{\rho,2m+1} -\partial_\rho B_{z,2m+1} - J_{\phi,2m+1},\label{eq:maxwellAM_Ephi}\\
\partial_t E_{z,2m}        &=  m\rho^{-1} B_{\rho,2m+1} +\rho^{-1}\partial_\rho (\rho B_{\theta,2m}) - J_{z,2m}  \label{eq:maxwellAM_Ez}
\end{align}
and half of the Faraday-equations are
\begin{align}
\partial_t B_{\rho_,2m+1}    &= \partial_z E_{\theta,2m+1} -m\rho^{-1}E_{z,2m}, \label{eq:maxwellAM_Br}\\
\partial_t B_{\theta,2m}  &=  -\partial_z E_{\rho,2m} +\partial_\rho E_{z,2m} ,\label{eq:maxwellAM_Bphi}\\
\partial_t B_{z,2m+1}     &=  m\rho^{-1} E_{\rho,2m} -\rho^{-1}\partial_\rho (\rho E_{\theta,2m+1}). \label{eq:maxwellAM_Bz}
\end{align}
The gradient, divergence operators, and Poisson's equation will be written as:
\begin{align}
\left[ \nabla \phi \right]_{2m} &= \left( \partial_\rho \phi_{2m} \right) {\bf \hat{e}_\rho} - \left( m \rho^{-1} \phi_{2m+1} \right) {\bf \hat{e}_\theta} + \left( \partial_z \phi_{2m} \right) {\bf \hat{e}_z}, \label{eq:maxwellAM_gradphi}\\
\left[ \nabla \cdot {\bf E} \right]_{2m}  &=  \rho^{-1}\partial_\rho (\rho E_{\rho,2m}) - m \rho^{-1} E_{\theta,2m+1} + \partial_z E_{z,2m}, \label{eq:maxwellAM_divE} \\
\left[ \nabla^2 \phi \right]_{2m}  &= \nabla_{\bot,m}^2 \phi_{2m} + \partial_z^2 \phi_{2m} = -\varrho_{2m}. \label{eq:maxwellAM_Poisson}
\end{align}
For completeness, we note that to get the other half of equations from Eqs. (\ref{eq:maxwellAM_Er})-(\ref{eq:maxwellAM_Poisson}) we just replace the indices $2m$ with $2m+1$ and vice versa, and flip the sign of $m$ if it is in a coefficient. Field modes with different $m$ do not couple to each other.

The transverse 2D Laplacian $\nabla_{\bot,m}^2$ is completely diagonal with respect to $m$:
\begin{align}
\nabla_{\bot,m}^2 &= \nabla_{\rho}^2 - m^2 \rho^{-2},\label{eq:maxwellAM_laplace2D} \\
 \nabla_{\rho}^2 &= \rho^{-1} \partial_\rho \left( \rho  \partial_\rho \right) = \partial_\rho^2+  \rho^{-1}\partial_\rho, \label{eq:maxwellAM_laplaceR}
\end{align}
where we denoted the radial Laplacian as  $\nabla^2_\rho$.

The formal form of the Maxwell-equations can be written independently according to each mode $m = 0, \ldots, M-1$ using all components in Eqs. (\ref{eq:maxwellAM_vecE}) and (\ref{eq:maxwellAM_vecB}) as:
\begin{equation} \label{eq:maxwellAM_formal}
 \partial_t \Psi_{2m}      = \op{H}_m \Psi_{2m} - {\bf J}_{2m},
\end{equation}
with
\begin{equation} \label{eq:maxwellAM_psi}
\Psi_{2m} =
\left( \begin{array}{c}
{\bf E}_{2m} \\
{\bf B}_{2m}
\end{array} \right)
,  \ \ 
\op{H}_m =
\left( \begin{array}{cc}
0          & \op{R}_m  \\
-\op{R}_m  & 0
\end{array} \right) , \ \ 
{\bf J}_{2m} \rightarrow 
\left( \begin{array}{c}
{\bf J}_{2m} \\
{\bf 0}
\end{array} \right)
\end{equation}
and
\begin{equation}  \label{eq:maxwellAM_H}
\op{R}_m =
\left( \begin{array}{cccccc}
0            & 0           & -\partial_z       &   0     & 0 &  -m\rho^{-1} \\
0            & 0           & 0       & -\partial_z &  m\rho^{-1}           & 0 \\
\partial_z   & 0           & 0       & 0 & -\partial_\rho  & 0           \\
0            & \partial_z  & 0 & 0          & 0         & -\partial_\rho          \\
0            & m\rho^{-1} &  \rho^{-1}\partial_\rho \rho & 0          & 0         & 0         \\
-m\rho^{-1}   & 0           & 0           & \rho^{-1}\partial_\rho \rho          & 0         & 0         \\
\end{array} \right) .
\end{equation}

The formal solution of Eq. (\ref{eq:maxwellAM_formal}) is written as:
\begin{multline} \label{eq:maxwellAM_solution}
\Psi_{2m}(t+\Delta t) =  \exp \left( \Delta t \op{H}_m \right)\Psi_{2m}(t)- \\
                \int_0^{\Delta t} 
                \exp \left((\Delta t-s)  \op{H}_m \right){\bf J}_{2m}(t+s) {\rm d}s
\end{multline}
which becomes second order accurate if we use the midpoint approximation of the integral:
\begin{multline} \label{eq:maxwellAM_solution2}
\Psi_{2m}(t+\Delta t) \approx \\
\exp \left( \Delta t \op{H}_m \right)\Psi_{2m}(t)- \exp \left(\Delta t  \op{H}_m /2 \right){\bf J}_{2m}(t+\Delta t /2)
\end{multline}
We then perform explicit Taylor-expansion of the exponentials \cite{castro2004kohnshampropagators, majorosi2026exponentialpic}. In such an explicit scheme only the matrix multiplication $\op{H}_m \Psi_{2m}$ needs to be defined, its full matrix form is not needed to be written out.

\subsection{Cylindrical discretization} \label{subsec:spatialAM_grid}

What follows are directly built on the 3D Cartesian formalism \cite{majorosi2026exponentialpic} with the radial coordinate $\rho$ taking the place of $x$ and in the $z$ coordinate every numerical method is kept the same as there. We need to focus on how to form a proper radial $\rho$ representation.

We discretize the cylindrical coordinates $\rho, \theta, z$ in a conventional form for scalar fields (primary grids):
\begin{equation} \label{eq:spatialAM_grid}
\rho_k = k \Delta \rho +\Delta \rho /2, \quad \theta_j = j\Delta \theta,  \quad z_i = i\Delta z+z_{0}, 
\end{equation}
where $\rho$ grid begins with $\Delta\rho/2$ to avoid the singularity at $\rho = 0$, and $z_{0}$ is a grid offset. The indices $k\in[0, M_\rho-1]$,  $j\in[0, M_\theta-1]$,  $i\in[0, M_z-1]$, where $M_\rho,M_\theta, M_z$ are the grid resolution per dimension. We also define an angular angular grid $\theta_j$ of cell size $\Delta \theta = 2\pi/M_\theta$ before we do the compatible azimuthal modes expansion Eq. (\ref{eq:maxwellAM_fourier}) with $M = M_{\theta}/2$. There is a special case of $M=M_{\theta}=1$ which implies full rotational symmetry.

Our real space solution necessitates the use of staggered (-) and non-staggered (+) radial grid variants:
\begin{equation} \label{eq:spatialAM_gridR}
\quad \rho_{(-),k} = \rho_k - \Delta \rho/2, \quad \rho_{(+),k} = \rho_k.
\end{equation}
All scalar quantities are defined on the primary $\rho_k$ grid. 

During the computational evaluation of the exponentials we utilize the banded diagonal matrix representation of differential and other operators.
The matrix product with such operators ${\rm A}_\rho$, ${\rm A}_z$ which act along the $\rho$ and $z$ directions are of the form for a mode $\phi_{2m,i,k}$  (same for all modes):
\begin{align}
\left( {\rm A}_{\rho} \phi \right)_{2m,i,k} &=  \sum_{k' = -N_{A}}^{N_{A}} {\rm A}_{\rho,k,k+k'}\phi_{2m,i,k+k'} \quad \rightarrow \quad  {\rm A}_\rho \phi_{2m}, \label{eq:bandedmatricesAM_Ar} \\
\left( {\rm A}_{z} \phi \right)_{2m,i,k} &=  \sum_{i' = -N_{A}}^{N_{A}} {\rm A}_{z,i,i+i'}\phi_{2m,i+i',k} \quad \ \ \rightarrow \quad  {\rm A}_z \phi_{2m}, \label{eq:bandedmatricesAM_Az} 
\end{align}
where $N_A$ denotes the half width of the banded diagonal matrix. The even $2m$ and odd $2m+1$ modal coefficients behave the same for these, only the angular derivative Eq. (\ref{eq:maxwellAM_dtheta}) couples them together.
In the above expressions, the terms with out of bounds indices $k+k'\not\in[0, M_\rho-1]$, $i+i'\not\in[0, M_z-1]$ are ignored. 

\subsection{Radial staggered finite differences} \label{subsec:spatialAM_deriv}

In our first attempt we tried to solve the cylindrical Maxwell equations using only a single grid Eq. (\ref{eq:spatialAM_grid}) similarly as in FBPIC \cite{lehe2016fbpic} with various (centered) high order finite differences. All of these turned out to be unstable or prone to artifacts in various ways due to the nonphysical spectral behaviour of such differences at the largest grid frequency $\kk_{\max}$. So the use of staggered derivatives and grids Eq. (\ref{eq:spatialAM_gridR}) are necessary for the vector fields.  The use of a spectral solution allows to use a single grid radially, shifting this sampling problem into the construction of the Bessel-basis.

In our solution we use high-order \emph{staggered} finite differences (of order $2N_D$) as we did in Cartesian coordinates \cite{majorosi2026exponentialpic} by placing their coefficients into the rows of the banded diagonal matrix ${\rm D}^{(\pm)}$ as 
 \begin{equation} \label{eq:spatial_derivativeSt}
\left( \partial_\rho   \phi  \right)_{k\pm1/2} \approx \sum_{k' = -N_{D}}^{N_D} {\rm D}^{(\pm)}_{k,k+k'} \phi_{k+k'},
\end{equation}
which stagger the result on the right hand side, similar to conventional Yee-codes \cite{lifschitz2009pic_fourier_decomposition, derouillat2018pic_smilei}. This form has a high frequency behavior that is regular and convergent with higher orders.
 We use the convention that downstaggering (-) changes the grid from $\rho_k$ to $\rho_{(-),k}$. We also note that Eq. (\ref{eq:spatial_derivativeSt}) is not applicable using Galilean-frame in the Maxwell-solver.

Next we write out the banded diagonal matrix form of the staggered radial divergence operator. It has the form:
\begin{equation} \label{eq:spatialAM_derivativeR_div}
\left( \rho^{ -1}\partial_\rho (\rho   \phi ) \right)_{k\pm\frac{1}{2}} \approx \rho_{(\pm),k}^{-1}\sum_{k' = -N_{D}}^{N_D} {\rm D}^{(\pm)}_{k,k+k'} 
\left( \rho_{(\mp)} \phi \right)_{k+k'}
\rightarrow { \rm D}_{\text{div},\rho}^{(\pm)}\phi.
\end{equation}
We note that the values of $\rho$ left and right to ${\rm D}^{(\pm)}$ are from the grids of different staggering. Furthermore, the matrix elements of the above operator at $\rho = 0$ is always zero for smooth functions.

We also use the following the discrete radial Laplacian Eq. (\ref{eq:maxwellAM_laplaceR}):
\begin{equation} \label{eq:spatialAM_laplaceR}
\nabla^2_\rho \approx {\rm D}^{2}_{\rho}+\rho^{-1} {\rm D}^{(C)}_{\rho} =   {\rm L_\rho},
\end{equation}
this contains the finite difference formulas for the second derivatives in ${\rm D}^2_{\rho}$, and non-staggered finite differences in ${\rm D}^{(C)}_{\rho}$. 
In the Cartesian direction ${\rm L}_z = {\rm D}^2_z$.

To use the staggered spatial representation, we also need to construct a banded diagonal matrix that staggers or destaggers the fields (grids) using high order Lagrange interpolation \cite{majorosi2026exponentialpic} in the form of:
\begin{equation} \label{eq:spatialAM_staggerS}
\phi_{k\pm1/2} \approx \sum_{k' = -N_{D}}^{N_D} {\rm S}^{(\pm)}_{k,k+k'} \phi_{k+k'}.
\end{equation}
Due to the high-order nature of the this interpolation, we can do this  with minimal accuracy loss (using the same order as the finite difference). The operators ${\rm D}^{(-)},{\rm S}^{(-)}$ stagger down from the primary grid Eq. (\ref{eq:spatialAM_grid}).

\subsection{Radial boundary conditions} \label{subsec:spatialAM_boundary}

\begin{table}[tp!]
\small
\begin{tabular*}{\linewidth}{@{\extracolsep{\fill}} c|c|ccc|ccc }

    & $\phi_{2m}$ & $E_{\rho,2m}$ & $E_{\theta,2m}$ & $E_{z,2m}$ & $B_{\rho,2m}$ & $B_{\theta,2m}$ & $B_{z,2m}$\tabularnewline \hline
stag. $\rho$ & 0 & $-1/2$ & 0 & 0 & 0 & $-1/2$ & $-1/2$ 
\tabularnewline
stag. $z$ & 0 & 0 & 0 & $-1/2$ & $-1/2$ & $-1/2$ & 0 
\tabularnewline  \hline
symm. & $mS$ & $mA$ & $mA$&  $mS$&  $mA$& $mA$ & $mS$
\tabularnewline \hline
$\rho\rightarrow0$ & $\rho^m$ & $\rho^{|m-1|}$ & $\rho^{|m-1|}$ & $\rho^m$ & $\rho^{|m-1|}$  & $\rho^{|m-1|}$ & $\rho^m$
\tabularnewline
$m = 0$ & const & $\rho$ & $\rho$ &  const &  $\rho$ & $\rho$ & const
\tabularnewline 
$m = 1$ & $\rho$ & const & const & $\rho$ & const  & const & $\rho$
\tabularnewline 
$m = 2$ & $\rho^2$ & $\rho$ & $\rho$ & $\rho^2$ & $\rho$  & $\rho$ & $\rho^2$
\tabularnewline  \hline
\end{tabular*}

\caption{Table of the staggering and  asymptotic behavior at $\rho = 0$ of selected scalar and vector field modal coefficients (cosine). The same is true for $2m+1$ modal indices (sine). Here $mS$ and $mA$ denote the $m$-dependent radial symmetry type.   \label{tab:spatialAM_boundary}}
\end{table} 

To form a correct solution it is necessary to specify how the modal coefficients of scalar $\phi_{2m}$ and vector fields ${\bf E}_{2m}$ (same for ${\bf B}_{2m}$, ${\bf J}_{2m}$) are handled near the cylindrical axis. The even ($2m$) and odd ($2m+1$) modal coefficients follow the same behavior with the same value of $m$ in this regard, so we omit the latter group here for simplicity. To get the out of boundary values in any radial banded diagonal matrix we take the symmetric ($S$) or antisymmetric ($A$) radial extension of the modes at $\rho \leq 0$ and express those values with the interior points of the domain.

The rules how to do this can be inferred from the Bessel basis for a sufficient smooth solution with mode index $m$, the properties of which we summarize in Table \ref{tab:spatialAM_boundary}. There are two groups regarding the near axis behavior: \emph{scalar like}  such as $\phi_{2m}$, ${E}_{z,2m}$ we denote by $mS$, and \emph{polar like}  such as ${E}_{\rho,2m}$, ${E}_{\theta,2m}$ we denote by $mA$. The latter group is a combination of two Cartesian modes $m-1$ and $m+1$.\footnote{In the complex formalism these two modes can be separated into vector components  ${E}_{\pm,m\mp1} =\left(E_{\rho,m}\pm\ii E_{\theta, m}\right)/2$. Then $E_{\rho,m}$ and $E_{\theta,m}$ must be on the same grid.} All modal coefficients with $m > 1$ are zero on the $\rho = 0$ axis.

We also need the boundary values at $\rho = 0$ of the staggered vector fields ($E_\rho$, $B_\theta$, $B_z$) for the same purpose ($E_\rho$ and $B_\theta$ has the same boundary). There are two possible cases: if a field value equals 0 on axis we include it straightforwardly in the matrix coefficients. If the field value should be constant  we express the edge value as the linear combination of the surrounding elements such as: 
\begin{equation}
{E}_{\rho,2m=2,0}=\sum_{k'=1}^{N_D}c_{k'} {E}_{\rho,2m=2,k'}, \quad  {B}_{z,2m=0,0}=\sum_{k'=1}^{N_D}c_{k'} {B}_{z,2m=0,k'},
\end{equation}
i.e. we calculate the interpolation coefficients $c_k$ on the high-order Lagrange-polynomial basis \cite{majorosi2026exponentialpic} as we described in the previous section for other operators, but also taking into account the symmetric boundaries. For example, to 10th order these coefficients are $5/3, -20/21, 5/14,-5/63,1/126$ for $k' = 1, 2, 3, 4, 5$.  We note that the only component in which the $\rho = 0$ value is required in the Maxwell-solver is ${B}_{z, 2m}$.

We introduce the following notation for the radial banded diagonal matrices of Section \ref{subsec:spatialAM_deriv} acting on  azimuthal mode pairs ($2m$ and $2m+1$)  which incorporate the proper $m$ dependent symmetric and anti symmetric boundary conditions at $\rho = 0$:
\begin{align} \label{eq:spatialAM_derivativeS}
{\rm D}^{(\pm, m S)}_{\rho} &= 
\begin{cases}
{\rm D}^{(\pm,S)}_{\rho},  \text{if } m \text{ is even}, \\
{\rm D}^{(\pm,A)}_{\rho},  \text{if } m \text{ is odd}, 
\end{cases}
\\
\label{eq:spatialAM_derivativeA}
{\rm D}^{(\pm, m A)}_{\rho} &= 
\begin{cases}
{\rm D}^{(\pm,S)}_{\rho},  \text{if } |m-1| \text{ is even}, \\
{\rm D}^{(\pm,A)}_{\rho},  \text{if } |m-1| \text{ is odd}, 
\end{cases} 
\end{align}
where ${\rm D}^{(\pm,S)}_{\rho}$ and ${\rm D}^{(\pm,A)}_{\rho}$ denotes the banded diagonal matrix variants that incorporate symmetric and anti symmetric boundary conditions at the $\rho = 0$ edge which may also up- or downstagger ($\pm$).  We also place the boundary values at $\rho = 0$ into the operators that upstagger (+). These boundary conditions are required for all radial banded diagonal matrices.

\subsection{Discrete Maxwell equations with AM decomposition} \label{subsec:spatialAM_maxwell}

We only present formulas for the $y$-symmetric modes of Eqs. (\ref{eq:maxwellAM_Er})-(\ref{eq:maxwellAM_Poisson}) in the following. If the modal coefficients of ${\bf E}^{(C)}$ and ${\bf B}^{(C)}$ are not staggered initially,  we stagger them using Eq. (\ref{eq:spatialAM_staggerS}) to be on the Yee-grid (downward)  with the following operation:
\begin{align} \label{eq:spatialAM_stagger_E}
E_{\rho, 2m} &= {\rm S}_\rho^{(-, mA)} E_{\rho,2m}^{(C)},  \quad E_{z,2m} = {\rm S}_{z}^{(-)} E_{z,2m}^{(C)}, \\
\quad B_{\rho, 2m+1} &= {\rm S}_z^{(-)} B_{\rho,2m+1}^{(C)},\quad   B_{z,2m+1} = {\rm S}_\rho^{(-, mS)}  B_{z,2m+1}^{(C)},  \label{eq:spatialAM_stagger_B} \\
 & \quad B_{\theta,2m} = {\rm S}_\rho^{(-, mA)}{\rm S}_z^{(-)} B_{\theta,2m}^{(C)}, \label{eq:spatialAM_stagger_Bth}
\end{align}
where the radial symmetric boundary of ${\rm S}_{\rho}^{(\pm,mA)}$ and ${\rm S}_{\rho}^{(\pm,mA)}$ at the $\rho = 0$ axis is defined according to Eqs. (\ref{eq:spatialAM_derivativeS}) and (\ref{eq:spatialAM_derivativeA}).  To (approximately) reverse the staggering we change all ${\rm S}^{(-)}$ to ${\rm S}^{(+)}$ and matrix multiply them with the Yee staggered components. We keep the modes of ${\bf E}, {\bf B}$ fields staggered for the Maxwell-solver and only destagger them in a separate instance when needed.

We always deposit the ${\bf J}_{2m}$ current density modes on the primary grids Eq. (\ref{eq:spatialAM_grid}), for technical reasons (see Section \ref{subsec:particlesAM_current}), therefore their $J_\rho$ component needs to be staggered.
Before the latter, however, we may filter the current with ${\rm F}_z$ banded diagonal filter matrices \cite{majorosi2026exponentialpic}, in the general form of:
\begin{align} \label{eq:spatialAM_filter_Jr}
\overline{J}_{\rho,2m} &=  {\rm S}_{\rho}^{(-,mA)}  {\rm F}_z  J_{\rho,2m} , \\  
\overline{J}_{\theta,2m+1} &=   {\rm F}_z  J_{\theta, 2m+1} ,\quad
\overline{J}_{z, 2m} =  {\rm F}_z {\rm F}_{z}^{(\text{div})} J_{z,2m},  \label{eq:spatialAM_filter_Jz}
\end{align}
where the Cartesian component may include an  optional integral filter ${\rm F}_{z}^{(\text{div})}$ to augment PIC charge conserving deposition in that direction \cite{majorosi2026exponentialpic}.

We discretize the real form of Maxwell-equations with azimuthal modes decomposition discussed in Section \ref{subsec:MaxwellAM} for $2m$ and $2m+1$ indices.  The Amp\'ere equations 
(\ref{eq:maxwellAM_Er})-(\ref{eq:maxwellAM_Ez}) become:
\begin{align}
\partial_t E_{\rho_,2m}     &= -{\rm D}_{z}^{(+)} B_{\theta,2m} - m\rho_{(-)}^{-1}B_{z,2m+1} - \overline{J}_{\rho,2m} \label{eq:spatialAM_maxwell_Er}\\
\partial_t E_{\theta,2m+1}  &=  {\rm D}_{z}^{(+)} B_{\rho,2m+1} -{\rm D}_{\rho}^{(+,mS)} B_{z,2m+1} - \overline{J}_{\phi,2m+1},\label{eq:spatialAM_maxwell_Ephi}\\
\partial_t E_{z,2m}        &=  m\rho^{-1} B_{\rho,2m+1} +{\rm D}_{\text{div},\rho}^{(+,mS)} B_{\theta,2m} - \overline{J}_{z,2m}.  \label{eq:spatialAM_maxwell_Ez}
\end{align}
The Faraday-equations  (\ref{eq:maxwellAM_Er})-(\ref{eq:maxwellAM_Ez}) are written as:
\begin{align}
\partial_t B_{\rho_,2m+1}    &= {\rm D}_{z}^{(-)}  E_{\theta,2m+1} -m\rho^{-1}E_{z,2m} \label{eq:spatialAM_maxwell_Br}\\
\partial_t B_{\theta,2m}  &=  -{\rm D}_{z}^{(-)}  E_{\rho,2m} +{\rm D}_{\rho}^{(-,mS)} E_{z,2m} ,\label{eq:spatialAM_maxwell_Bphi}\\
\partial_t B_{z,2m+1}        &=  m\rho_{(-)}^{-1} E_{\rho,2m} -{\rm D}_{\text{div},\rho}^{(-,mS)}  E_{\theta,2m+1}. \label{eq:spatialAM_maxwell_Bz}
\end{align}
 The extra difficulty here that we have to keep track of both the $m$ dependent radial boundary conditions Eq. (\ref{eq:spatialAM_derivativeS}) and the staggering. The operator $\rho$ just multiplies with the respective grid values. We note that all field values needed at $\rho = 0$ are eliminated in the numerical system by the boundary conditions of Section \ref{subsec:spatialAM_boundary}.

The gradient, divergence operators, and Poisson's equation Eqs. (\ref{eq:maxwellAM_gradphi})-(\ref{eq:maxwellAM_laplace2D}) are discretized as:
\begin{align}
\left[ \nabla \phi \right]_{2m} &= \left( {\rm D}_{\rho}^{(-,mS)} \phi_{2m} \right) {\bf \hat{e}_\rho} - \left( m \rho^{-1} \phi_{2m+1} \right) {\bf \hat{e}_\theta} + \left( {\rm D}_z^{(-)} \phi_{2m} \right) {\bf \hat{e}_z} \label{eq:spatialAM_gradphi}\\
\left[ \nabla \cdot {\bf E} \right]_{2m}  &=  {\rm D}_{\text{div},\rho}^{(+,mS)} E_{\rho,2m} - m \rho^{-1} E_{\theta,2m+1} + {\rm D}_z^{(+)}  E_{z,2m} \label{eq:spatialAM_divE} \\
\left[ \nabla^2 \phi \right]_{2m}  &= {\rm L}_{\rho}^{(mS)} \phi_{2m} - m^2 \rho^{-2} \phi_{2m}+ {\rm L}_{z} \phi_{2m} = -\varrho_{2m}, \label{eq:spatialAM_Poisson}
\end{align}
As we noted earlier these are half of the equations, to get the other half we have to interchange the basis function indices $2m$ and $2m+1$ and flip the sign of $m$ in the coefficients. We also note that real coordinate stretching can be incorporated into the radial banded diagonal matrices (see \cite{majorosi2026exponentialpic}). Absorbing layers can be included the same way as we discussed with our Cartesian solver. 

\emph{Our exponential solution can be summarized as follows:} At the initialization step we calculate all banded diagonal matrices needed in Eqs. (\ref{eq:spatialAM_stagger_E})-(\ref{eq:spatialAM_Poisson})
in accordance with the descriptions in Sections \ref{subsec:spatialAM_deriv}-\ref{subsec:spatialAM_boundary} at a selected spatial order and boundaries, and we do the azimuthal modes expansion and staggering Eqs. (\ref{eq:maxwellAM_fourier}), (\ref{eq:spatialAM_stagger_E})-(\ref{eq:spatialAM_stagger_Bth}). To step forward in time with step $\Delta t$, we compute the midpoint solution
Eq. (\ref{eq:maxwellAM_solution2}) with the filtered current sources Eqs. (\ref{eq:spatialAM_filter_Jr})-(\ref{eq:spatialAM_filter_Jz}) using 4th, 8th,
12th or higher order Taylor-expansion of the exponentials as $\sum_{n = 0}^{N}\frac{\Delta t ^n}{n!} H^{n}_m \Psi_{2m}$.
The definition of $H_m$ follows from  Eqs. (\ref{eq:spatialAM_maxwell_Er})-(\ref{eq:spatialAM_maxwell_Bz}). Any matrix-vector
product we encounter we compute in accordance with Eqs. (\ref{eq:bandedmatricesAM_Ar})-(\ref{eq:bandedmatricesAM_Az}).
We call this method finite difference exponential time domain (FDETD),
or finite difference exponential solution. 

We implemented this algorithm alongside our Cartesian code in C++ language using OpenMP threads on CPU and using Nvidia CUDA on GPU. Computationally, this method mirrors our Cartesian algorithm closely (see \cite{majorosi2026exponentialpic}), with the same parallelization and domain decomposition properties. Since no radial transformation to the Bessel basis (complexity $\sim M_\rho^2$) is required in it (complexity $\sim N_{D} M_\rho$) high resolution radial grids are supported. 

The disadvantage is that nearby the $\rho = 0$ axis the solution is not perfectly smooth. Also, the matrix norm of $H_m$ increases with larger $m$, which means that larger $\kk_\bot$ Bessel eigenmodes appear in modal coefficients localized near the axis. This worsens the temporal stability, but can be compensated by increasing the order of the exponential Taylor-expansion. But in principle arbitrary accuracy can be achieved spatially and temporally.

\newpage



\section{Particle-in-cell method in AM representation} \label{sec:particlesAM}

\subsection{Particle push}
We store and push the ${\bf r}_\pp$ particle positions and ${\bf u}_\pp$ reduced momentum vectors in 3D Cartesian coordinates  identical to our Cartesian code \cite{majorosi2026exponentialpic}. Notable point of difference is that we have to convert the cylindrical field components to Cartesian coordinates after interpolation,  and we have to convert the particle velocity ${\bf v}_\pp$ components to cylindrical coordinates during current deposition. For simplicity, the index $\pp$ always refers to a particle of species $\sss$. We denote charge and mass of these particles with $\qq_\sss$ and $\mm_\sss$, respectively.

\subsection{Angular particle shape} \label{subsec:particlesAM_angular}
\newcommand{\sone}{(1)}

We define the angular shape as Dirac-delta function $\delta(\theta-\theta_p)$ at angular coordinate $\theta_\pp = \arctan \left(y_{\pp} /x_{\pp} \right)$ of a particle $\pp$ - which is numerically done by trigonometric interpolation \cite{BOOK_NUMERICAL_RECIPIES}. In the framework of the azimuthal modes decomposition for even and odd modal coefficients this shape is written as \cite{lifschitz2009pic_fourier_decomposition, davidson2015OSIRIS_RZ}:
\begin{equation}
\hat{S}_{m'}^{(l)} (\theta_\pp) =
\begin{cases}
 (2\pi)^{-l} \delta_{m = 0} +  \pi^{-l}\delta_{m \neq 0}  \cos(m\theta_\pp), &\text{if } m'=2m  \\
 \pi^{-l}\delta_{m \neq 0}  \sin(m\theta_\pp),  &\text{if } m'=2m+1
\end{cases}
\label{eq:shapesAM_M}
\end{equation}
where $\delta$ denotes the Kronecker delta symbol, which equals 1 if the relation in the subscript is true, 0 otherwise.  The angular shape is to be evaluated for all  modal coefficients $m'\in [0, 2M-1]$ resulting an array of size $2M$.
We also introduced the superscript $(l)$ to select proper normalized form of the angular shape during interpolation ($l=0$) or deposition $(l=1)$.

To optimize the numerical evaluation of trigonometric functions for higher $m$s  we use the addition theorems of $\cos(\alpha+\theta_\pp)$ and $\sin(\alpha+\theta_\pp)$ with $\alpha = (m-1)\theta_\pp$ as others did before \cite{davidson2015OSIRIS_RZ}.

\subsection{Angular sampling} \label{subsec:shapesAM_sampleth}

There is also a nontrivial issue regarding the ordered placement of the particles angularly. Let us denote the position indices of a placed particle $\pp$ along $z,\theta,\rho$ as $i''$,$j''$, $k''$ respectively as they can be placed more densely than the nodes of the primary grid. Let us place $M_{\theta}'$ particles in a given $i'', k''$ ring, then the ordered angular placement reads:
\begin{equation}
\theta_{i'',j'', k''} = \theta_{i'',k''}^{(0)} + \left( j'' + \frac{1}{2} \right) \frac{2\pi}{M_\theta'}, \quad j''\in \left[0, M_{\theta}'-1 \right],
\label{eq:shapesAM_sampleth}
\end{equation}
where we can choose an angle offset $\theta_{i'',k''}^{(0)} \in \left[ -\pi/M_{\theta}', \pi/M_{\theta}' \right]$. We implemented a scheme in order to allow better sampling of the angular phase space, in which we can choose $\theta_{i'',k''}^{(0)}$ randomly from uniform distribution for each $i'', k''$.\footnote{To our knowledge FBPIC and other codes assume that this random angle offset is the same within a cell such that $\theta_{i'',k''}^{(0)}=\theta_{i,k}^{(0)}$.} By using this sampling method we can improve the angular sampling with the same $M_\theta'$ just by increasing the resolution or by increasing the number of particles in the cells along $z,\rho$, while we retain noiseless density initially. By increasing the number of particles this way we can reduce artifacts related to the  undersampling of the 3D space. A direct analogy of Eq. (\ref{eq:shapesAM_sampleth}) can also be used in 3D along a chosen Cartesian direction to reduce ordered grid artifacts.

\subsection{Field interpolation} \label{subsec:particlesAM_interpolation}

Field interpolation steps follow the same main steps as our Cartesian code \cite{majorosi2026exponentialpic}. First, we destagger the modes of the ${\bf E}$, ${\bf B}$ fields to the primary grids Eq. (\ref{eq:spatialAM_grid}) in direct analogy to Eqs. (\ref{eq:spatialAM_stagger_E})-(\ref{eq:spatialAM_stagger_Bth}):
\begin{align} \label{eq:shapesAM_destagger_E}
E_{\rho, m'}^{(C)} &\approx {\rm S}_\rho^{(+, mA)} E_{\rho,m'},  \quad E_{z,m'}^{(C)} \approx {\rm S}_{z}^{(+)} E_{z,m'}, \\
B_{\rho, m'}^{(C)} &\approx {\rm S}_z^{(+)} B_{\rho,m'},\quad\quad  B_{z,m'}^{(C)} \approx {\rm S}_\rho^{(+, mS)}  B_{z,m'},  \label{eq:shapesAM_destagger_B} \\
\quad B_{\theta,m'}^{(C)} &\approx {\rm S}_\rho^{(+, mA)} {\rm S}_z^{(+)} B_{\theta,m'}, \label{eq:shapesAM_destagger_Bth}
\end{align}
Our supersampling scheme \cite{majorosi2026exponentialpic} which doubles the resolution of the fields before interpolation is also supported in $z$ and $\rho$ directions.

The interpolation formula for all the $\phi_{m',i,k}$ field modes reads as follows:
\begin{equation} \label{eq:shapesAM_interpolate}
{\phi}_{\pp} = \Delta z \Delta \rho \sum _{m'=0}^{2M-1} \sum_{k} \hat{S}^{(n)} (z_i-z_\pp) \hat{S}^{(n)} (\rho_k-\rho_\pp) \hat{S}_{m'}^{(0)}(\theta_\pp) \phi_{m',i,k},
\end{equation}
where $\rho_\pp = \sqrt{x_\pp^2+y_\pp^2}$ is the particle's radial position and $\hat{S}^{(n)}$ denotes a standard PIC particle shape of order $n$. We use 3rd order particle shapes as default both in $z$ and $\rho$ directions, therefore we assume from now on that these are the same. During interpolation we ignore the variations of the local cell volume. Thus, during interpolation we treat the radial shape as if it were a Cartesian shape. The interpolation algorithm here has the same parallelization properties as the one in our Cartesian code.

To interpolate the ${\bf E}$, ${\bf B}$ field modes the formula Eq. (\ref{eq:shapesAM_interpolate}) has to be repeated 6 times for destaggered fields Eqs. (\ref{eq:shapesAM_destagger_E})-(\ref{eq:shapesAM_destagger_Bth}). However, we need to take into account the $mS$ or $mA$ symmetric boundary behavior of their modes by extending them to $\rho < 0$ by a couple of cells in accordance with Table \ref{tab:spatialAM_boundary}. Then we perform  the  $k$ sum in Eq. (\ref{eq:shapesAM_interpolate}) on the corresponding extended grid.

\subsection{Density deposition} \label{subsec:particlesAM_density}
For the density deposition in cylindrical coordinates the local cell volume must be taken into account. This has nontrivial consequences regarding the a particle's conservation of charge at arbitrary positions.

We define the normalized radial density shape of a particle as the following on the primary grid Eq. (\ref{eq:spatialAM_grid}):
\begin{equation} \label{eq:shapesAM_R}
\hat{S}_\rho (\rho_k,\rho_\pp) = \mathcal{N}_\pp^{-1} \left| \rho ^{\epsilon-1}_k \right| \hat{S}^{(n)}({\rho_k-\rho_\pp}),
\end{equation}
with $\mathcal{N}_\pp$ normalization factor
\begin{equation} \label{eq:shapesAM_RN}
\mathcal{N}_\pp = \sum_{k} h_k  \left| \rho^{\epsilon}_k \right| \hat{S}^{(n)}({\rho_k-\rho_\pp}) \Delta \rho_k,
\end{equation}
where the $\epsilon$ parameter fine tunes the particle shape ($0 \leq \epsilon \leq 1$). If a coordinate transformation is applied to the radial coordinate then $\Delta \rho_k = g_k \Delta \rho$, where $g_k$ is the discretized transformation function.  We use the same $\hat{S}^{(n)}$ particle shape as during interpolation.

The factor $\mathcal{N}_\pp$ is responsible for conserving the charge with the quadrature coefficients $h_k$ of the integral $\int_{0}^{\infty} {\rm d}\rho$. The usual choice \cite{lifschitz2009pic_fourier_decomposition, lehe2016fbpic} of $h_k = 1$ is only \emph{second order accurate} at $\rho = 0$. This is consistent with the second order Yee-solvers, but not for high-order solution - it breaks the charge conservation on the density level. We managed to improve this by fitting 4th order polynomials on points $\rho = 0, \Delta \rho /2,  3\Delta \rho /2,  5\Delta \rho /2$ yielding the quadrature coefficients:
\begin{equation} \label{eq:shapesAM_Rh}
h_0 = 7/8, \quad h_1 = 73/72, \quad h_{k>1} = 1,
\end{equation}
which is a 4th order accurate quadrature. If we look closely at the above coefficients versus $h_k=1$ the charge of a particle moving to $\rho = 0$ axis fluctuates more than $10$\% with the latter. The above formula yields two orders of magnitude improvement for this - but it is still not exact on the high order Lagrange-polynomial basis.

 \begin{figure}[t]
\centering
\includegraphics[trim={20 10 60 0},clip, height=6.5cm]{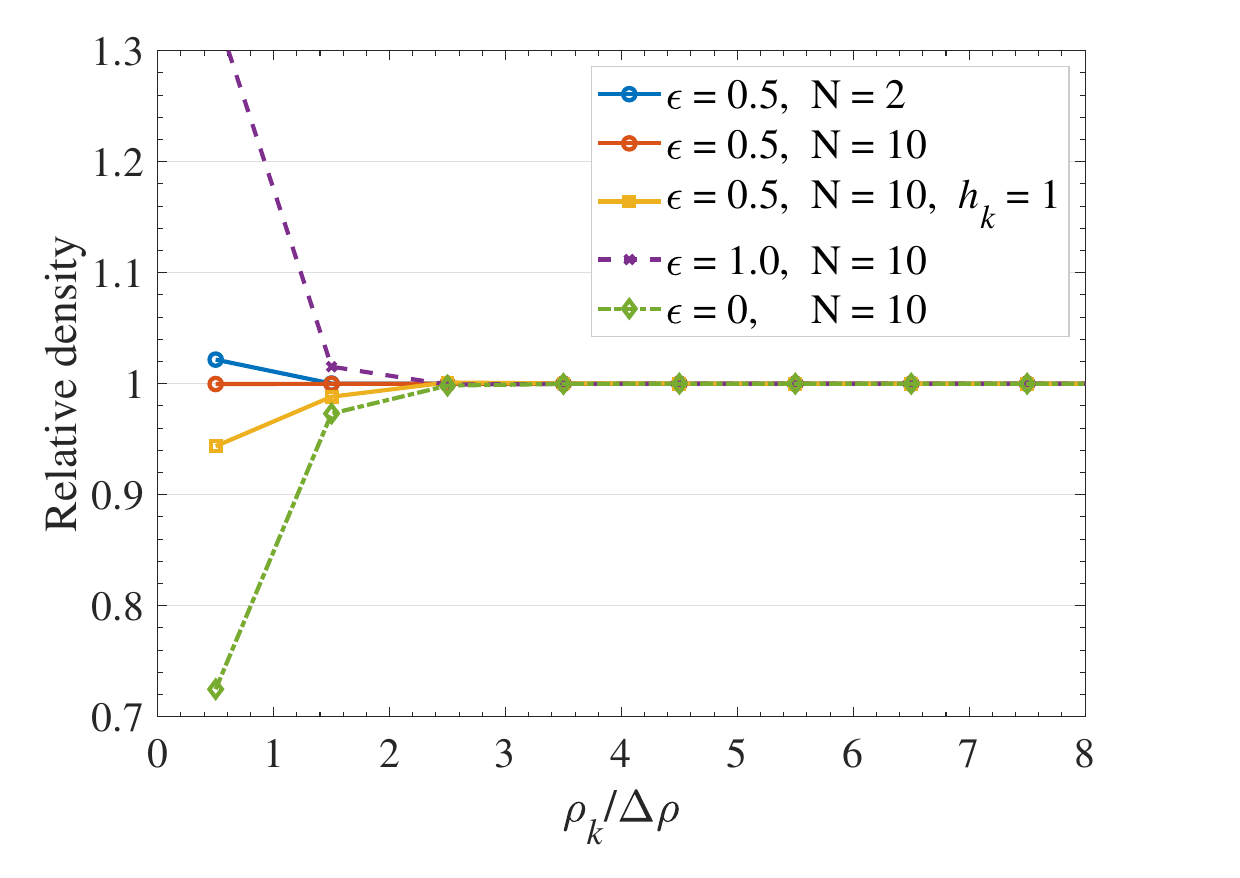}
\caption{
Results of  density depositions of a radially uniform density $\varrho(\rho_k) = 1$ using Eq. (\ref{eq:shapesAM_R}) with different $\epsilon$ parameters, and $N$ particles per radial cell. For the orange curve we used 2nd order quadrature $h_k = 1$ instead of Eq. (\ref{eq:shapesAM_Rh}).
\label{fig:particlesAM_densityR}}
\end{figure}

A tangentially related problem for the latter is that when we uniformly sample  a uniform density with particles radially and then we deposit the particles' radial density using the shape Eq. (\ref{eq:shapesAM_R}), the result is not uniform - contrary to Cartesian coordinates. This issue occurs at couple of grid points near the $\rho = 0$ axis. As we found choosing $\epsilon=1/2$ in Eq. (\ref{eq:shapesAM_R}) and using the quadrature Eq. (\ref{eq:shapesAM_Rh}) solves this issue of the particle representation with 3rd order shape. We summarize the results of depositing uniform density in Fig. \ref{fig:particlesAM_densityR}. We can see that choosing $\epsilon=0$ (as others did in quasi 3D codes \cite{lifschitz2009pic_fourier_decomposition, lehe2016fbpic} and fix this issue another way) yields suboptimal results with more than $20$\%  error in the density. Our improved radial shape is also not perfect but it can yield orders of magnitude improvement when we increase the number of particles per cell radially.\footnote{We speculate that $\epsilon=1/2$ is optimal because the radial Laplacian can be made symmetric with the $\nabla_\rho^2 = \rho^{-1/2}\left( \partial_\rho^2+\rho^2/4 \right)\rho^{1/2}$ transform.}


Combining the angular shape Eq. (\ref{eq:shapesAM_M}) and the radial shape (\ref{eq:shapesAM_R}) results in the following deposition formula for the charge density:
\begin{equation} \label{eq:shapesAM_deposit}
{\varrho}_{m',i,k} =  \sum_{\pp}  \qq_\sss   w_\pp \hat{S}{(z_i-z_\pp)} \hat{S}_\rho (\rho_k,  \rho_\pp) \hat{S}_{m'}^{\sone}(\theta_\pp).
\end{equation}
There is another peculiar property of the cylindrical density deposition near the  $\rho = 0$ axis: the particle shape Eq. (\ref{eq:shapesAM_R}) are also deposited at $\rho < 0$ and then the latter are reflected back according to the $mS$ boundary conditions of Section \ref{subsec:spatialAM_boundary}  (and $mA$ for the $J_\rho,J_\theta$ current) - as if there were particles mirrored on the other side. This means that the angular and the radial density shapes are not separable at $\rho=0$ axis due to the boundary conditions. We also choose for all $m>1$ modes antisymmetric reflection so that the deposited density of a particle at $\rho_\pp=0$ is zero in any $m > 1$ mode to improve angular smoothness.

\subsection{Current deposition} \label{subsec:particlesAM_current}

We deposit $J_\rho$, $J_\theta$ current density components based on the density formula Eq. (\ref{eq:shapesAM_deposit}) as:
\begin{equation} \label{eq:shapesAM_depositJr}
{J}_{\rho,m',i,k} =  \sum_{\pp}  \qq_\sss   w_\pp  {v}_{\rho,\pp} \hat{S}{(z_i-z_\pp)} \hat{S}_\rho (\rho_k,  \rho_\pp) \hat{S}_{m'}^{\sone}(\theta_\pp),
\end{equation}
\begin{equation} \label{eq:shapesAM_depositJth}
{J}_{\theta,m',i,k} = \sum_{\pp}  \qq_\sss   w_\pp  {v}_{\theta,\pp} \hat{S}{(z_i-z_\pp)} \hat{S}_\rho (\rho_k,  \rho_\pp) \hat{S}_{m'}^{\sone}(\theta_\pp).
\end{equation}
We found that the above  $J_\rho,J_\theta$ cylindrical currents  are not well-defined enough because of the radial-angular coupling at $\rho = 0$ to have enough accuracy for LWFA simulations - they could induce spurious divergence near the axis.
We wrote this current deposition method such that the divergence corrections in $x,y$ are separated (see Section \ref{subsec:particlesAM_correction}) from the Cartesian direction.

We adapt modified Esirkepov current deposition algorithm for $z$ component as in \cite{majorosi2026exponentialpic}. First we need to introduce the Cartesian particle offset positions:
\begin{equation} \label{eq:shapesAM_positionES}
z_\pp^{(\pm)}  = z_\pp \pm \frac{\Delta t}{2} {v_{z,\pp}}, 
y_\pp^{(\pm)}  = y_\pp \pm \frac{\Delta t}{2} {v_{y,\pp}},
x_\pp^{(\pm)}  = x_\pp \pm \frac{\Delta t}{2} {v_{x,\pp}},
\end{equation}
\begin{equation} \label{eq:shapesAM_positionESAM}
\theta_\pp^{(\pm)} = \arctan \left(  y_\pp^{(\pm)}/x_\pp^{(\pm)} \right) , \quad
\rho_\pp^{(\pm)}  = \sqrt{\left( x_\pp^{(\pm)} \right)^2+\left(y_\pp^{(\pm)} \right)^2}.
\end{equation}

The staggered current component $J_z$ is recovered from the second order continuity equation in direction $z$ as:
\begin{multline} \label{eq:shapesAM_depositZ}
{J}_{z,m',i-\frac{1}{2},k} =   \sum_{\pp}  \frac{1}{2}  \qq_\sss  w_\pp \delta \hat{S}_{z} (z_i,z_\pp) \times \\
\left[ 
\hat{S}_\rho\left(\rho_k, \rho_\pp^{(+)}\right)\hat{S}_{m'}^{\sone}\left(\theta_\pp^{(+)}\right)+ 
\hat{S}_\rho\left(\rho_k, \rho_\pp^{(-)}\right)\hat{S}_{m'}^{\sone}\left(\theta_\pp^{(-)}\right)
\right],
\end{multline}
where we introduced the following integrated one dimensional differential  particle shape:
\begin{equation} \label{eq:shapesAM_shapeJZ}
\delta \hat{S}_{z} (z_i,z_\pp) = \frac{1}{\Delta t} \sum_{i' = 0}^{i-1} \left[ \hat{S}\left(z_{i'}-z_\pp^{(+)} \right)-\hat{S}  \left(z_{i'}-z_\pp^{(-)} \right)  \right] \Delta z.
\end{equation}
This is a discretized integral using second order midpoint quadrature, which also staggers down the result by $-\Delta z/2$. For high order divergence corrections this ${J}_z$ needs to be multiplied with a banded diagonal matrix ${\rm F}_z^{(\text{div})}$ 
that contains higher order integral quadrature coefficients. This solves the integral continuity equation in high order in the same way as in our Cartesian code.

\subsection{Current correction} \label{subsec:particlesAM_correction}

We found that we need to do a current correction step to improve of the imperfect cylindrical $J_{\rho},J_{\theta}$ currents near the $\rho = 0$ axis. Unlike FBPIC \cite{lehe2016fbpic} we do not want to solve the full Poisson's equation to correct divergence of $\bf E$, we must avoid nonlocal numerical corrections whenever it is possible. 

We propose an  Esirkepov like deposition to satisfy the continuity equations for $\bf J$ in the Cartesian $z$ direction and in the $x,y$ plane in a separated manner. We gave the deposition formula of $J_z$ in the previous Section with this in mind. To do the current  correction in the transverse plane, we use the same Cartesian method.

First, we deposit $J_\rho$ and $J_\theta$ according to Eqs. (\ref{eq:shapesAM_depositJr}), (\ref{eq:shapesAM_depositJth}). We also need to deposit the following density differential in the $x,y$ plane:
\begin{multline} \label{eq:shapesAM_currentES_XY}
{\varrho}_{xy,m',i,k} =  \Delta t^{-1} \sum_{\pp}  \frac{1}{2}  \qq_\sss  w_\pp \left[ \hat{S}\left(z_i-z_\pp^{(+)} \right)+\hat{S}  \left(z_i-z_\pp^{(-)} \right)  \right] \times \\
\left[ 
\hat{S}_\rho\left(\rho_k, \rho_\pp^{(+)}\right)\hat{S}_{m'}^{\sone}\left(\theta_\pp^{(+)}\right)- 
\hat{S}_\rho\left(\rho_k, \rho_\pp^{(-)}\right)\hat{S}_{m'}^{\sone}\left(\theta_\pp^{(-)}\right)
\right].
\end{multline}
We must include the contribution from all species $\sss$  into $\varrho_{xy}$ and $\bf J$. Then we have to apply the current filters to the modes of $\bf J$ and $\varrho_{xy}$ in accordance with Eqs. (\ref{eq:spatialAM_filter_Jr})-(\ref{eq:spatialAM_filter_Jz}). We denote these after filtering as $\overline{\bf J}$ and $\overline{\varrho}_{xy}$.

In the following the $i$ indices play no role, we omit them for simplicity. The planar divergence error that needs to be corrected can be computed from the expression $ \delta \tilde{\varrho} = \overline{\varrho}_{xy}-\nabla_\bot \cdot \overline{\bf J}$ as:
\begin{align} \label{eq:shapesAM_currentRHS2m}
\delta \tilde{\varrho}_{2m} &= \overline{\varrho}_{xy,2m} - {\rm D}^{(+, mS)}_{\text{div},\rho} \overline{J}_{\rho,2m} + m \rho^{-1} \overline{J}_{\theta,2m+1}, \\
\delta \tilde{\varrho}_{2m+1} &= \overline{\varrho}_{xy,2m+1} - {\rm D}^{(+, mS)}_{\text{div},\rho} \overline{J}_{\rho,2m+1} - m \rho^{-1} \overline{J}_{\theta,2m}. \label{eq:shapesAM_currentRHS2m1}
\end{align}

We solve the planar Poisson's equation for all mode indices $2m$ and $2m+1$ to recover the $\tilde{\phi}$ scalar potential of the divergence correction in $x,y$ plane as:
\begin{align} \label{eq:shapesAM_laplace2m}
\left[ {\rm L}_{\rho}^{(mS)} -m^2 \rho^{-2} \right]  \tilde{\phi}_{2m}  &= \delta \tilde{\varrho}_{2m}, \\
\left[ {\rm L}_{\rho}^{(mS)} -m^2 \rho^{-2} \right]  \tilde{\phi}_{2m+1}  &= \delta \tilde{\varrho}_{2m+1}.  \label{eq:shapesAM_laplace2m1}
\end{align}

Within the azimuthal modes decomposition the above equation separates according to the mode index $m$, which means that the LHS coefficient matrix of the above equation becomes a one dimensional banded diagonal matrix. Then we can efficiently compute $\tilde{\phi}_{2m}$, $\tilde{\phi}_{2m+1}$ using the LU decomposition of this matrix. For the best numerical stability we use the radial Lagrange operator that is calculated from the finite differences of second order derivatives Eq. (\ref{eq:spatialAM_laplaceR}).

The corrected current components are recovered using $\nabla_\bot\tilde{\phi}$:
\begin{align} \label{eq:shapesAM_currentJrho}
\tilde{J}_{\rho,2m} &= \overline{J}_{\rho,2m} + {\rm D}^{(-, mS)}_{\rho} \tilde{\phi}_{2m}, \\
\tilde{J}_{\rho,2m+1} &= \overline{J}_{\rho,2m+1} + {\rm D}^{(-, mS)}_{\rho} \tilde{\phi}_{2m+1},\\
\tilde{J}_{\theta,2m  } &= \overline{J}_{\theta,2m  } - m \rho^{-1} \tilde{\phi}_{2m+1}, \\
\tilde{J}_{\theta,2m+1} &= \overline{J}_{\theta,2m+1} + m \rho^{-1} \tilde{\phi}_{2m}. \label{eq:shapesAM_currentJth}
\end{align}
These steps conclude our current correction algorithm. This method works because the deposited charge density in Eq. (\ref{eq:shapesAM_currentES_XY}) is  better defined near the $\rho = 0$ axis than the cylindrical currents Eq. (\ref{eq:shapesAM_depositJr})-(\ref{eq:shapesAM_depositJth}), thus this correction method improves the physics there. The use of the  improved quadrature formula Eq. (\ref{eq:shapesAM_Rh}) in the former is also necessary to accomplish this.

With this method we introduced (possibly) nonlocal corrections due to the Poisson's equations (\ref{eq:shapesAM_laplace2m})-(\ref{eq:shapesAM_laplace2m1}) but these are such only in the $\rho$ direction. This also means that domain decomposition is no longer feasible at this dimension (because of the matrix inversion) and we can only perform the above algorithm concurrently according to $i,m$ indices. In all deposited quantities
we omit the "imaginary part" of $m = 0$ coefficient in Eq. (\ref{eq:maxwellAM_fourier}) entirely, since no charge conservation is satisfied there.

\subsection{Current filtering} \label{subsec:particlesAM_filter}

Filtering of the deposited current $\bf J$ is necessary to improve smoothness, and to suppress numerical Cherenkov radiation. In the Cartesian direction we use the same filtering methods as \cite{godfrey2013PIC_stability, majorosi2026exponentialpic} in Eqs. (\ref{eq:spatialAM_filter_Jr})-(\ref{eq:spatialAM_filter_Jz}) to suppress these effects. Radially, however, these are not applicable, because they could break the charge conservation. The suggestion of Lehe et al. \cite{lehe2016fbpic} is to use binomial smoothing on the Bessel basis of eigenvalues $\kk_\bot$. This is not applicable in real space - so we did not include such current filtering. One possibility is to use a diffusive operator $\exp \left(\alpha \nabla^2_{\bot,m}\right)$ from Eqs. (\ref{eq:maxwellAM_laplace2D}) and (\ref{eq:spatialAM_laplaceR}) to suppress high $\kk_\bot$ eigenfunctions. The peculiarity of the real space solution is that for $m>0$ highest $\kk_\bot$ functions appear localized at $\rho_0 = \Delta\rho/2$. By using the knowledge of the asymptotic behaviors in Table \ref{tab:spatialAM_boundary} we can smooth these modes by enforcing these asymptotes. For example, by setting $\varrho_{2m,i,0}\approx 3^{-m}\varrho_{2m,i,1}$, $J_{\rho,2m,i,0}\approx 3^{-|m-1|} J_{\rho,2m,i,1}$  etc. at $k=0$ (as $\rho_0/\rho_1=1/3$) we can achieve a mild radial-angular smoothing effect.


\section{Laser potential envelope model} \label{sec:laserA}

\subsection{Propagation of a laser potential} \label{subsec:laserA}

 It is a useful modeling tool to propagate the vector potential $\bf A$ of the laser field instead of the Maxwell-equations - the common use cases are field propagation in wave guides and the laser envelope model within plasmas \cite{esarey2009laser_plasma_accelerators}, the latter discussed later in this paper.
    First let us introduce $A$ as the scalar "laser potential" assuming a fixed transverse polarization vector $\hat{\bf p}$ of the laser field propagating in the $z$ direction as:
\begin{equation} \label{eq:laserA}
 {\bf A} = \hat{\bf p} A,  \quad {\rm and \quad } \hat{\bf p} = p_x \hat{\bf e}_x + p_y \hat{\bf e}_y,
\end{equation}
with any real or complex components as long as $|\hat{\bf p}| = 1$. From a real laser potential $A$ we can get the transverse part of the $\bf E$ electric and $\bf B$ magnetic fields as
\begin{equation} \label{eq:laserA_EB}
{\bf E} = -\hat{\bf p} \partial_t {A},  {\quad \rm and \quad } {\bf B} = \hat{\bf p} \times \nabla A.
\end{equation}
For completeness, the solution of $\nabla \cdot \bf E = \varrho$ as Poisson's equation is required for the longitudinal part of $\bf E$. Depending on the problem we may neglect this latter term or use an approximate analytical solution \cite{li2016fields_tightly_focused, majorosi2023tightlyfocused}, which yields for the longitudinal component $E_z \approx - c^{-1} \left[ \hat{\bf p}\cdot \nabla \right]A$ in vacuum.

This potential will propagate according to the D'Alembert wave equation, which we consider in the following form:
\begin{equation} \label{eq:laserA_equation}
\partial_t^2 A = \nabla^2 A - \chi A,
\end{equation}
where $\chi = \chi({\bf r}, t)$ is the (plasma) susceptibility which depends on self-consistently on the value of $A$, for example, through the PIC particles.

Next we introduce the laser envelope of a carrier wave with wave number $\kk_0=2\pi/\lambda$, which is propagating in the same direction according to
\begin{equation} \label{eq:laserA_envelope}
A({\bf r}, t) = \exp( - \ii \kk_0 (t- z) ) a({\bf r},t).
\end{equation}
Using Eq. (\ref{eq:laserA_envelope}) we can rewrite Eq. (\ref{eq:laserA_equation}) to  become the complex laser envelope equation:
\begin{equation} \label{eq:laserA_envelope_equation}
\partial_t^2 a = \nabla^2 a + 2 \ii \kk_0 \left ( \partial_t a + \partial_z a \right) - \chi a.
\end{equation}
Even though Eq. (\ref{eq:laserA_envelope_equation}) can be the basis of further approximations, we aim to solve it directly with an exponential operator.

To solve Eq. (\ref{eq:laserA_envelope_equation}) we will rewrite it as two coupled first order PDE-s with a new auxiliary variable $\tilde{a}$:
\begin{align}
\partial_t a &=  \tilde{a} ,\label{eq:laserA_envelope_equation_A}\\
\partial_t \tilde{a} &= \nabla^2 a + 2 \ii \kk_0 \partial_z a + 2 \ii \kk_0 \tilde{a} - \chi a.\label{eq:laserA_envelope_equation_At}
\end{align}
Next we introduce $\psi_a$ which is a real 4-component vector that describes the evolution of this complex scalar envelope equation. Then, it formally becomes:
\begin{equation} \label{eq:laserA_envelope_formal}
 \partial_t \psi_a      = \op{H}_a \psi_a,
\end{equation}
where
\begin{equation} \label{eq:laserA_envelope_psi}
\psi_a =
\left( \begin{array}{cccccc}
a_r & a_i & \tilde{a}_r & \tilde{a}_i
\end{array} \right)  ^T ,
\end{equation}
\begin{equation}  \label{eq:laserA_envelope_H}
\op{H}_a =
\left( \begin{array}{cccccc}
0                  & 0      & 1       &      0       \\
0    & 0                   & 0       &  1  \\
 \nabla^2-  \chi & 2\kk_0  \partial_z                   & 0       & 2 \kk_0 \\
-2\kk_0  \partial_z                  &  \nabla^2-  \chi & -2\kk_0       & 0                  \\
\end{array} \right) .
\end{equation} 
The complex auxiliary variable $\tilde{a}=\tilde{a}_r+\ii \tilde{a}_i$ does play an important physical role in the evolution of the electromagnetic state. It describes the transverse electric field envelope via Eq. (\ref{eq:laserA_EB}) such that
\begin{equation} \label{eq:laserA_envelope_E}
|{\bf E}| = \left| \tilde{A} \right| = \left|\tilde{a} -\ii \kk_0 a  \right|,
\end{equation}
where $\tilde{A}=\partial_t A$ is the auxiliary variable for the corresponding wave equation. This is an important distinction in plasma, where only $\left| \tilde{A} \right|$ has direct physical meaning, not $|a|$. We note that if we exclude the non-diagonal  terms that contain $\kk_0$ in Eq. ({\ref{eq:laserA_envelope_H}}) we can solve the original wave equation (\ref{eq:laserA_equation}) with the same method.

The formal solution of Eq. (\ref{eq:laserA_envelope_formal}) in any geometry is \cite{leforestier1991tdsecomparisom, castro2004kohnshampropagators, dijk2014tdsesource}
\begin{equation} \label{eq:laserA_envelope_solution}
\psi_a(t+\Delta t) = \exp \left( \Delta t \op{H}_a \right)\psi_a(t),
\end{equation}
which is exact if $\op{H}_a$ is time-independent. If it is not, then $\op{H}_a$ should be replaced with its temporal average between $t$ and $t+\Delta t$, for example with $\op{H}_a(t+\Delta t/2)$, then Eq. (\ref{eq:laserA_envelope_formal}) becomes second order accurate in $\Delta t$. We then apply explicit Taylor-expansion for the exponential to get arbitrary accurate propagation in vacuum.
In practice, the rapidness of the temporal evolution of $\chi$ will limit the accuracy in laser-plasma simulations.  Of course, higher-order solutions also possible \cite{blanes2009review_magnus_expansion, bandrauk2013splitting}.

\subsection{Discrete laser potential propagation} \label{subsec:spatialAM_potential}
Let us also discuss the discretization of the laser potential equations of Section \ref{subsec:laserA} in the following. The laser potential is a scalar mode discretized on the primary grid Eq. (\ref{eq:spatialAM_grid}), and with radial boundary designation $mS$  in Table \ref{tab:spatialAM_boundary}.

If we write the cylindrical Eqs. (\ref{eq:laserA_envelope_equation_A})-(\ref{eq:laserA_envelope_equation_At}) with azimuthal modes decomposition, then $\chi$ becomes a field matrix that couples different $m$ modes together. This would make using the latter suboptimal, it could be better to use the angular discretization directly. As for our use cases we are only interested to use cylindrically symmetric systems with the laser potential model. In accordance with this, we denote $\chi = \chi_0$. 


We discretize the laser envelope operator Eq. (\ref{eq:laserA_envelope_H}) by substituting the derivatives with centered second and first order difference matrices as $\nabla^2 \approx {\rm L}_\rho + {\rm L}_z$ via Eq. (\ref{eq:spatialAM_laplaceR}) and $\partial_z \approx {\rm D}_z^{(C)}$, which yields for the $m = 0$ mode:
\begin{equation}  \label{eq:spatialAM_envelope_H}
{\rm H}_a =
\left( \begin{array}{cccccc}
0                  & 0      &1        &      0       \\
0    & 0                   & 0       &   1\\
\left( {\rm L}_\rho + {\rm L}_z \right)- \chi_0 & 2\kk_0  {\rm D}_{z}^{(C)}                   & 0       & 2 \kk_0 \\
-2\kk_0  {\rm D}_{z}^{(C)}                  &   \left( {\rm L}_\rho + {\rm L}_z \right)-  \chi_0 & -2\kk_0       & 0                  \\
\end{array} \right) .
\end{equation}
Other modes can also be propagated if we include $-m^2 \rho^{-2}$ term in the discrete Laplacian as long as $\chi_0$ is cylindrically symmetric. We use centered differences ${\rm D}^{(C)}_{z}$ because the propagation of this equation is relatively smooth, it does not interact directly with PIC current sources.

Using Eq. (\ref{eq:spatialAM_envelope_H}) turns the $\op{H}_a$ operator into a matrix. Then exponential solution Eq. (\ref{eq:laserA_envelope_solution}) is expanded using high order Taylor expansion. If $\chi_0$ depends on time, we substitute it with $\chi_0(t+\Delta t/2)$ to form a second order accurate midpoint solution. We can also add absorbing layers to this with couple of extra steps (\ref{subsubsec:absorbing_layerA}). Using this method very high accuracy can be achieved consistent with the Maxwell-solver, even without slowly varying envelope approximation. However, this comes with an increased computational cost.

\subsection{Laser envelope PIC model} \label{subsec:particlesAM_envelope}

We implement the laser envelope model \cite{esarey2009laser_plasma_accelerators, cowan2011envelope_model, benedetti2018envelope_solver, terzani2019envelope} in cylindrical symmetry. Here in addition to the Maxwell-fields, we propagate the laser envelope vector $\psi_a$ with the  exponential method (see Section \ref{subsec:laserA}) both temporally offset by $+\Delta t/2$ compared to the particle properties at $t_n = n\Delta t$. In the laser envelope model the particle dynamics in the laser field is replaced with cycle averaged motion such that the plasma only interacts with the intensity envelope  $|a|^{2}$ which is called ponderomotive guiding center approximation. This leads to substantial performance improvement compared to the usual AM simulations when the laser potential $a$ is cylindrically symmetric, because it induces such a smooth ``averaged`` particle motion that remains cylindrically symmetric. The required longitudinal cell size can also be increased along with the time step since only the $|a|^{2}$ pulse envelope needs to be resolved.

In the following we list the extra steps that we need to incorporate into our PIC method. First the definition of gamma is altered to include the averaged motion:
\begin{equation} \label{eq:envelopeAM_gamma}
\tilde{\gamma}_\pp = \sqrt{1+{\bf u}_\pp^2+ \Phi_\pp}, \quad \text{ with } \quad \Phi_\pp = \frac{\qq^2_\sss}{2 \mm_\sss^2} |a|^2_\pp
\end{equation}
where the interpolated $\Phi_\pp$ is the ponderomotive potential for particle $\pp$ which includes the particle's charge over mass ratio in normalized units, and $|a|^2_\pp$ is the interpolated value of the $|a|^2$ intensity envelope at position ${\bf r}\pp$. It also alters the velocity values as ${\bf v}_\pp = \tilde{\gamma}^{-1}_\pp {\bf u}_\pp$. Implementation wise there is nontrivial consequence: the particle acquires a virtual reduced momentum component $u_{a,\pp} = \sqrt{\Phi_\pp}$, which depends both on the particle position and the laser field. Since we propagate the particles and the fields with leapfrog staggering (at $t_n$ and $t_{n+1/2}$, respectively) the value of $u_{a,\pp}$ must be stored and also predicted in future time indices when interpolation is not available. 

During the deposition step, the cylindrical symmetric density of the susceptibility $\chi_0$ also needs to be deposited along with the charge and current according to the formula:
\begin{equation} \label{eq:envelopeAM_deposit}
{\chi}_{0,i,k,n+1} =   \sum_{\pp} \frac{\qq^2_\sss}{\mm_\sss}  w_\pp \tilde{\gamma}_{\pp,n+1}^{-1}  \hat{S}{(z_i-z_\pp)} \hat{S}_\rho (\rho_k,  \rho_\pp),
\end{equation}
where in $\tilde{\gamma}_{\pp,n+1}$ the value $\Phi_{\pp,n+1}$ also need to be predicted as the fields are at  the temporal midpoints $n+\nicefrac{1}{2}$. We propose the second order predictor: 
\begin{equation}  \label{eq:envelopeAM_uapred}
\Phi_{\pp, n+1} = 2 \Phi_{\pp, n+\nicefrac{1}{2}}-\Phi_{\pp, n}.
\end{equation}
Substituting Eq. (\ref{eq:envelopeAM_deposit}) in the propagator Eq. (\ref{eq:laserA_envelope_solution}) the $\psi_a$ envelope vector can be propagated in second order in $\Delta t$.

For the particle dynamics, a new ponderomotive force field $\tilde{\bf A}=-\nabla |a|^2$ is introduced with components:
\begin{equation} \label{eq:envelopeAM_Afield}
\tilde{A}_\rho = -{\rm D}^{(-,mS)}_{\rho} |a|^2, \quad \tilde{A}_z = -{\rm D}^{(-)}_{z} |a|^2.
\end{equation}
During the field interpolation step the components of $\tilde{\bf A}$ and  $|a|^2$ also have to be interpolated at the particle positions  alongside with the $\bf E$ and $\bf B$ fields.  It can be formally written as an electric field ${\bf E}_{\Phi,\pp}$ :
\begin{equation} \label{eq:envelopeAM_Fpond}
{\bf E}_{\Phi,\pp, n+1/2} = \frac{\qq_\sss}{4 \mm_\sss}  \tilde{\bf A}_{\pp, , n+1/2},
\end{equation}
and  ${\bf F}_\pp = \qq_\sss {\gamma}^{-1}_{\pp} {\bf E}_{\Phi,\pp}$ is the ponderomotive force.


To do a particle push by $\Delta t$ in accordance with the equations of motion, we also need to approximate the temporal average of ${\gamma}^{-1}_{\pp,n}$.
Let us also introduce the following averaged values of the fields for the particle push as:
\begin{equation} \label{eq:envelopeAM_avgEF}
\overline{\bf E}_{\pp,n} = \frac{\qq_\sss \Delta t}{2 \mm_\sss} {\bf E}_{\pp,n+1/2}, \quad
\overline{\bf F}_{\pp,n} = \frac{\qq_\sss \Delta t}{2 \mm_\sss} {\bf E}_{\Phi,\pp,n+1/2}. \quad
\end{equation}

The simplest approach we use is to predict the particles' reduced momenta ${\bf u}_{-}$ at the temporal midpoint by
\begin{equation} \label{eq:envelopeAM_predict}
{\bf u}_{-} = {\bf u}_{\pp,n}+ \overline{\bf E}_{\pp,n}+\gamma_{0}^{-1} \overline{\bf F} \quad \text{and} \quad \gamma_{-} = \sqrt{1+{\bf u}_{-}^2+\Phi_{p,n+1/2}},
\end{equation}
where $\gamma_0 = \sqrt{1+{\bf u}^2_{\pp,n}+\Phi_{p,n+1/2}}$. Then we proceed with a regular particle push such that we add $\gamma_{-}^{-1}{\bf E}_{\Phi,\pp, n+1/2}$ to the electric field and we use the formula of the ponderomotive gamma Eq. (\ref{eq:envelopeAM_gamma}) throughout. For the Boris pusher we show an alternative method in \ref{subsubsec:BorisA}.


\newpage

\section{Benchmarks} \label{sec:BenchmarkAM}

To determine the accuracy characteristics of our PIC method we performed a series of tests, which are heavily inspired by FBPIC paper \cite{lehe2016fbpic}, such as vacuum propagation and linear wake fields. We perform the tests using real azimuthal modes decomposition of the Maxwell-fields (AM), and propagating cylindrical symmetric wave laser potentials (A) and true laser envelope laser potentials (Envelope), in each case. We primarily focus on laser interaction with underdense plasma, where the laser propagation direction is $z$, and is linearly polarized in the $x$ direction. For the laser we use \emph{Gaussian beams} and \emph{Gaussian pulses}\footnote{Gaussian pulse length is not FWHM.} for simplicity. Usually, we also use a moving window that moves in direction $z$ with speed of light $c$. We also solve the Poisson equation for the initial laser fields to get them as correct as numerically possible. We indicate if we use stretched coordinates tranversally \footnote{Stretched coordinates in $\rho$  for some of the tests (see \cite{majorosi2026exponentialpic}): inner zone width without stretching is $10\mum$, transition zone width is $15\mum$,  stretching factor of 2 in the outer zone. \label{foot:trf1}}.   We also use "lowpass A" digital filtering for the current (see \cite{majorosi2026exponentialpic}). We set the initial temperature of the plasma to zero and use ordered sampling for the initial particle positions.

\subsection{Laser pulse propagation in vacuum} \label{subsubsec:test_am_vacuum}

\begin{figure*}[t]
\centering
\includegraphics[trim={5 0 5 0},clip,height=6.4cm]{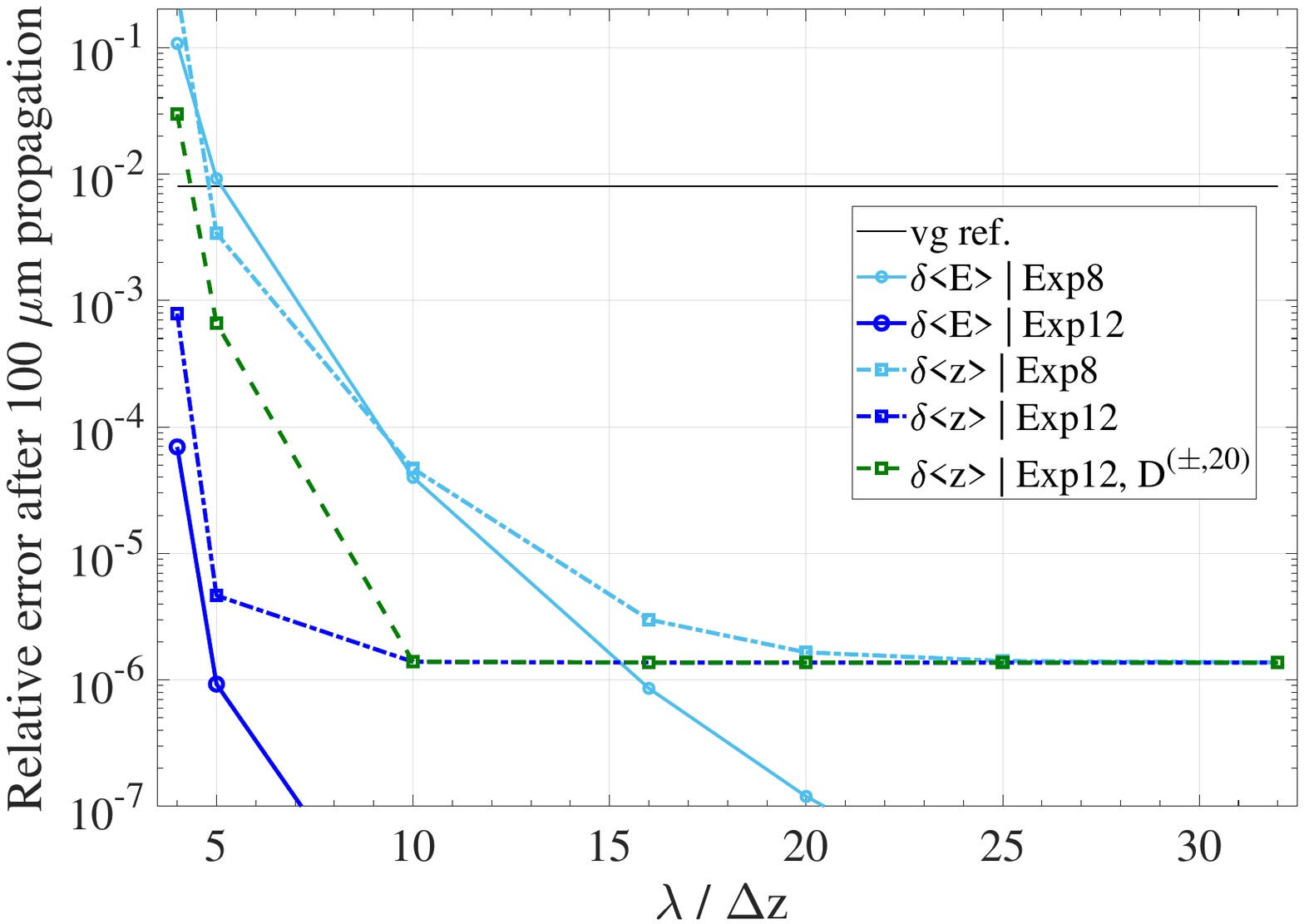}
\includegraphics[trim={5 0 5 0},clip,height=6.4cm]{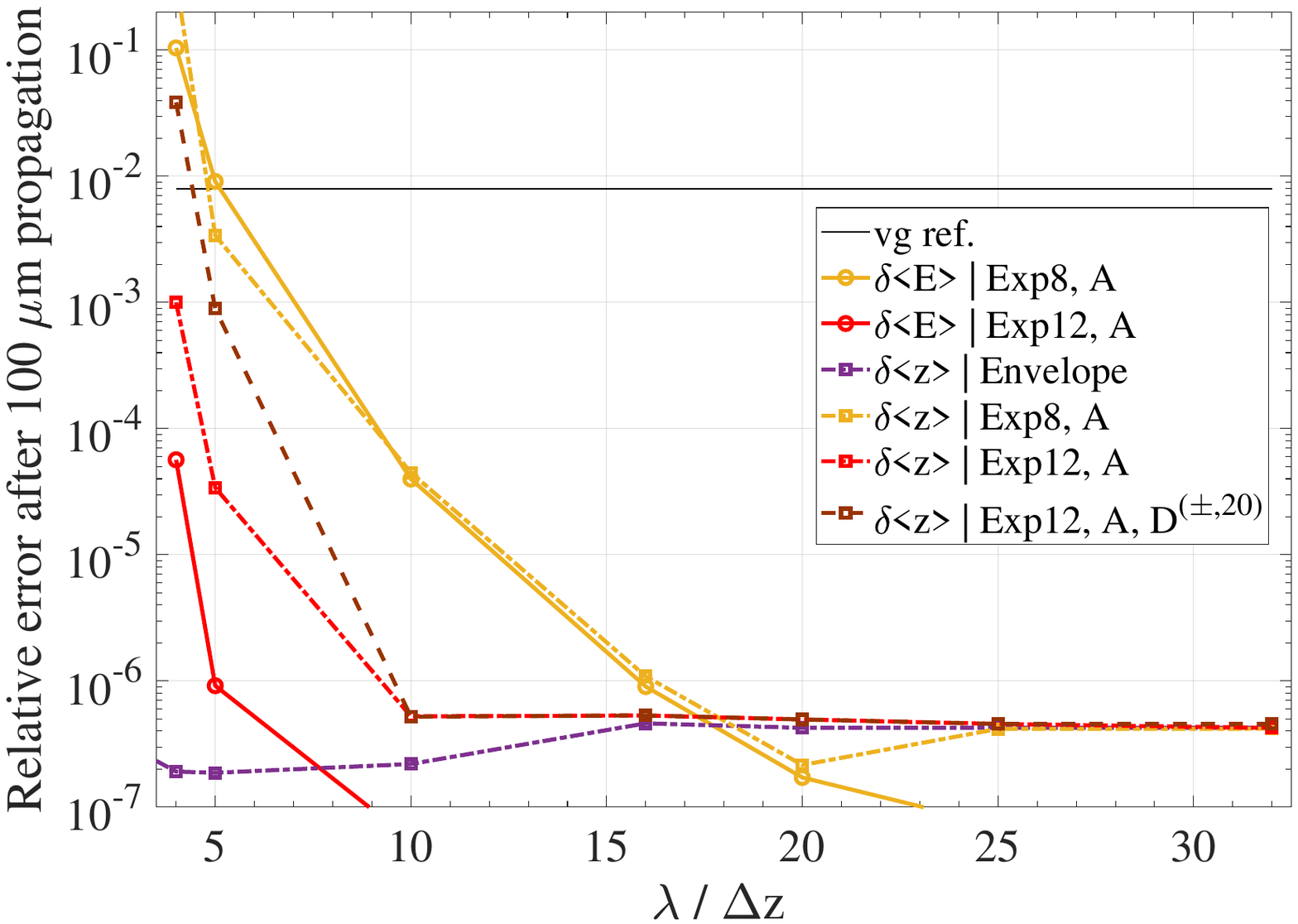}
\caption{Error characteristics of a laser pulse of  $\lambda=0.8\mum$ wavelength in vacuum after $100\mu m$ propagation in cylindrical geometry. We show the relative energy loss $\delta\left<E \right>$ (solid lines, circle markers) and the mean distance error $\delta\left<z\right>$ of the pulse centroid (in $\lambda$ units, dash-dotted lines, square markers) versus the $N = \lambda/\Delta z$ resolution with 8th  and 12th  order exponentials at  $\Delta t = \Delta z /c$ using 30th order finite differences. (We also show the results using 20th order differences with 12th order exponential with dashed lines.) For reference, we plot the value of the analytical centroid shift $t_{\max}(c-v_g)/\lambda$ of the laser pulse with solid lines ($0.008\lambda$).  \emph{On the left} we show error characteristics of the azimuthal mode Maxwell-propagation. \emph{On the right} we show same error characteristics for  laser potential wave propagation. The purple curve shows the corresponding distance error for the laser envelope propagation using  8th order differentials, 4th order exponentials with $\Delta t = 0.5 \Delta z/c$. The energy loss of the latter was below y axis range.  
\label{fig:test_am_vacuum}} 
\end{figure*}

In our first test we propagate a laser pulse of $\lambda=0.8\mum$ carrier wavelength in vacuum over a temporal duration of $100\mum/c$, we place this pulse such that it is initially at $z_0=-50\mum$ from its focus position ($z=0$) and after the propagation it should be at position $-z_0$.  We performed numerical simulations  with varying $N=\lambda/\Delta z$ resolution in a $140\mu m\times60\mum$ moving box $(\rho,z)$ with $w_0=16\mum$ laser spot size at the focus, and pulse length of $\tau=10\mum/c$. We used fixed transverse resolution $\Delta r = 0.2\mum$ with stretched coordinates\footref{foot:trf1}, the presence of the latter did not affect the accuracy. We choose $\Delta t = \Delta z/c$ unless otherwise stated.

In this test we analyze our Maxwell solver's accuracy properties by computing the  $\left< E \right>$ pulse energy and the error of its conservation property $\delta\left< E \right>$:
\begin{equation} \label{eq:test_am_vacuum_E}
\left< E \right> = \int \left(  \left| {\bf E} \right|^2 + c^2\left| {\bf B} \right|^2  \right) {\rm d}V,  \quad
\delta \left< E \right> = \frac{\left< E \right>-\left< E \right>_0}{\left< E \right>_0}
\end{equation}
and for the dispersion error the position of the $\left<z \right>$ pulse centroid and its error $\delta \left<z \right>$:
\begin{equation} \label{eq:test_am_vacuum_Z}
\left< z \right> = \left< E \right> ^{-1} \int z \left(  \left| {\bf E} \right|^2 + c^2 \left| {\bf B} \right|^2 \right)  {\rm d}V, \ \
\delta \left< z \right> = \frac{\left< z \right>+\left< z \right>_0}{\lambda},
\end{equation}
where $\left< z \right>_0$ is the initial position of the pulse centroid.
In the case of propagating the laser potential (A, Envelope) we integrate over the $|\tilde{A}|^2$ complex electric field envelope  Eq. (\ref{eq:laserA_envelope_E}).

It is worth noting that there exist an analytical solution to the $v_g$ group velocity \cite{esarey1995tightly_focused_pulses} for the pulse centroid:
\begin{equation} \label{eq:test_am_vacuum_vg}
\frac{c-v_g}{c} = \left(\frac{\lambda}{2\pi w_0} \right)^2 \quad  \text{in 3D},
\end{equation}
which is approximately $-6.33\times10^{-5}$ in our case,  that means $0.008\lambda$ centroid delay for this $100\mum$ propagation test. We note that initially the pulse centroid $\left< z \right>_0$ is not exactly at $-50\mum$, it is such that $v_g \approx -2c \left<z \right>_0/100\mu{\rm m}$ with negligible error - therefore $\delta \left<z \right>$ expresses the difference from this analytical result.

On Fig. \ref{fig:test_am_vacuum} we summarize the $\delta \left<E \right>$, $\delta \left<z \right>$ error characteristics of our methods in this test scheme, and the errors shown scale linearly with the propagation distance (for example, $10^2\times$ for 1cm). The errors follow precisely the same characteristics that we outlined in our Cartesian code \cite{majorosi2026exponentialpic}. We can see the errors of the exponential expansions with 30th order finite differences versus spatial resolution $N  = \lambda/\Delta z$ (and the temporal step $\Delta t = \Delta z /c$). We can see that the error curves of the $\delta\left< E\right>$ energy loss and the relative pulse centroid delay $\delta\left<z\right>$ is of the same order for a given exponential. Unfortunately, in this frequency range our primary concern is the $\delta\left< E\right>$ energy loss (solid lines). \emph{On the left} panel we can see this for the Maxwell-AM decomposition, \emph{on the right} for laser potential simulations. We can see that in both cases the error characteristics are the same despite the differing representations. For reference we show the value of the analytical centroid delay $0.008\lambda$ as horizontal lines, if any of the $\delta\left< z \right>$ error curves go below the latter this effect can be numerically resolved.

The true laser Envelope simulations (on Fig. \ref{fig:test_am_vacuum}, right panel, purple curve), were run with lower numerical requirements: 4th order exponential, 8th order finite differences and $\Delta t = 0.5\Delta z /c$. Using these parameters we achieved negligible $\delta\left< E\right>$, $\delta\left<z\right>$ errors, however, this is the usual expectation. We would like to emphasize that this is due to the fact is that the laser envelope is devoid of the laser oscillations, and the envelope itself has a very low wavelength, which is the cause of its high accuracy in vacuum. However, if the plasma or medium introduces frequency modulation in the complex laser envelope $a$, the accuracy could worsen as in the wave laser potential representation.

Overall, to assess the dispersion and norms errors with specific exponential and differential orders we need to take the largest error value on the corresponding panel of Fig. \ref{fig:test_am_vacuum} and multiply it with a chosen propagation distance. By properly selecting these, negligible dispersion and norm loss can be achieved for long distances. For these both in AM, and A representation, similar accuracy to FBPIC can be achieved for the laser fields.

\subsection{Linear propagation in plasma} \label{subsubsec:test_am_plasma}
 
 \begin{figure}[ht]
\centering
\includegraphics[trim={10 0 0 0},clip,height=6.4cm]{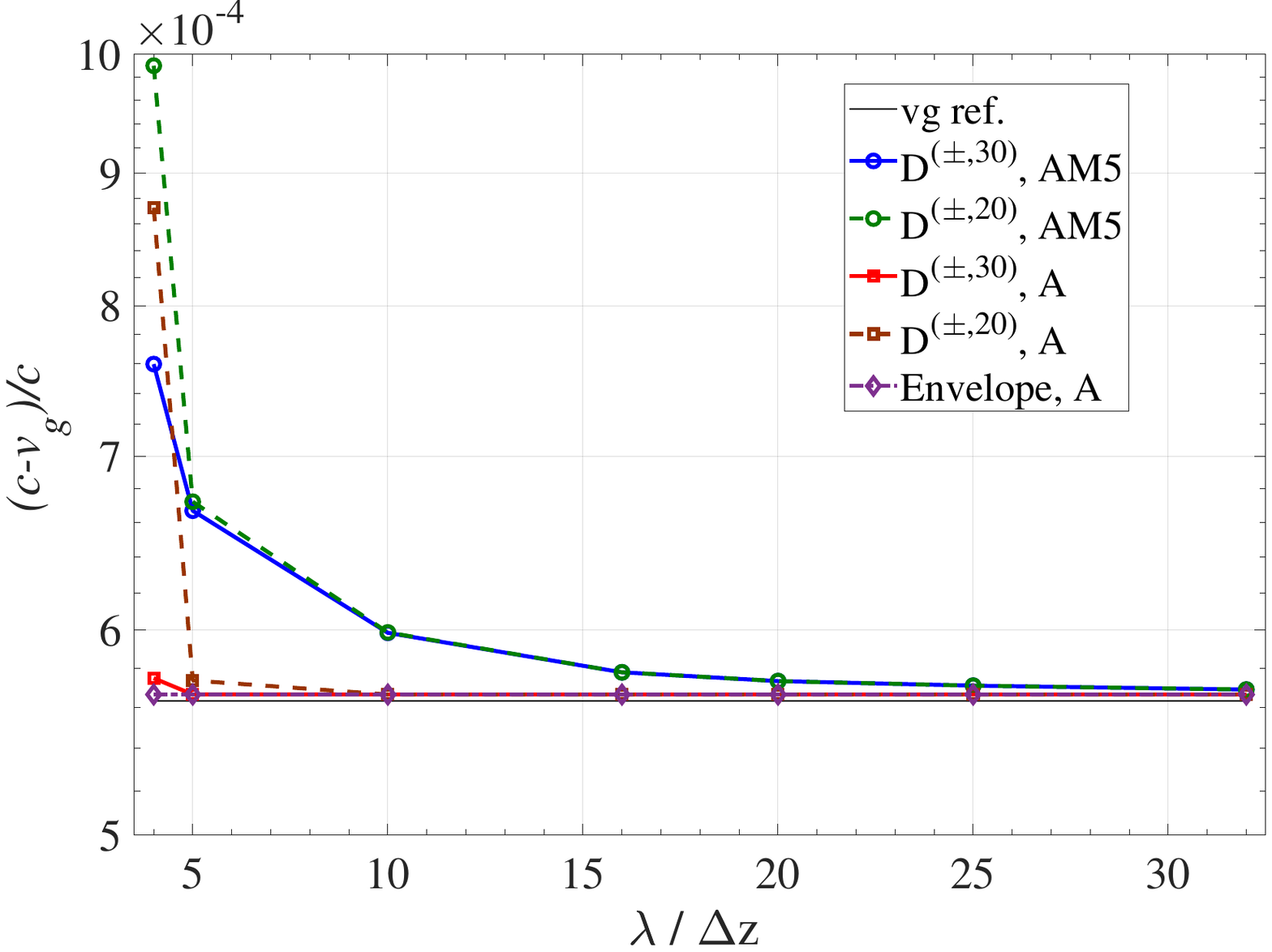}
\caption{ Relative difference between $c$ and the group velocity $v_g$ of the pulse centroid in a plasma of $10^{-3}n_c$ density for different $N = \lambda/\Delta z$ resolutions in cylindrical geometry. The  black line represents the analytical prediction ($5.634\times10^{-4}$), the blue, green lines (circle markers)  correspond to simulations using azimuthal modes decomposition with 5 basis functions, the rest propagates the laser potentials (A, Envelope). We show the results using 30th order (solid lines) and 20th order finite differences (dashed lines) in each case. We also show the results of the true laser Envelope model (purple dashed line) which we ran using 8th order finite differences.  
\label{fig:test_am_plasma}}
\end{figure}

In this test we propagate a laser pulse of $\lambda=0.8\mum$ carrier wavelength  with $a_0 = 0.01$ in an underdense plasma of  $n_0= 1.75\times10^{18} {\rm cm}^{-3} = 10^{-3} n_c$ density, where $n_c$ is the critical density. We do this between $z=0$ and $z=50\mum$, where the focus position of the laser is at $z=0$. We perform this test with varying $N = \lambda/\Delta z$ resolution.  We performed numerical simulations in $140\mu m\times60\mum$ moving box, the laser had $w_0=16\mum$ spot size at the focus, and Gaussian pulse length $\tau=10\mum/c$. We used fixed transverse resolution $\Delta r = 0.2\mum$ with stretched coordinates\footref{foot:trf1}, the presence of the latter did not affect the accuracy. We choose $\Delta t = 0.5\Delta z/c$ unless otherwise stated, and for the field solver we use 30th order finite differences and 12th order exponentials. 

There exists an analytical formula for the $v_g$ group velocity for the pulse centroid motion  \cite{esarey1995tightly_focused_pulses}, which is
\begin{equation} \label{eq:test_am_plasma_vg}
\frac{c-v_g}{c} = \frac{n_0}{2 n_c}+ \left(\frac{\lambda}{2\pi w_0} \right)^2 \quad  \text{in 3D}.
\end{equation}
This yields approximately $-5.63\times10^{-4}$  in this test case. We calculate this using the time derivative of the  $\left< z \right>$ pulse centroid, see Eq. (\ref{eq:test_am_vacuum_Z}). Despite this result is analytical, it is not exact.

We summarize our results in Fig. \ref{fig:test_am_plasma} versus the resolution $N  = \lambda/\Delta z$. We use 3rd order particle shapes with $2\times$ interpolation supersampling.  We show the Maxwell-azimuthal modes decomposition with 5 basis vectors with blue and green curves. We can see that due to the high accuracy of the field solver our results are not far from the analytical prediction, and have similar accuracy as those simulated with FBPIC \cite{lehe2016fbpic} and our Cartesian code \cite{majorosi2026exponentialpic}. Using our methods, satisfying accuracy can be reached at $\lambda/16$ resolution.

We also show on Fig. \ref{fig:test_am_plasma} the laser potential simulation using laser envelope PIC model. These curves almost reproduce the analytical result independent of the simulation resolution. (The analytical solution also uses the laser envelope model.) Using 20th order finite differences with wave simulations (A) only seems to impact high frequency ($\Delta z = \lambda/4$) results due to the wave dispersion effect. The true laser Envelope simulations (dashed purple) retain analytical precision even at $\lambda/4$ resolution using only 8th order finite differences. 

Overall, our PIC implementation in cylindrical geometry reproduces the analytical group velocity reasonably well. Simulations using laser envelope models, however, exhibit superior accuracy for the group velocity in plasma.

\subsection{Linear laser-wakefield} \label{subsubsec:test_am_linear}
 
 \begin{figure*}[t]
\centering
\includegraphics[trim={100px 0 140px 30},clip,height=5.0cm]{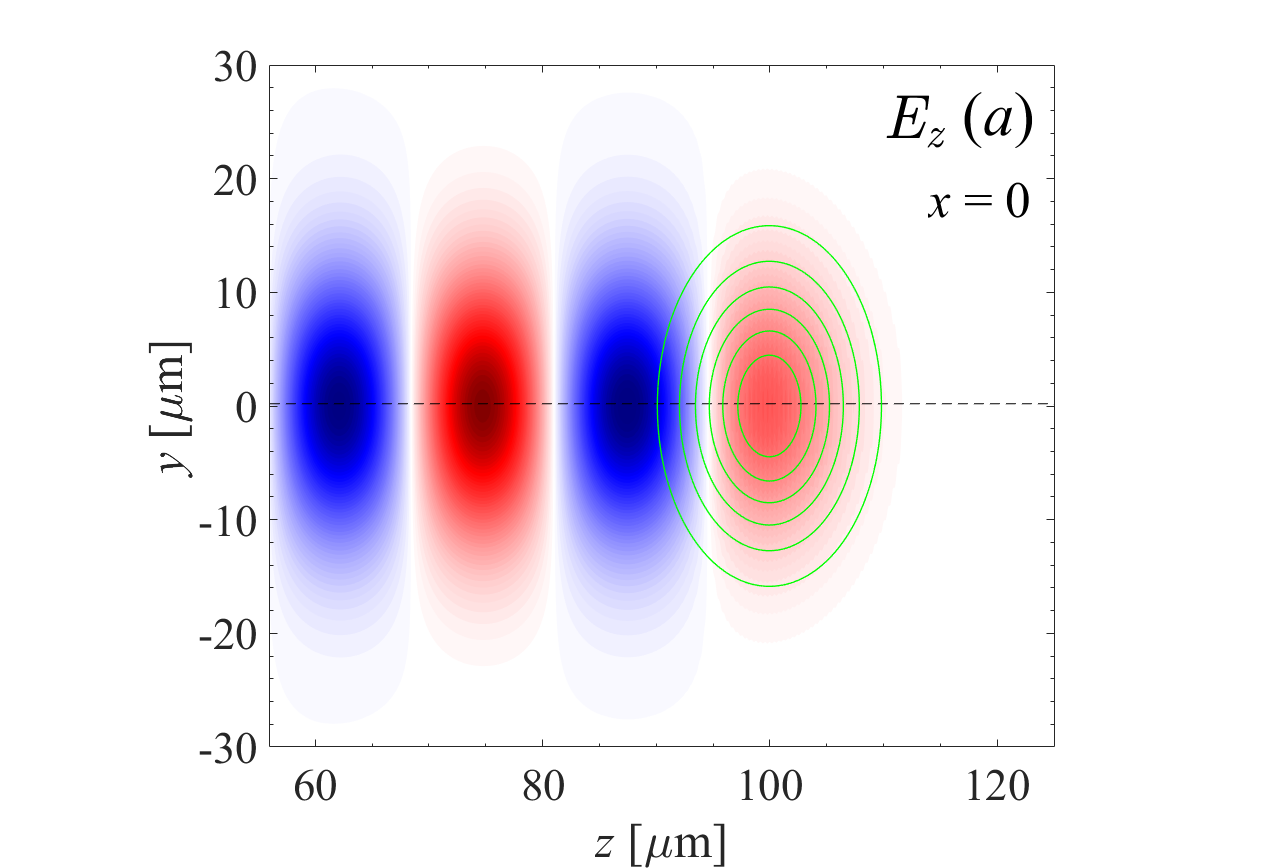}
\includegraphics[trim={0px 0 0px 0},clip,height=5.05cm]{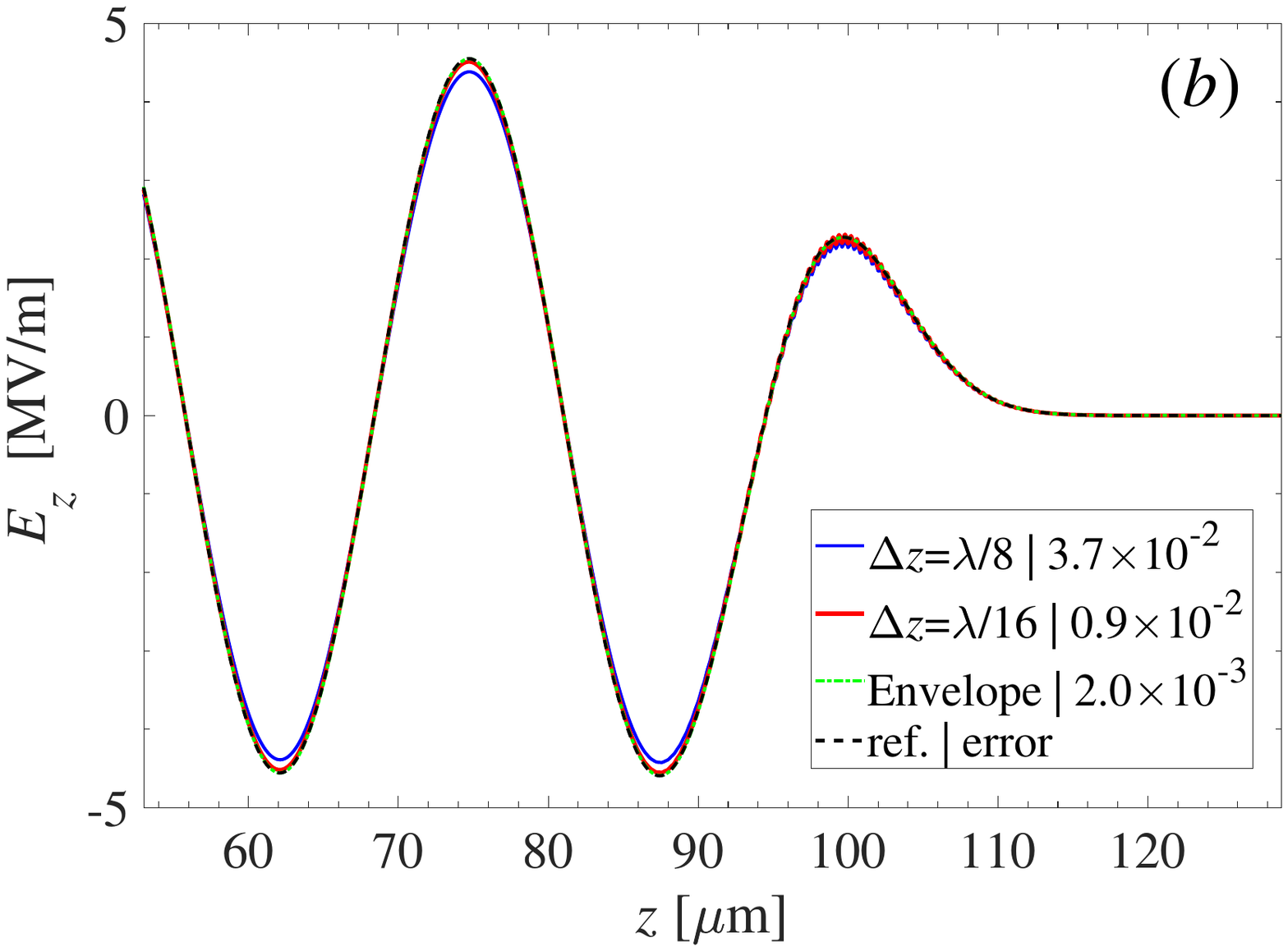}
\linebreak
\includegraphics[trim={100px 0 140px 30},clip,height=5.0cm]{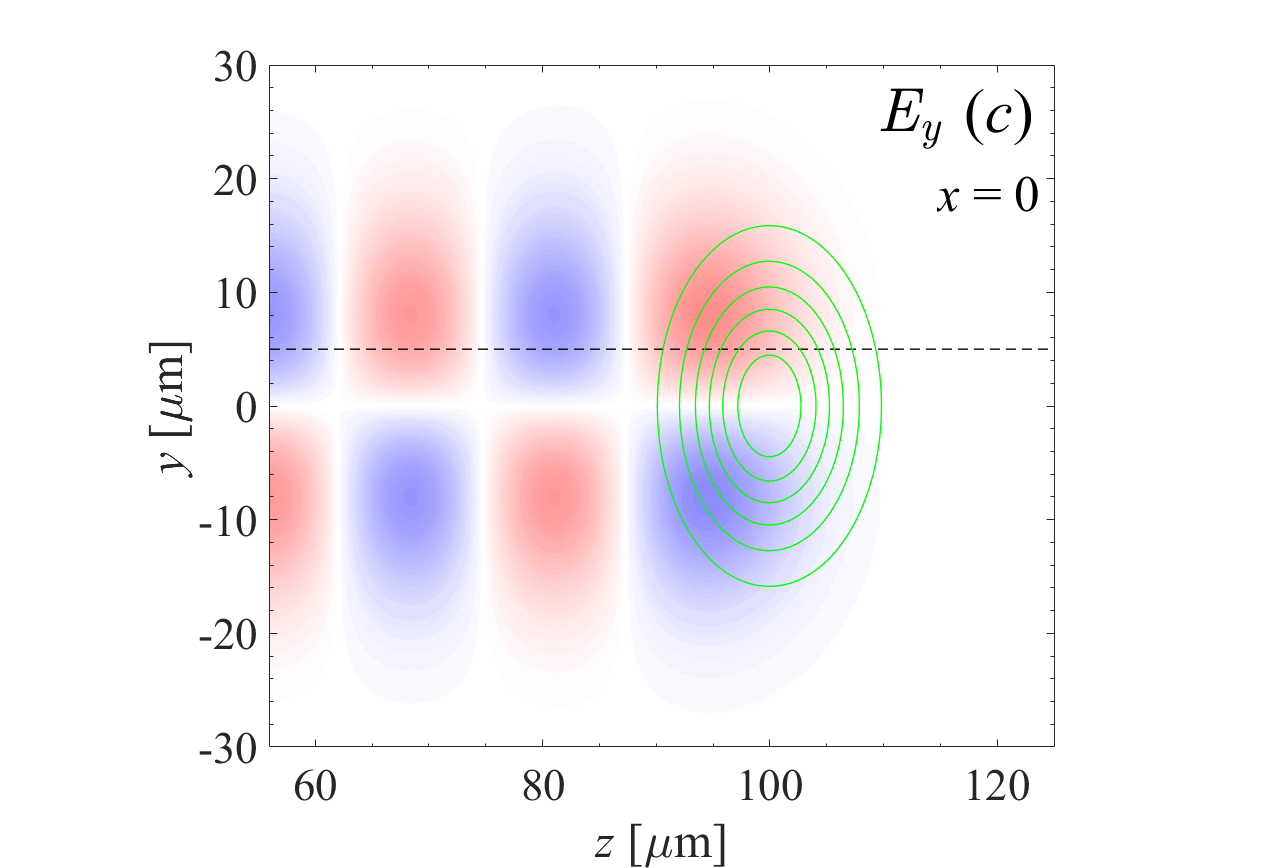}
\includegraphics[trim={0px 0 0px 0},clip,height=5.00cm]{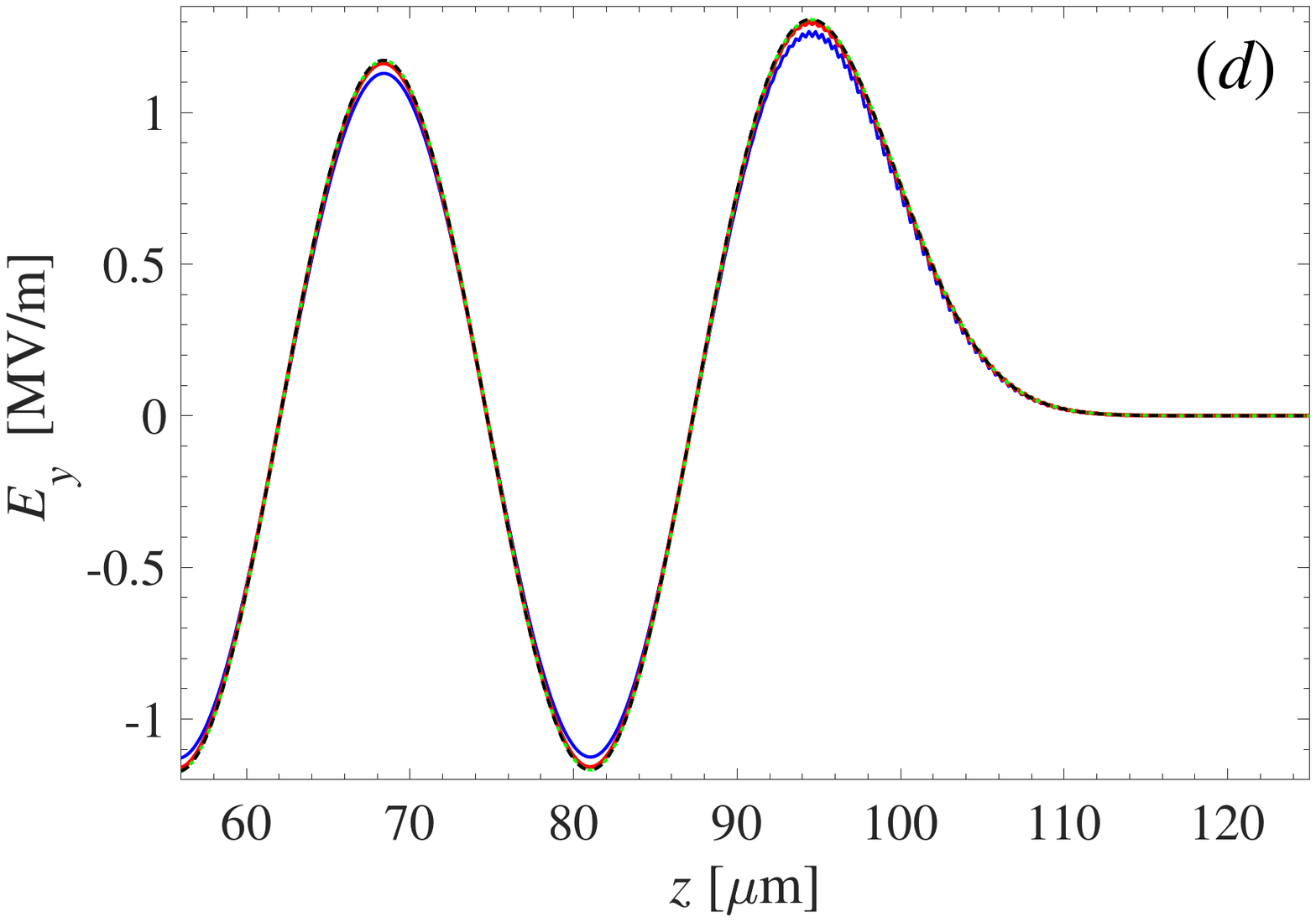}
\caption{Plot of the linear wakefields $E_z$ (top panels), $E_y$ (bottom panels) generated by a laser pulse of $\lambda = 0.8\mu$m after $100\mu$m propagation in an underdense plasma in cylindrical geometry. We show the  colormap of these fields in $(a)$, $(c)$ in the $x=0$ plane where we also have indicated the position of the laser envelope and a line cross section. We show the latter on panels $(b)$, $(d)$ to compare the accuracy of the wakefields, where the dashed lines show the expected analytical results, the blue and red lines show the results using Maxwell azimuthal modes decomposition (AM) at $\Delta z = \lambda/8$ and $\lambda/16$ resolutions with $\Delta t = 0.5 \Delta z/c$, for the green lines we used the laser Envelope model at $\Delta z = \lambda/5$ resolution with $\Delta t = 0.25\Delta z/c$. We also indicated the approximate relative error of the wakefields. 
\label{fig:test_am_wake}} 
\end{figure*}

Similar to the previous Section, we propagate a laser pulse of $\lambda=0.8\,\mum$ carrier wavelength with $a_0 = 0.01$ in an underdense plasma of density $n_0= 1.75\times10^{18} {\rm cm}^{-3}$. We perform this test at $\Delta z = \lambda/8, \lambda/16$ resolutions and we analyze the resulting wakefields.   We performed numerical simulations in a $140\mu m\times60\mum$ moving box, the laser had $w_0=16\,\mum$ spotsize at the focus (at $z=0$), and Gaussian pulse length of $\tau=10\,\mum/c$. We used fixed transverse resolution $\Delta r = 0.2\,\mum$ with stretched coordinates\footref{foot:trf1}, the presence of the latter did not affect the accuracy. We choose the temporal step $\Delta t = 0.5\Delta z/c$ such that it provides a temporally convergent result, and for the field solver we use 30th order finite differences and 12th order exponentials. We enabled our $2\times$ field  interpolation supersampling feature for better accuracy.  For the laser envelope model simulations we use 8th order finite differences, $\Delta z = \lambda/5$, $\Delta t = 0.25\Delta z/c$ resolution.

We propagate the laser pulse in a constant density between $z=0$ and $z=100\,\mum$ and at the latter position of the laser pulse we extract the simulated $E_y$, $E_z$ wakefields at the $x = 0$ planar cross-section. Since this is a low intensity linear propagation, the generated laser wakefields can be given analytically within the laser envelope model, which involves the numerical integration of the ponderomotive potential of the laser  with the plasma oscillations. The respective formulas of which can be found in e.g. \cite{esarey2009laser_plasma_accelerators, lehe2016fbpic}. We extracted the approximate longitudinal envelope from the simulations to get the best match with the analytical prediction. The laser envelope simulations directly correspond to the analytical model while in the AM-Maxwell PIC scheme the wakefields are created from the relativistic oscillations of the electrons. 

We summarize our results on Fig. \ref{fig:test_am_wake}, where we show the planar cross sections on Fig. \ref{fig:test_am_wake} $(a)$, $(c)$ for the $E_z$, $E_y$ electric fields respectively. We took the axial lineouts corresponding to the horizontal dashed lines of these cross sections (on $(b)$, $(d)$ panels) and show them at $\Delta z = \lambda /8$ (blue curves) and $\Delta z = \lambda /16$ resolutions (red curves) and the laser envelope model (green curves). 

The results are really similar to the ones we acquired with our Cartesian code \cite{majorosi2026exponentialpic} and to those from FBPIC \cite{lehe2016fbpic}. We can state that our numerical PIC scheme underestimates the ponderomotive force felt by the electrons resulting wake field errors on the order of a couple of percent. The laser envelope simulation, however, provided much more accurate wakefields (error $\sim10^{-3}$) even using lower resolution - this, again, is the usual expectation for this model. If we double the spatial resolution the wakefield errors decrease by a factor of $1/4$ in all simulation types. For the azimuthal modes simulations the magnitude of wakefield errors are nearly the same as Cartesian code \cite{majorosi2026exponentialpic}, which suggest they are subject to the same wakefield behavior. Our $2\times$ field interpolation supersampling is as effective in cylindrical coordinates for reducing wakefield errors as in Cartesian coordinates.

Overall, our PIC simulations in cylindrical geometry could reproduce the analytical expectation reasonably well, but the Maxwell-AM representation underestimates the resulting ponderomotive wake fields on the order of $10^{-2}$ as our Cartesian method. To achieve the best accuracy for these at modest resolutions our $2\times$ field interpolation supersampling \cite{majorosi2026exponentialpic} seems to be required. The laser envelope simulations, however, yield superior accuracy for these wakefields.

\subsection{Laser wakefield acceleration in the bubble regime} \label{subsubsec:test_am_inject}

\begin{figure*}[t]
\includegraphics[trim={22px 0px 50px 35px}, clip, height=4.5cm]{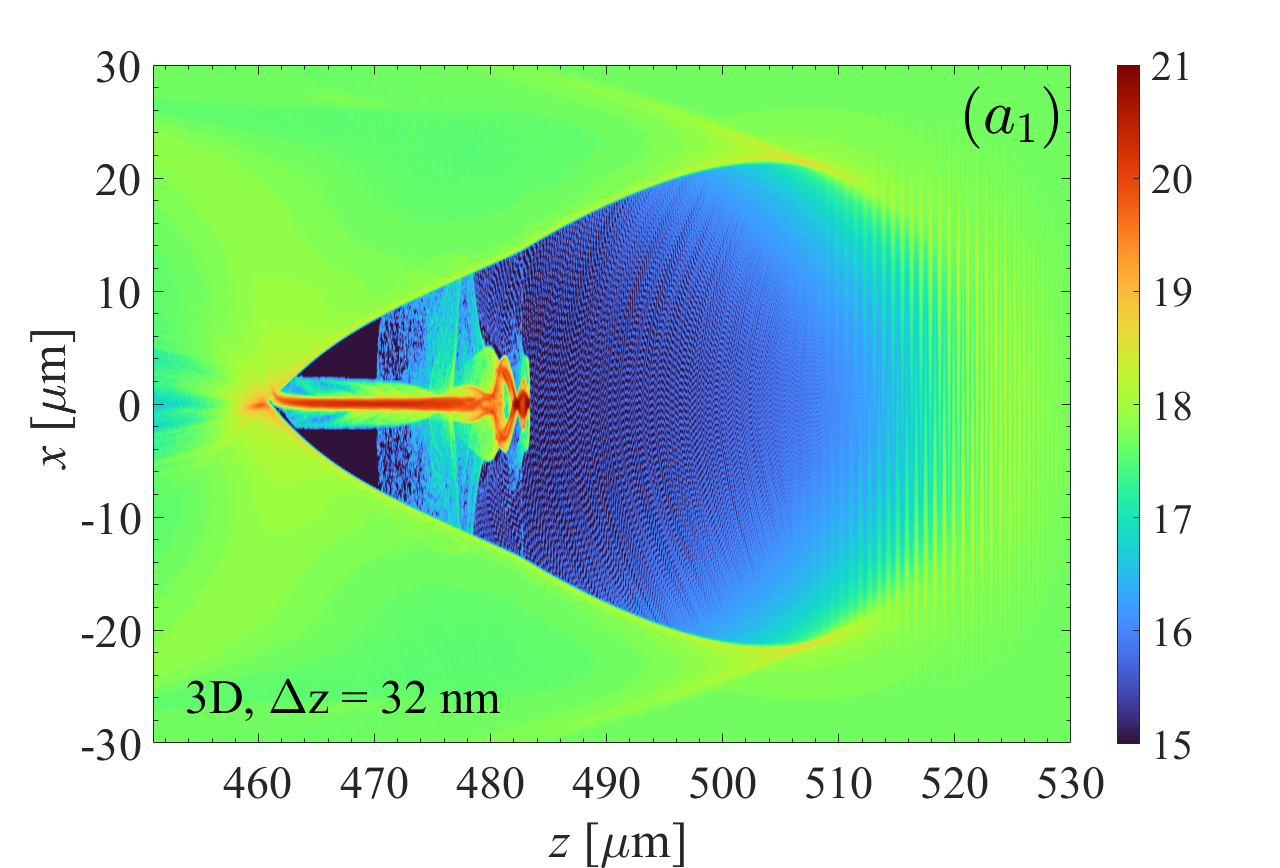}
\includegraphics[trim={25px 0px 210px 35px}, clip, height=4.5cm]{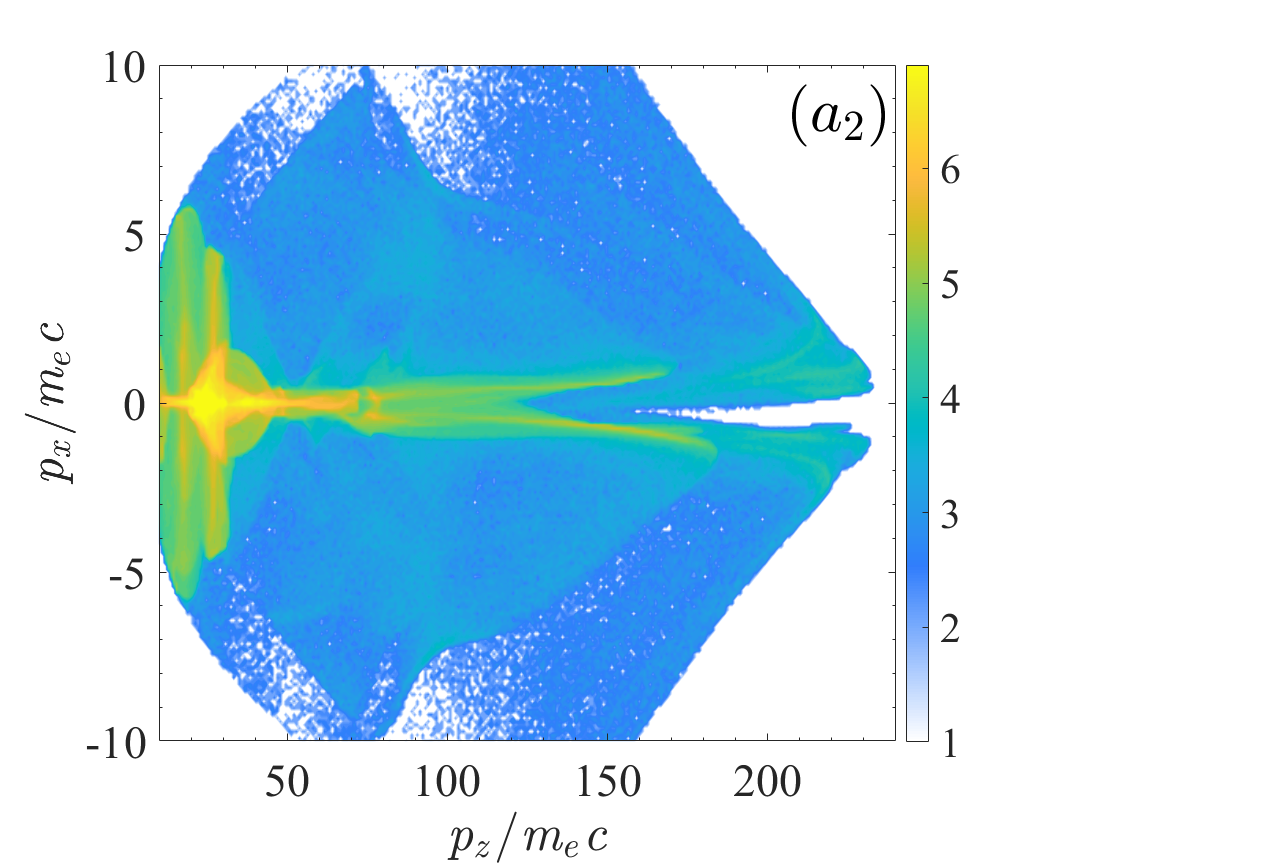}
\includegraphics[trim={25px 0px 210px 35px}, clip, height=4.5cm]{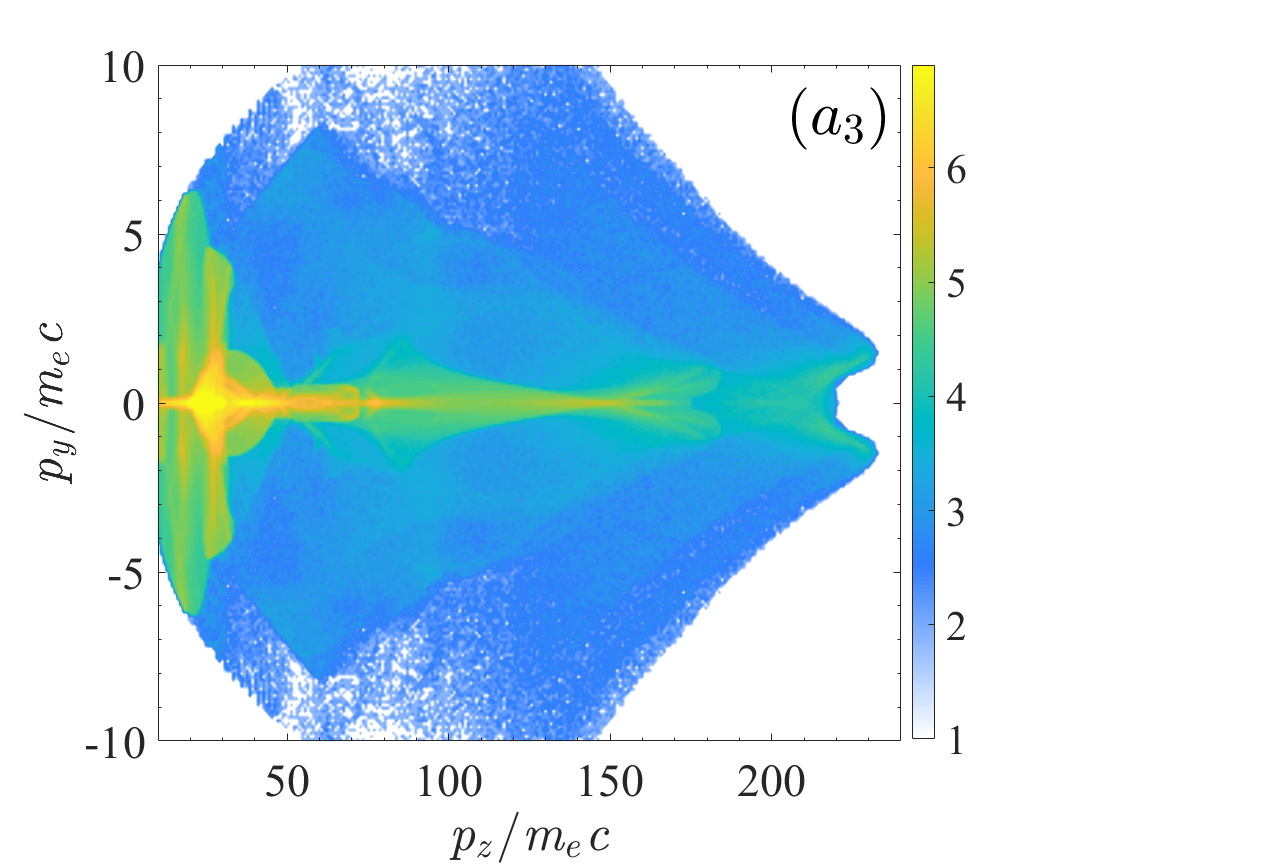}
\\
\includegraphics[trim={22px 0px 50px 35px}, clip, height=4.5cm]{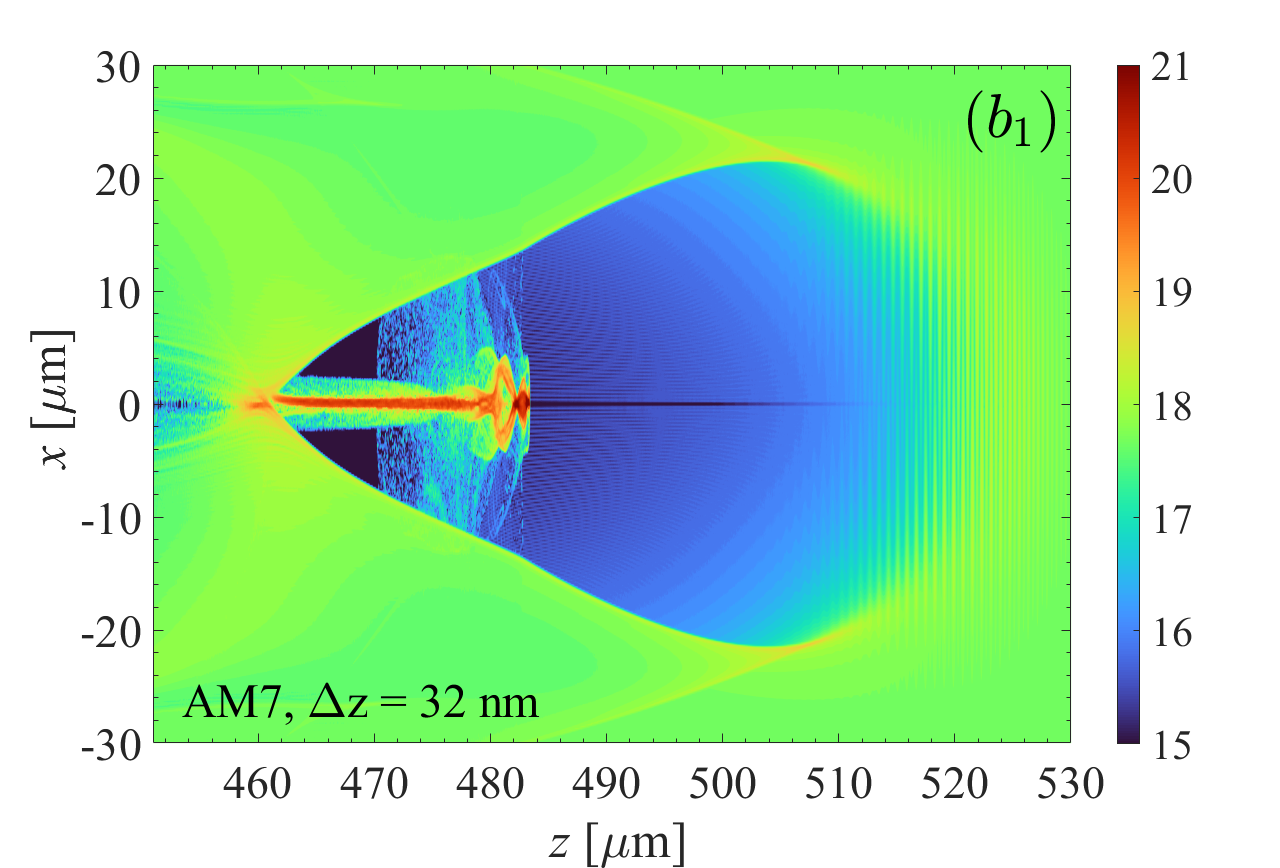}
\includegraphics[trim={25px 0px 210px 35px}, clip, height=4.5cm]{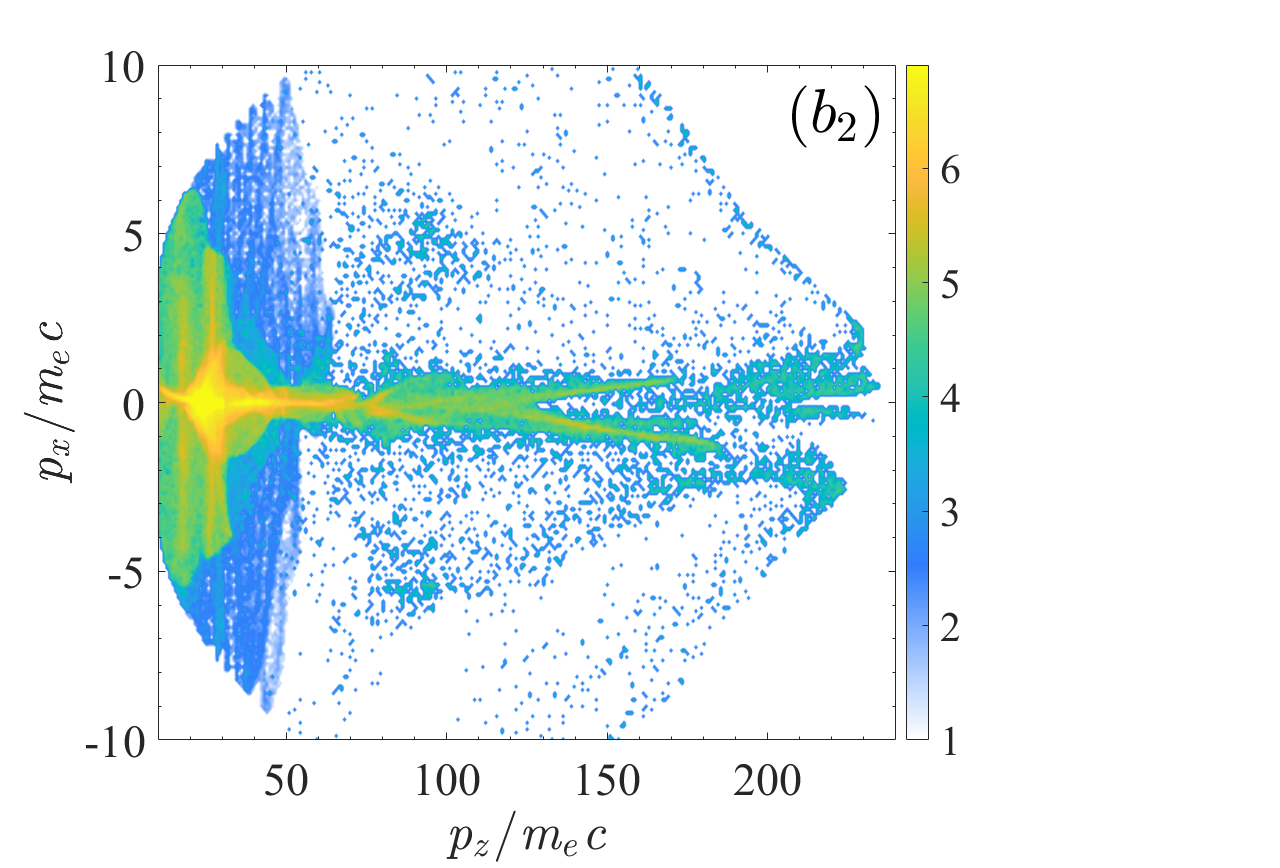}
\includegraphics[trim={25px 0px 210px 35px}, clip, height=4.5cm]{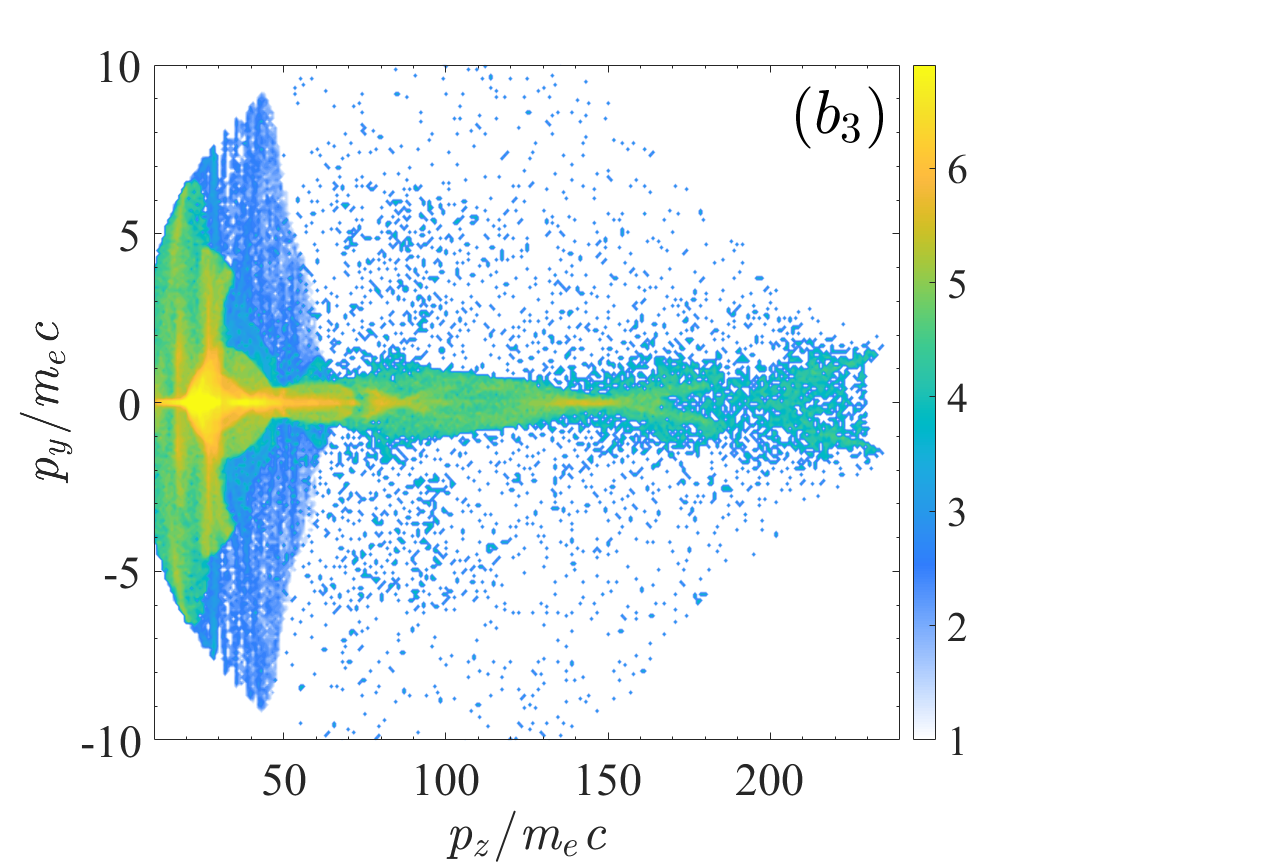} 
\\
\includegraphics[trim={22px 0px 50px  35px}, clip, height=4.5cm]{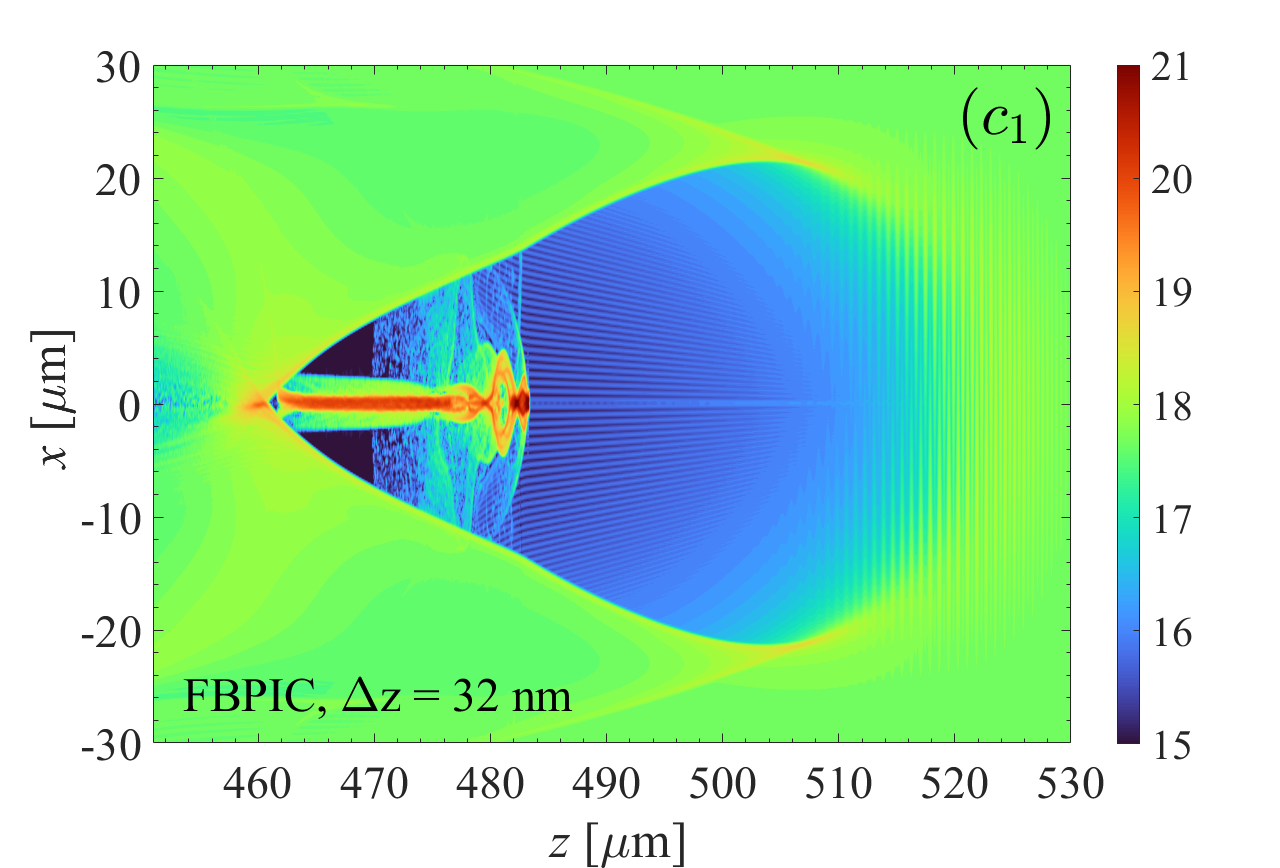}
\includegraphics[trim={25px 0px 210px 35px}, clip, height=4.5cm]{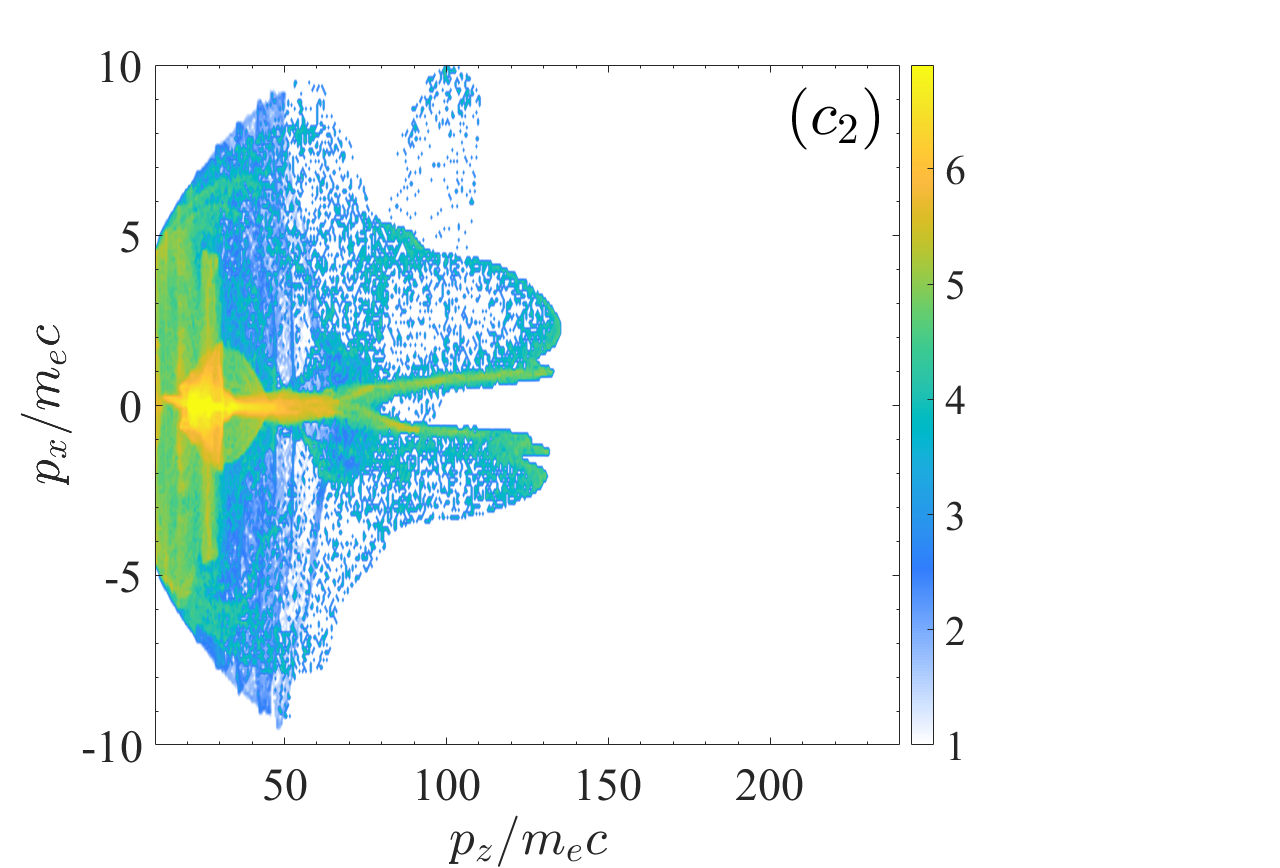} 
\includegraphics[trim={25px 0px 210px 35px}, clip, height=4.5cm]{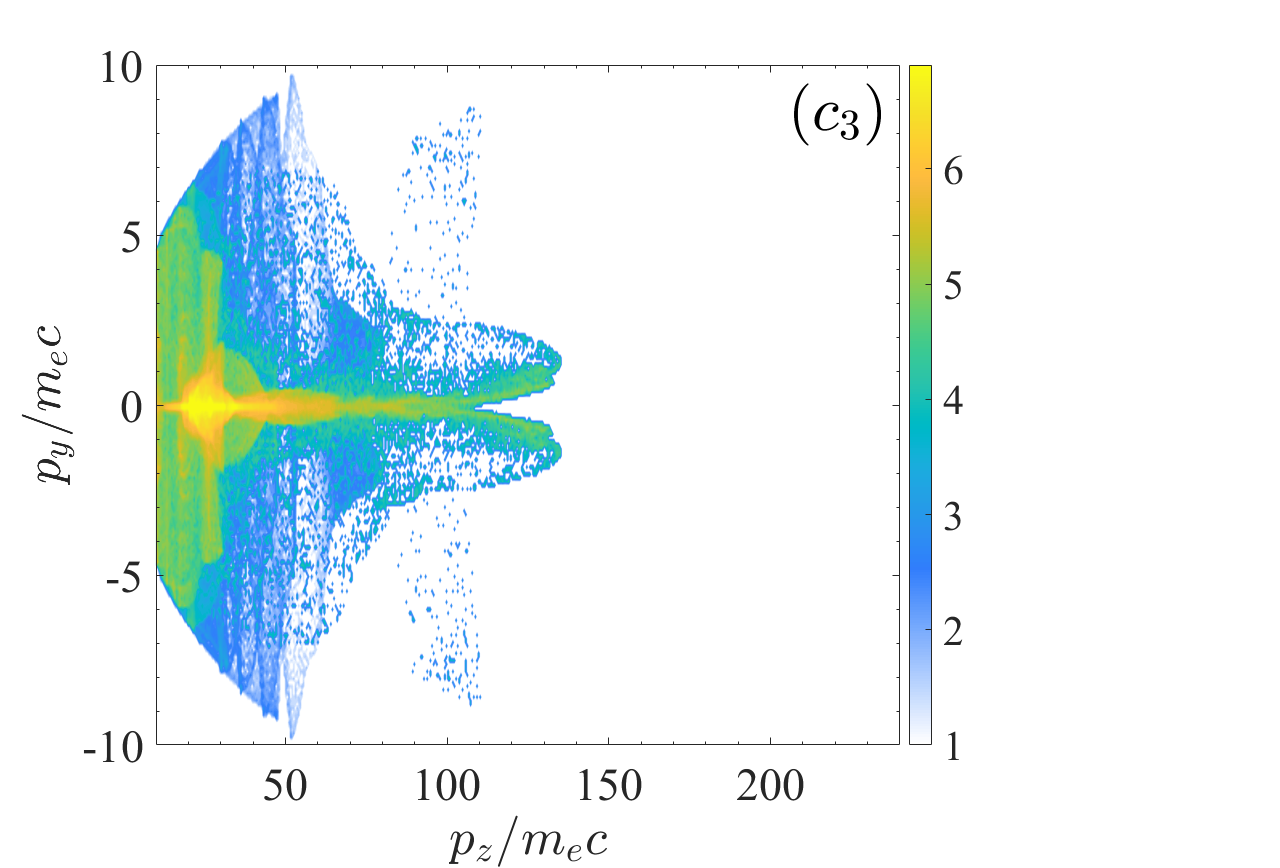}
\\
\includegraphics[trim={22px 0px 50px  35px}, clip, height=4.5cm]{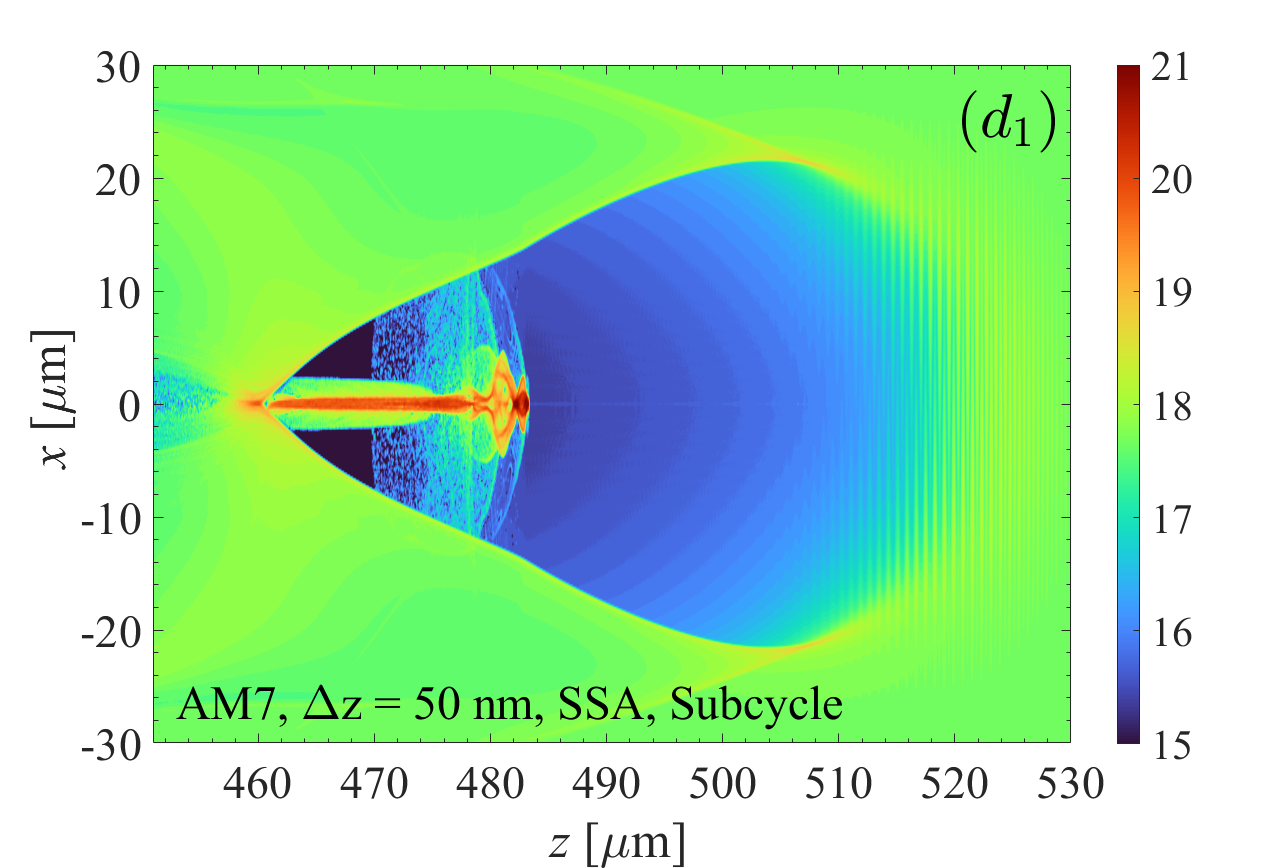}
\includegraphics[trim={25px 0px 210px 35px}, clip, height=4.5cm]{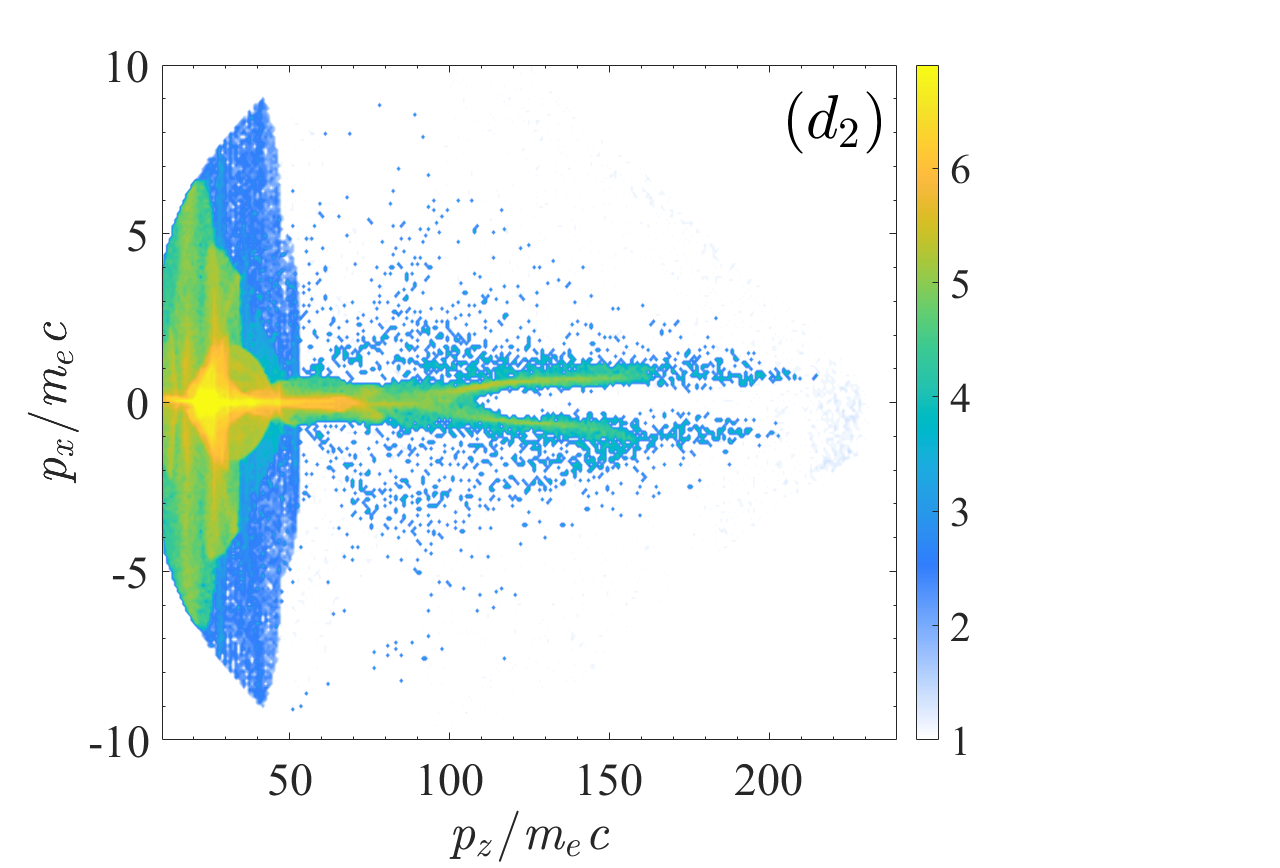}
\includegraphics[trim={25px 0px 210px 35px}, clip, height=4.5cm]{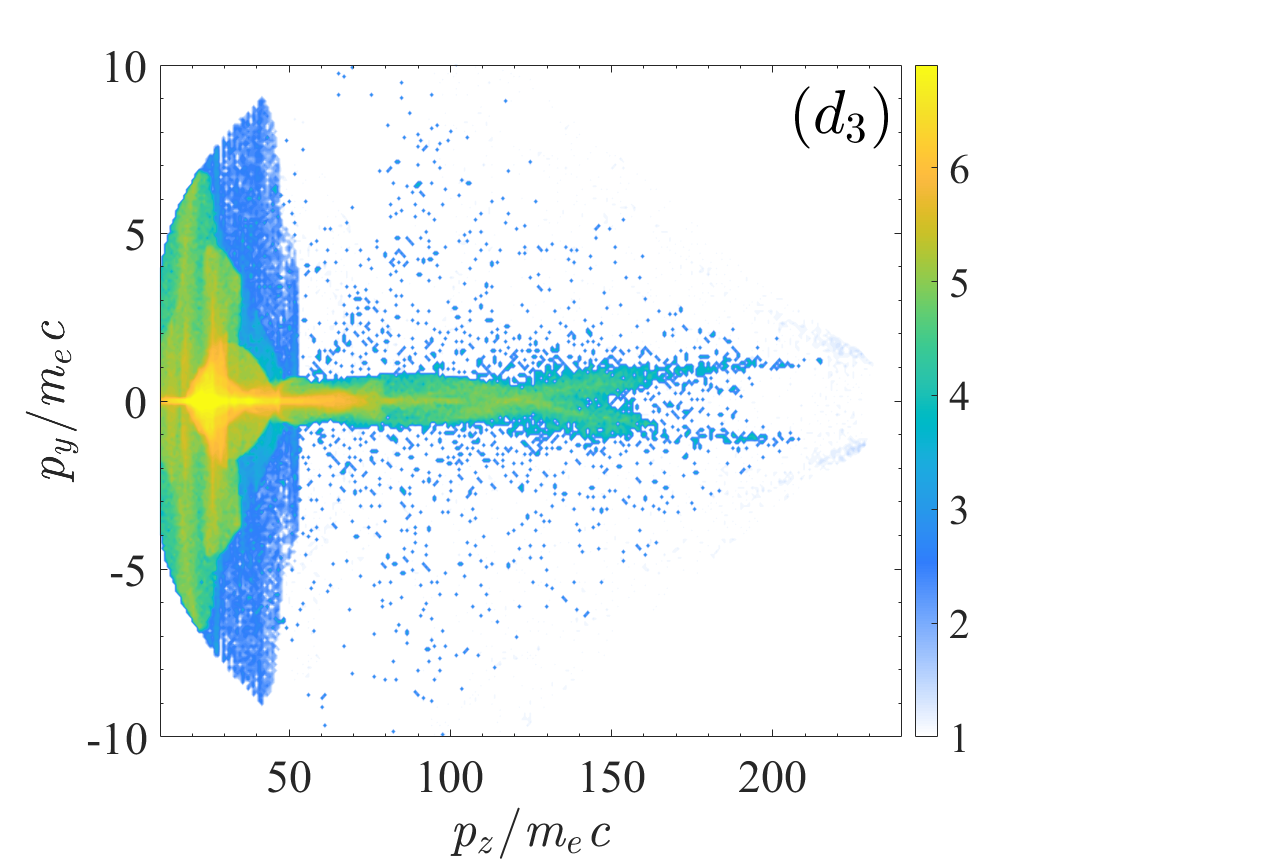}
\\
\includegraphics[trim={22px 0px 50px  35px}, clip, height=4.5cm]{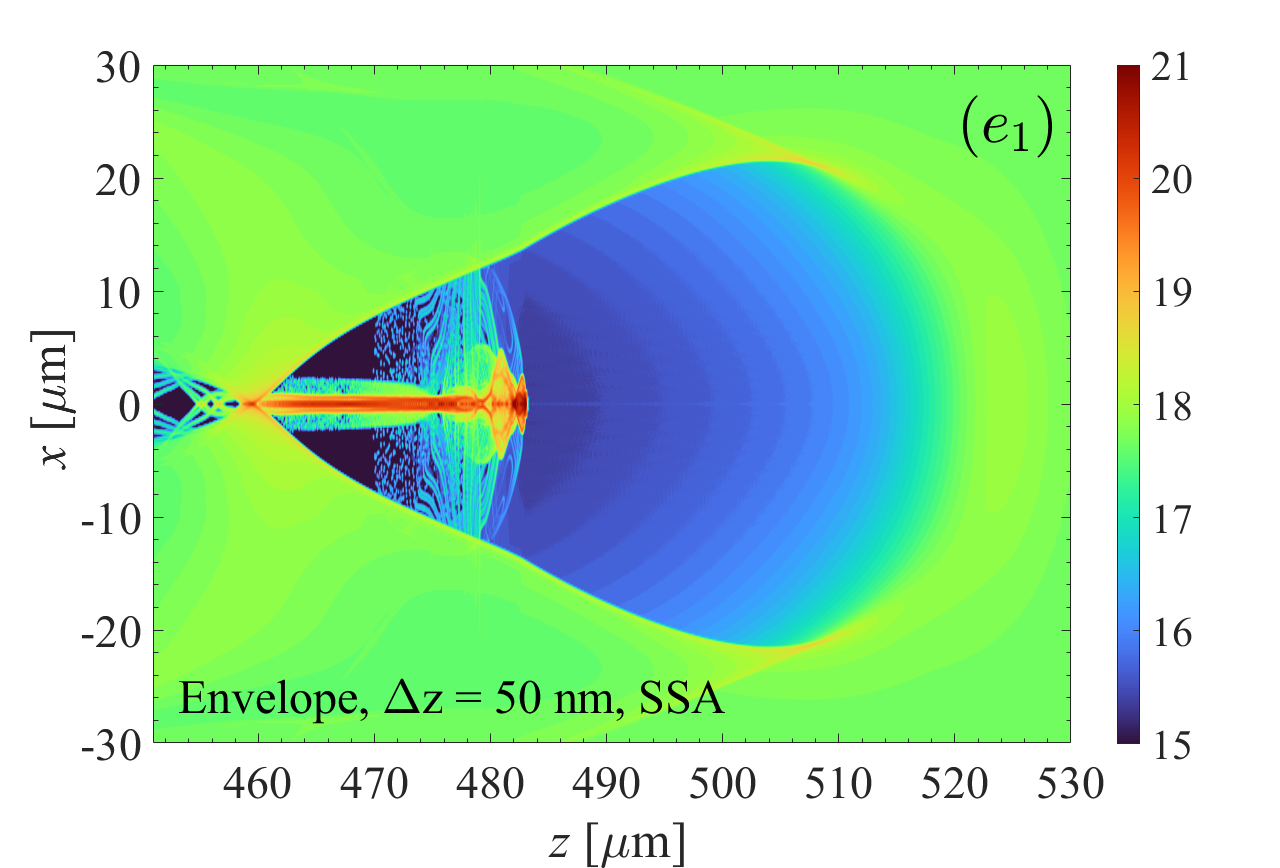}
\includegraphics[trim={25px 0px 180px 35px}, clip, height=4.5cm]{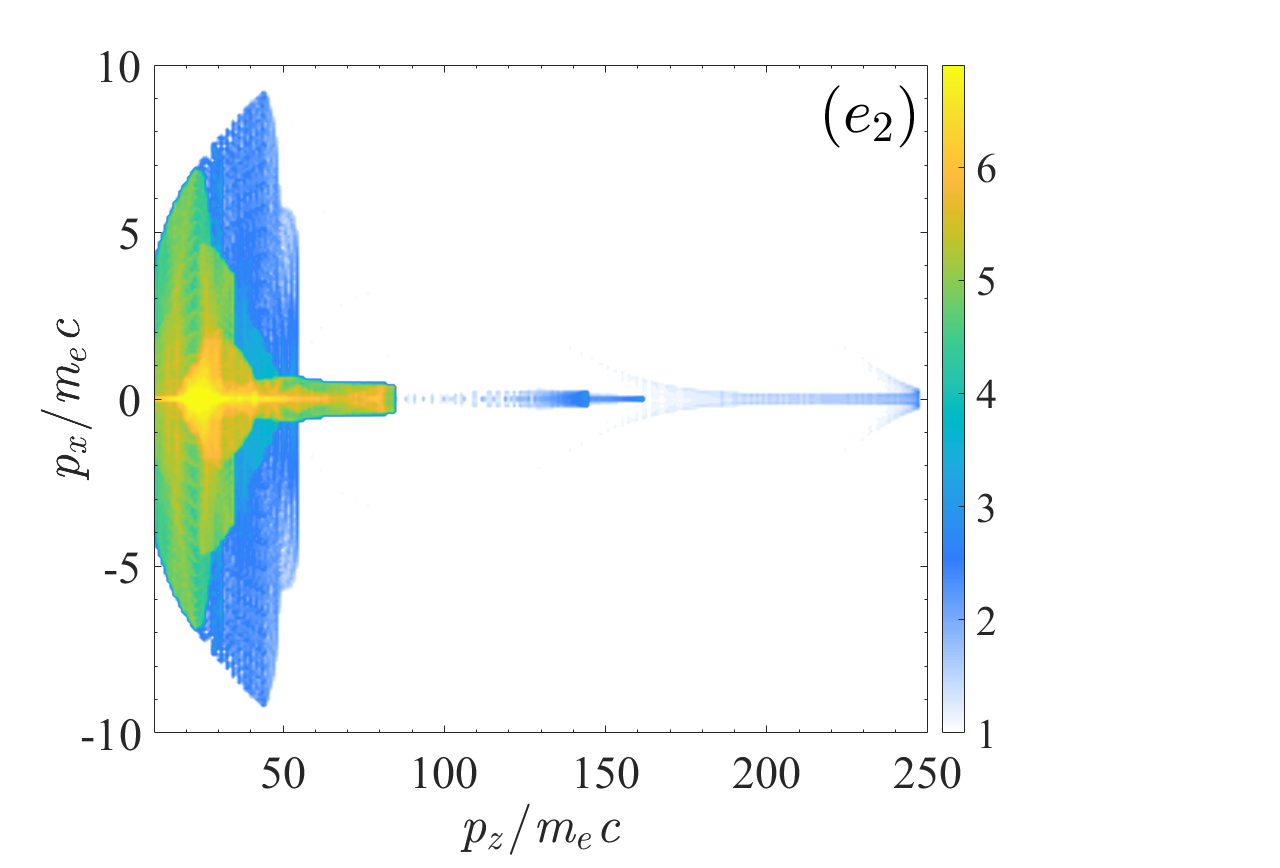}
\includegraphics[trim={20px 0px 20px 5px}, clip, height=4.5cm]{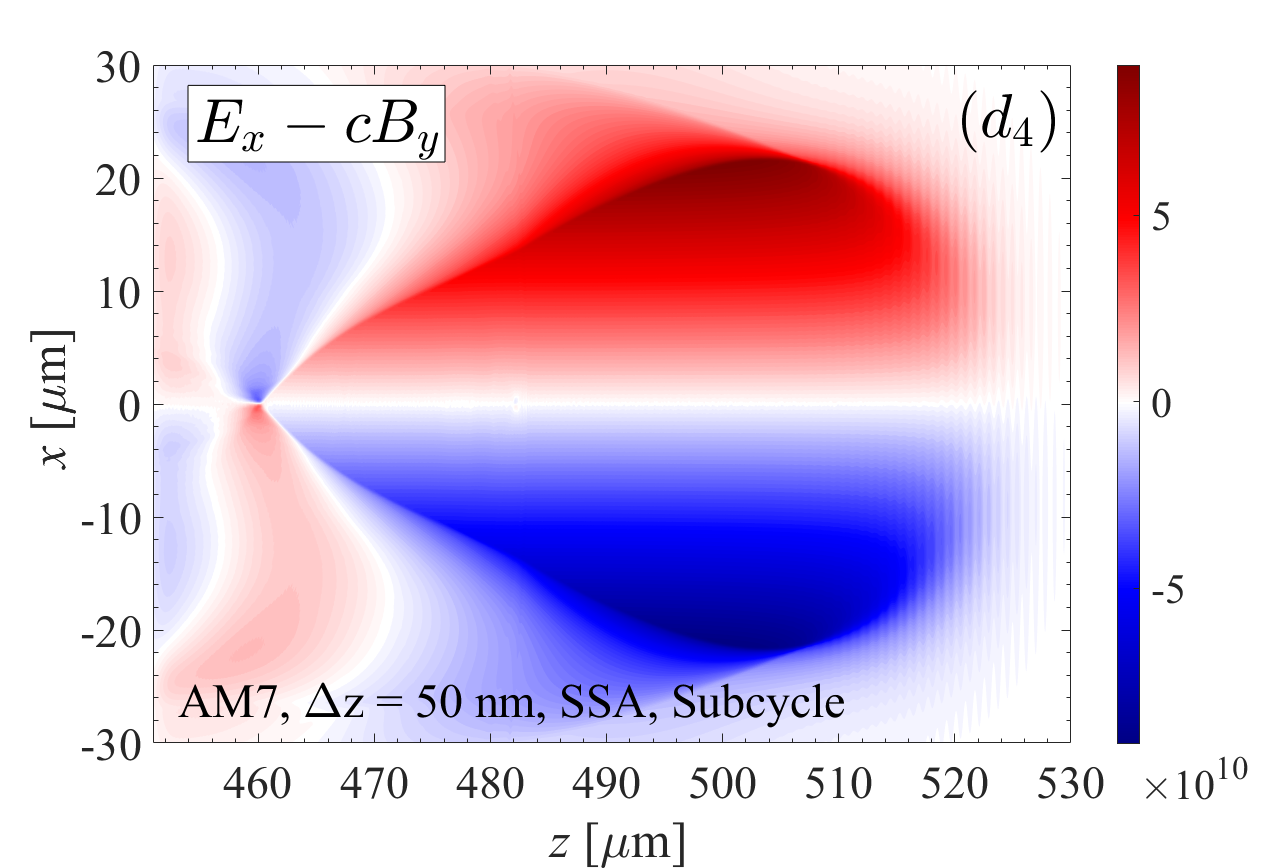}  

\caption{Comparison of simulations with different configurations $(a)$-$(e)$. In the first column  the electron density distributions are shown with logarithmic color scale ($log_{10}[n_e(cm^{-3})]$), in the second  and third columns the $p_z$-$p_x$ and $p_z$-$p_y$ distributions are presented, in the plane of laser's polarization plane and in the orthogonal plane, respectively. We also show the  $E_x-cB_y$ field on $(d_4)$  where the numerical Cherenkov radiation is present if we use a low order solver. }
\label{fig:test_LWFA} 
\end{figure*}

In this test we recreate the "Suppression of zero-order numerical Cherenkov effect" test case from Ref. \cite{lehe2016fbpic}. We compare the results simulated by our 3D Cartesian code (3D), our cylindrical symmetric PIC envelope model (Envelope), our azimuthal modes decomposition PIC (using 7 basis vectors, AM7), and FBPIC ($M=4$). This test case features electron self-injection into a plasma bubble at a plasma density down-ramp and laser wakefield acceleration afterwards (LWFA). In this test case we have to get the near axis behavior as right as possible with our cylindrical solution.

We propagate a laser pulse of $\lambda=0.8\,\mum$ carrier wavelength with $a_0 = 4$,  $w_0=16\,\mum$ waist and  $\tau=10\,\mum/c$ length. The plasma was preionized, it had $100\,\mum$ rising density gradient at its front, followed by a $200\mum$ long density plateau of $2n_0$, followed by $100\,\mum$ downramp, so as to finally reach density of $n_0 = 0.5\times10^{18}{\rm cm}^{-3}$. The focus position is set at the beginning of plateau ($z=0$). The physical electron injection process is extremely nonlinear and very sensitive to numerical effects even in 3D. Our goal is to reproduce the results of the 3D simulation using the AM (quasi-3D) version of the code, applying the similar numerical parameters. 

 We ran this test on the GPU nodes of Hungary's DKF Komondor HPC. The quasi-3D simulations were run on a single 40GB NVIDIA A100 server-class GPU. Our code turned out to be a little bit faster ($\lesssim 10\% $)  than FBPIC on this hardware, and the simulation times were primarily bottle-necked ($\gtrsim 60\%$ of time) by the GPU density and current deposition algorithms.

We ran five kind of simulations in one of the three categories: 
\begin{itemize}
\item High-resolution 3D simulation $(a)$: $\Delta x = \Delta y = 0.16\mum$, $\Delta z = 0.032\mum$ with $2\times2\times2$ ordered particles per cell ($x,y,z$). Along $x$ we employed the sampling strategy of Section \ref{subsec:shapesAM_sampleth} to improve quality. We also use $2\times$ stretched coordinates along $x$ and $y$ outside a $\pm 25 \mum$ inner zone.

\item High-resolution AM7 $(b)$ and FBPIC $(c)$ simulations with  $\Delta \rho = 0.16\mum$, $\Delta z = 0.032\mum$ using FBPIC and our AM code. We set the number of particles per cell ($z,\theta,\rho$) as $2\times8\times4$ in our AM solver, while it is $2\times16\times2$ in the FBPIC simulation. This is due to the different sampling strategies between the two codes (see Section \ref{subsec:shapesAM_sampleth}).

\item Low-resolution AM7 $(d)$ and Envelope $(e)$ simulations with  $\Delta \rho = 0.16\mum$, $\Delta z = 0.05\mum$.  We used $2\times$ interpolation supersampling in the $z$ direction, and $\Delta t/2$ substeps for particle dynamics in $(d)$. We set the number of particles per cell ($z,\theta,\rho$) as $4\times8\times4$ in the AM solver, $4\times1\times4$ in the Envelope solver.
\end{itemize}

 We used 3rd order particle shapes for all simulations and absorbing layers at the outer edges and $\Delta t = 0.5\Delta z$. We ran the simulations with 24th and 30th order finite differences, in $\rho$, $z$ respectively, and 9th order exponentials and FBPIC's true spectral solver. As discussed in Section \ref{subsec:particlesAM_filter} using some kind of current filter is mandatory for numerical reasons, so we use ``lowpass A`` filter in Cartesian directions. In FBPIC we use the built-in binomial current filters (which is the default recommended by the developers).

We summarize these simulations in Fig. \ref{fig:test_LWFA},  where we compare not only the snapshots of density distributions, but also the distributions of the high energy electrons in the momentum phase-space. All simulations are essentially free from the numerical Cherenkov-radiation, which we illustrated with Fig. \ref{fig:test_LWFA}$(d_4)$  showing $E_x-cB_y$ fields, where these are typically visible. So, the suppression of zero-order numerical Cherenkov effect was accomplished in real space.

All simulations are also resulted in the same wakefield dynamics, wakefield cavity, and not only these, the spatial density of the injected electron bunches were also similar in all runs (first column of Fig. \ref{fig:test_LWFA}). The second parts of the injected electron bunches have longer spatial extension (between $z=460\,\mum$ and $z=480\,\mum$ in Fig. \ref{fig:test_LWFA}) which originates from the density down-ramp and it manifests as a strong peak in the momentum distribution with a center near $(p_z=25m_ec, p_x=p_y=0)$. Since this process is less nonlinear it is well resolved in all configurations.  If we compare our 3D and AM7 simulation densities (on Fig. \ref{fig:test_LWFA}$(a_1)$ and $(b_1)$) we can see that our AM7 simulation reproduces all features of the 3D bunch agreeing quantitatively - just with less particles. The FBPIC simulation resulted in a noticeably smoother density and less noise on Fig. \ref{fig:test_LWFA}$(c_1)$.

Despite the fact that the density distributions are almost identical, the momentum distributions $p_z$-$p_x$ and $p_z$-$p_y$ of the high energy electrons can be very different on Fig. \ref{fig:test_LWFA}. The electron's cut-off energy strongly depends on the longitudinal resolution, a fact which was recognized already in \cite{majorosi2026exponentialpic} as well. Here, the maximum Lorentz factor ($\gamma$) of electrons increases by 50 \% after decreasing the grid size from $\Delta z=50$ nm to $\Delta z=32$ nm. It is a consequence of high sensitivity of the process on the numerical parameters, which is wave breaking (or self-injection) in this case. We have verified that the electron injection starts already in the plateau region, $z\approx 270\,\mum$, before the down-ramp is reached by the laser pulse. This event is responsible for the high energy tail of the energy spectrum and it is extremely challenging to resolve numerically.

For instance, using the Envelope solver (Fig. \ref{fig:test_LWFA}$(e)$) the observed cut-off is generated differently than in the high-resolution AM case - which is not surprising since the former is fully cylindrically symmetric. It does have a high energy electron bunch around  $150m_ec$ and above on Fig. \ref{fig:test_LWFA}$(e_2)$ (compare to Fig. \ref{fig:test_LWFA}$(d_2)$). For the very high energy electrons to appear at $p_z > 200 m_ec$ in these figures we had to use 4 particles per cell in the $z$  direction. Overall, the Envelope model can be used for a quick-and-approximate simulations. A slightly surprising outcome of this comparison is that the FBPIC code produces an energy cut-off much lower than the expected value at this high resolution (see Fig. \ref{fig:test_LWFA}$(c)$).  It is even lower than our low-resolution AM7 result in Fig. \ref{fig:test_LWFA}$(d)$. We mostly blame the binomial current filter along $z$ for the former results.

Finally, if we compare these momentum distributions at high resolutions between our 3D and AM7 code (Fig. \ref{fig:test_LWFA}$(a)$ vs Fig. \ref{fig:test_LWFA}$(b)$). We can see that these show good agreement between the two runs. Although our AM cylindrical solution heavily undersamples the 3D space, the generated physics of the 3D self-injection are reproduced very well.

\section{Conclusion}
 
Following on our previous work \cite{majorosi2026exponentialpic} dealing with Cartesian coordinates, we derived the same exponential solution for the propagation of the Maxwell fields ${\bf E}, {\bf B}$, and for the propagation of the laser potential $A$ in cylindrical geometry with real azimuthal modes decomposition. We developed the spatial representation on cylindrical Yee-grids using very high order (8th-32th) staggered finite differences to suppress numerical wave dispersion effects, like the numerical Cherenkov radiation. We adapted the cylindrical quasi-3D particle-in-cell method and combined it with our exponential solutions, which included modifications to support the laser envelope PIC model. We outlined a local and separable current correction method to avoid field divergence errors near the cylindrical axis. All important operations are reduced to matrix-vector multiplications with banded diagonal matrices, which provide highly accurate and local solutions in real space, suitable for thread-based parallelization.

We have verified our results in multiple benchmarks. For our cylindrical Maxwell solver and scalar laser potential propagator we found that during vacuum propagation and during linear propagation in underdense plasma, the accuracy characteristics were exactly the same as our Cartesian solution \cite{majorosi2026exponentialpic}, so it can be made arbitrarily accurate in vacuum. We also found that the use of very high-order finite differences is also required to simulate accurate wave propagation approaching the accuracy of the spectral codes. Notable exception for the latter was the laser envelope solver, which provides very high accuracy and good dispersion properties even with lower-order numerical methods due to the laser envelope's smoothness. Interestingly, the simulations with the laser envelope PIC model also yielded better pulse group velocity and generated more accurate linear wakefields in underdense plasma compared to the Maxwell-azimuthal modes decomposition or 3D simulations. 

Our solution was benchmarked against simulations involving laser wakefield acceleration and self-injection of electrons. We compared the results obtained in different scenarios, namely the 3D exponential solution, laser envelope simulations, our AM simulations and the results of FBPIC \cite{lehe2016fbpic}. The results of a high-resolution 3D simulation were used as a reference, resulting in converging momentum and spatial density distributions. All cylindrical simulations resulted in nearly the same spatial density distribution of the injected electrons. However, the reference (3D) momentum distribution was not reproduced using the truly spectral solver of FBPIC. This could be explained with its binomial current smoothing which is to suppress artifacts in the density distribution near the axis, a problem which we tackle using more advanced charge-conserving schemes locally. Numerical Cherenkov radiation could not be observed in any simulations.

We have to note that there are many possibilities to extend this work with enhanced methods, physics modules, and parallelization with domain decomposition. Our primary goal is to implement the quantum-electrodynamics module, which will allow us to self-consistently model process like betatron radiation, radiation reaction, or even electron-positron pair creation within the frame of laser wakefield acceleration. The field ionization of atoms and ions is already on the way, but it is beyond the scope of this paper, and it will be published elsewhere. We are hopeful that the method shown in this paper will be useful for various relativistic laser-plasma simulations in the future.

\section*{Acknowledgments}
The ELI ALPS project (GINOP-2.3.6-15-2015-00001) is supported by the European Union and co-financed by the European Regional Development Fund. We acknowledge DKF for awarding us access to HPC resource based in Debrecen, Hungary.\footnote{\url{https://ncc.dkf.hu/en.html}}

 
\appendix

\section{Scalar absorbing layer} \label{subsubsec:absorbing_layerA}

We have developed a boundary layer for the scalar wave propagation that can absorb outgoing waves and which is compatible with our exponential formalism. The exact implementation of PML is possible when we break the wave equation into couple of first order transport equations \cite{johnson2021PMLnotes}. Unfortunately, for the second order wave equation we used in Section \ref{subsec:laserA} this turns out to be very complicated. Instead, we start from such a wave equation that includes strongly absorbing layers, it has the following form in the frequency domain \cite{barucq2007maxwell_absorbing}:
\begin{equation} \label{eq:absorbingA_wave_omega}
 (i\omega)^2 \left(1+(i\omega)^{-1} \sigma +(i\omega)^{-2} \sigma ^2 \right) A      =  \nabla^2 A,
\end{equation}
where $\sigma = \sigma_x(x)+\sigma_y(y)+\sigma_z(z)$ are the absorbing layer functions defined similarly as in the PML formalism, (see \cite{majorosi2026exponentialpic}). In the temporal domain we have:
\begin{equation} \label{eq:absorbingA_wave}
 \partial_t^2 A + \sigma \partial_t A +\sigma^2 A      =  \nabla^2 A,
\end{equation}
We want to reduce this wave equation to two coupled first order PDEs using $\tilde{A}$ as a new variable. We define the latter from $\partial_t A$  as:
\begin{equation} \label{eq:absorbingA_Adt}
 \partial_t A  = h^{-1} \tilde{A} - \sigma A /2, 
\end{equation}
Then, the wave equation Eq. (\ref{eq:absorbingA_wave}) will have the form:
\begin{equation} \label{eq:absorbingA_waveA}
 \partial_t \tilde{A}  = h \nabla^2 \tilde{A} - \sigma \tilde{A} /2-h \sigma^2 A,
\end{equation}
where $h$ is a scaling factor of choice between the two equations. In the main text we set $h = 1$. 

We describe the evolution of the scalar wave using the vector $\tilde{\bf A} = (A, \tilde{A}) $ to get  the formal equation:
\begin{equation} \label{eq:absorbingA_formal}
 \partial_t \tilde{\bf A}  = \left( \op{H}_{A} - \sigma/2 \right) \tilde{\bf A},
\end{equation}
where $\op{H}_{A}$ is the wave operator from Eqs. (\ref{eq:absorbingA_Adt}) and (\ref{eq:absorbingA_waveA}) with an extra term of $\sigma^2 A$.
We can use the split operator formula (see \cite{bandrauk2013splitting,majorosi2026exponentialpic}) to get the second order solution for $\tilde{\bf A}$ as
\begin{equation} \label{eq:absorbingA_solution2}
\tilde{\bf A}(t+\Delta t)   \approx \exp (-\sigma \Delta t/4) \exp ( \op{H}_{A} \Delta t) \exp (-\sigma \Delta t/4) \tilde{\bf A} .
\end{equation} 

This is the reason why we have chosen the particular symmetric form of Eqs. (\ref{eq:absorbingA_Adt}) and (\ref{eq:absorbingA_waveA}): the solution Eq. (\ref{eq:absorbingA_solution2}) will have exceptional stability, limited by the central exponential. The extra term $\sigma^2 A$ can be merged with the D'Alembert susceptibility (at temporal midpoints) : $\tilde{\chi}_{1/2} = \chi_{1/2}+\sigma^2$. This latter has only a small effect: it can be regarded as narrow wave guide that slightly slows down reflected waves from the layer. The resulting method Eq. (\ref{eq:absorbingA_solution2}) does absorb outgoing waves by orders of magnitude. Unfortunately,
waves propagating with glazing incidence at the layers will  be subject to diffraction.

\section{Laser envelope Boris pusher} \label{subsubsec:BorisA}

Here we derive a modified Boris pusher for the laser envelope PIC model (Section \ref{subsec:particlesAM_envelope}). To start we need Eqs. (\ref{eq:envelopeAM_Afield})-(\ref{eq:envelopeAM_avgEF}). We need to approximate the average of ${\gamma}^{-1}_{\pp,n}$ in the ponderomotive force during a Boris push over $\Delta t$.

To do that, we write out a predictor momentum push without the ponderomotive force: 
\begin{equation} \label{eq:envelopeAM_boris0}
{\bf u}_{-} = {\bf u}_{\pp,n}+ \overline{\bf E}_{\pp,n} \quad \text{and} \quad \gamma_{-} = \sqrt{1+{\bf u}_{-}^2+\Phi_{p,n+1/2}}.
\end{equation}
We then add the ponderomotive force to this, which will include all electric field like components in the modified Boris push:
\begin{equation} \label{eq:envelopeAM_borisF}
\tilde{\bf u}_{-} = {\bf u}_{-} +\overline{\gamma}^{-1} \overline{\bf F}_{\pp,n}, \quad \text{and} \quad
\overline{\gamma} \approx \sqrt{1+\tilde{\bf u}_{-}^2+\Phi_{\pp,n+1/2}}
\end{equation}
We then approximate the value of $\overline{\gamma}$ using $\tilde{\bf u}_{-}$, which is the gamma of the  Boris step. This leads to following  equation for the values of $\overline{\gamma}$:
\begin{equation} \label{eq:envelopeAM_borisgamma}
\overline{\gamma}^{2}- \gamma_-^2 - 2 \overline{\gamma}^{-1} \overline{\bf F}\cdot {\bf u}_{-}
- \overline{\gamma}^{-2} \left| \overline{\bf F} \right| ^2 = 0,
\end{equation}
where one usually neglects the last term leading to a cubic equation that needs to be solved \cite{terzani2019envelope}.
We use the following low order explicit formula for the cubic form with Hayley-iteration \cite{BOOK_NUMERICAL_RECIPIES}:
\begin{equation} \label{eq:envelopeAM_borisgamma2}
\overline{\gamma} \approx \gamma_{-}+
\frac{\overline{\bf F}\cdot {\bf u}_{-}}
{\gamma_{-}^{2}+\frac{3}{2} \gamma_{-}^{-1}\left( \overline{\bf F}\cdot {\bf u}_{-} \right)}.
\end{equation}

To overview, we can use Eq. (\ref{eq:envelopeAM_borisgamma2}) to calculate the approximate value of  $\overline{\gamma}$, which we substitute it into Eq. (\ref{eq:envelopeAM_Fpond}). We then proceed by combining the electric and ponderomotive fields as  ${\bf E}_{\pp}+\overline{\gamma}^{-1}{\bf E}_{\Phi,\pp}$, and finally  we  do a regular Boris push using Eq. (\ref{eq:envelopeAM_borisF}). 
This override tactic works for Boris pushers in any exponential form, but is not straightforwardly applicable with pushers that conserve ${\bf E}\times{\bf{B}}$ in the force free case, like the Higuera-Cary \cite{higuera2017pusher} pusher.



  
\bibliographystyle{elsarticle-num}
\bibliography{0Introduction.bib, 0BIB_pic_theory.bib, 0BIB_laser_theory.bib, 0BIB_laser_plasma.bib}

\begin{thebibliography}{10}
\expandafter\ifx\csname url\endcsname\relax
  \def\url#1{\texttt{#1}}\fi
\expandafter\ifx\csname urlprefix\endcsname\relax\def\urlprefix{URL }\fi
\expandafter\ifx\csname href\endcsname\relax
  \def\href#1#2{#2} \def\path#1{#1}\fi

\bibitem{Strickland85}
D.~Strickland, G.~Mourou,
  \href{https://www.sciencedirect.com/science/article/pii/0030401885901518}{Compression
  of amplified chirped optical pulses}, Optics Communications 55~(6) (1985)
  447--449.
\newblock \href {https://doi.org/https://doi.org/10.1016/0030-4018(85)90151-8}
  {\path{doi:https://doi.org/10.1016/0030-4018(85)90151-8}}.
\newline\urlprefix\url{https://www.sciencedirect.com/science/article/pii/0030401885901518}

\bibitem{Tajima79}
T.~Tajima, J.~M. Dawson,
  \href{https://link.aps.org/doi/10.1103/PhysRevLett.43.267}{Laser electron
  accelerator}, Phys. Rev. Lett. 43 (1979) 267--270.
\newblock \href {https://doi.org/10.1103/PhysRevLett.43.267}
  {\path{doi:10.1103/PhysRevLett.43.267}}.
\newline\urlprefix\url{https://link.aps.org/doi/10.1103/PhysRevLett.43.267}

\bibitem{esarey2009laser_plasma_accelerators}
E.~Esarey, C.~B. Schroeder, W.~P. Leemans,
  \href{https://link.aps.org/doi/10.1103/RevModPhys.81.1229}{Physics of
  laser-driven plasma-based electron accelerators}, Rev. Mod. Phys. 81 (2009)
  1229--1285.
\newblock \href {https://doi.org/10.1103/RevModPhys.81.1229}
  {\path{doi:10.1103/RevModPhys.81.1229}}.
\newline\urlprefix\url{https://link.aps.org/doi/10.1103/RevModPhys.81.1229}

\bibitem{lu2007electron_laser_wakefield}
W.~Lu, M.~Tzoufras, C.~Joshi, F.~S. Tsung, W.~B. Mori, J.~Vieira, R.~A.
  Fonseca, L.~O. Silva,
  \href{https://link.aps.org/doi/10.1103/PhysRevSTAB.10.061301}{Generating
  multi-{G}e{V} electron bunches using single stage laser wakefield
  acceleration in a 3d nonlinear regime}, Phys. Rev. ST Accel. Beams 10 (2007)
  061301.
\newblock \href {https://doi.org/10.1103/PhysRevSTAB.10.061301}
  {\path{doi:10.1103/PhysRevSTAB.10.061301}}.
\newline\urlprefix\url{https://link.aps.org/doi/10.1103/PhysRevSTAB.10.061301}

\bibitem{albert2016applications_lwfa}
F.~Albert, A.~G.~R. Thomas,
  \href{https://doi.org/10.1088/0741-3335/58/10/103001}{Applications of laser
  wakefield accelerator-based light sources}, Plasma Physics and Controlled
  Fusion 58~(10) (2016) 103001.
\newblock \href {https://doi.org/10.1088/0741-3335/58/10/103001}
  {\path{doi:10.1088/0741-3335/58/10/103001}}.
\newline\urlprefix\url{https://doi.org/10.1088/0741-3335/58/10/103001}

\bibitem{BOOK_PLASMA_SIMULATION}
C.~K. Birdsall, A.~B. Langdon, Plasma Physics via Computer Simulation, CRC
  press, 1991.
\newblock \href {https://doi.org/https://doi.org/10.1201/9781315275048}
  {\path{doi:https://doi.org/10.1201/9781315275048}}.

\bibitem{fonseca2002OSIRIS}
R.~A. Fonseca, L.~O. Silva, F.~S. Tsung, V.~K. Decyk, W.~Lu, C.~Ren, W.~B.
  Mori, S.~Deng, S.~Lee, T.~Katsouleas, J.~C. Adam, Osiris: A
  three-dimensional, fully relativistic particle in cell code for modeling
  plasma based accelerators, in: P.~M.~A. Sloot, A.~G. Hoekstra, C.~J.~K. Tan,
  J.~J. Dongarra (Eds.), Computational Science --- ICCS 2002, Springer Berlin
  Heidelberg, Berlin, Heidelberg, 2002, pp. 342--351.

\bibitem{pritchett2003pic_tutorial}
P.~L. Pritchett, Particle-in-cell simulation of plasmas—a tutorial, Space
  Plasma Simulation (2003) 1--24.

\bibitem{arber2015pic_epoch}
T.~D. Arber, K.~Bennett, C.~S. Brady, A.~Lawrence-Douglas, M.~G. Ramsay, N.~J.
  Sircombe, P.~Gillies, R.~G. Evans, H.~Schmitz, A.~R. Bell, C.~P. Ridgers,
  \href{https://doi.org/10.1088/0741-3335/57/11/113001}{Contemporary
  particle-in-cell approach to laser-plasma modelling}, Plasma Physics and
  Controlled Fusion 57~(11) (2015) 113001.
\newblock \href {https://doi.org/10.1088/0741-3335/57/11/113001}
  {\path{doi:10.1088/0741-3335/57/11/113001}}.
\newline\urlprefix\url{https://doi.org/10.1088/0741-3335/57/11/113001}

\bibitem{belayev2015PICsar}
M.~A. Belyaev,
  \href{https://www.sciencedirect.com/science/article/pii/S1384107614001407}{{PIC}sar:
  A 2.5d axisymmetric, relativistic, electromagnetic, particle in cell code
  with a radiation absorbing boundary}, New Astronomy 36 (2015) 37--49.
\newblock \href {https://doi.org/https://doi.org/10.1016/j.newast.2014.09.006}
  {\path{doi:https://doi.org/10.1016/j.newast.2014.09.006}}.
\newline\urlprefix\url{https://www.sciencedirect.com/science/article/pii/S1384107614001407}

\bibitem{derouillat2018pic_smilei}
J.~Derouillat, A.~Beck, F.~Pérez, T.~Vinci, M.~Chiaramello, A.~Grassi,
  M.~Flé, G.~Bouchard, I.~Plotnikov, N.~Aunai, J.~Dargent, C.~Riconda,
  M.~Grech,
  \href{https://www.sciencedirect.com/science/article/pii/S0010465517303314}{Smilei
  : A collaborative, open-source, multi-purpose particle-in-cell code for
  plasma simulation}, Computer Physics Communications 222 (2018) 351--373.
\newblock \href {https://doi.org/https://doi.org/10.1016/j.cpc.2017.09.024}
  {\path{doi:https://doi.org/10.1016/j.cpc.2017.09.024}}.
\newline\urlprefix\url{https://www.sciencedirect.com/science/article/pii/S0010465517303314}

\bibitem{lifschitz2009pic_fourier_decomposition}
A.~Lifschitz, X.~Davoine, E.~Lefebvre, J.~Faure, C.~Rechatin, V.~Malka,
  \href{https://www.sciencedirect.com/science/article/pii/S0021999108005950}{Particle-in-cell
  modelling of laser–plasma interaction using {F}ourier decomposition},
  Journal of Computational Physics 228~(5) (2009) 1803--1814.
\newblock \href {https://doi.org/https://doi.org/10.1016/j.jcp.2008.11.017}
  {\path{doi:https://doi.org/10.1016/j.jcp.2008.11.017}}.
\newline\urlprefix\url{https://www.sciencedirect.com/science/article/pii/S0021999108005950}

\bibitem{davidson2015OSIRIS_RZ}
A.~Davidson, A.~Tableman, W.~An, F.~Tsung, W.~Lu, J.~Vieira, R.~Fonseca,
  L.~Silva, W.~Mori,
  \href{https://www.sciencedirect.com/science/article/pii/S0021999114007529}{Implementation
  of a hybrid particle code with a {PIC} description in r–z and a gridless
  description in phi into osiris}, Journal of Computational Physics 281 (2015)
  1063--1077.
\newblock \href {https://doi.org/https://doi.org/10.1016/j.jcp.2014.10.064}
  {\path{doi:https://doi.org/10.1016/j.jcp.2014.10.064}}.
\newline\urlprefix\url{https://www.sciencedirect.com/science/article/pii/S0021999114007529}

\bibitem{lehe2016fbpic}
R.~Lehe, M.~Kirchen, I.~A. Andriyash, B.~B. Godfrey, J.-L. Vay,
  \href{https://www.sciencedirect.com/science/article/pii/S0010465516300224}{A
  spectral, quasi-cylindrical and dispersion-free particle-in-cell algorithm},
  Computer Physics Communications 203 (2016) 66--82.
\newblock \href {https://doi.org/https://doi.org/10.1016/j.cpc.2016.02.007}
  {\path{doi:https://doi.org/10.1016/j.cpc.2016.02.007}}.
\newline\urlprefix\url{https://www.sciencedirect.com/science/article/pii/S0010465516300224}

\bibitem{majorosi2026exponentialpic}
S.~Majorosi, N.~A. Hafz, Z.~Lécz,
  \href{https://www.sciencedirect.com/science/article/pii/S0010465526000366}{High-order
  exponential solver method for particle-in-cell simulations}, Computer Physics
  Communications 322 (2026) 110054.
\newblock \href {https://doi.org/https://doi.org/10.1016/j.cpc.2026.110054}
  {\path{doi:https://doi.org/10.1016/j.cpc.2026.110054}}.
\newline\urlprefix\url{https://www.sciencedirect.com/science/article/pii/S0010465526000366}

\bibitem{leforestier1991tdsecomparisom}
C.~Leforestier, R.~Bisseling, C.~Cerjan, M.~Feit, R.~Friesner, A.~Guldberg,
  A.~Hammerich, G.~Jolicard, W.~Karrlein, H.-D. Meyer, N.~Lipkin, O.~Roncero,
  R.~Kosloff,
  \href{https://www.sciencedirect.com/science/article/pii/002199919190137A}{A
  comparison of different propagation schemes for the time dependent
  {S}chrödinger equation}, Journal of Computational Physics 94~(1) (1991)
  59--80.
\newblock \href {https://doi.org/https://doi.org/10.1016/0021-9991(91)90137-A}
  {\path{doi:https://doi.org/10.1016/0021-9991(91)90137-A}}.
\newline\urlprefix\url{https://www.sciencedirect.com/science/article/pii/002199919190137A}

\bibitem{castro2004kohnshampropagators}
A.~Castro, M.~A.~L. Marques, A.~Rubio,
  \href{https://doi.org/10.1063/1.1774980}{Propagators for the time-dependent
  {K}ohn–{S}ham equations}, The Journal of Chemical Physics 121~(8) (2004)
  3425--3433.
\newblock \href
  {http://arxiv.org/abs/https://pubs.aip.org/aip/jcp/article-pdf/121/8/3425/19313892/3425\_1\_online.pdf}
  {\path{arXiv:https://pubs.aip.org/aip/jcp/article-pdf/121/8/3425/19313892/3425\_1\_online.pdf}},
  \href {https://doi.org/10.1063/1.1774980} {\path{doi:10.1063/1.1774980}}.
\newline\urlprefix\url{https://doi.org/10.1063/1.1774980}

\bibitem{bandrauk2013splitting}
A.~D. Bandrauk, H.~LU, Exponential propagators (integrators) for the
  time-dependent {S}chrödinger equation, Journal of Theoretical and
  Computational Chemistry 12~(06) (2013) 1340001.
\newblock \href {https://doi.org/https://doi.org/10.1142/S0219633613400014}
  {\path{doi:https://doi.org/10.1142/S0219633613400014}}.

\bibitem{cowan2011envelope_model}
B.~M. Cowan, D.~L. Bruhwiler, E.~Cormier-Michel, E.~Esarey, C.~G. Geddes,
  P.~Messmer, K.~M. Paul,
  \href{https://www.sciencedirect.com/science/article/pii/S0021999110005036}{Characteristics
  of an envelope model for laser–plasma accelerator simulation}, Journal of
  Computational Physics 230~(1) (2011) 61--86.
\newblock \href {https://doi.org/https://doi.org/10.1016/j.jcp.2010.09.009}
  {\path{doi:https://doi.org/10.1016/j.jcp.2010.09.009}}.
\newline\urlprefix\url{https://www.sciencedirect.com/science/article/pii/S0021999110005036}

\bibitem{benedetti2018envelope_solver}
C.~Benedetti, C.~B. Schroeder, C.~G.~R. Geddes, E.~Esarey, W.~P. Leemans,
  \href{https://dx.doi.org/10.1088/1361-6587/aa8977}{An accurate and efficient
  laser-envelope solver for the modeling of laser-plasma accelerators}, Plasma
  Physics and Controlled Fusion 60~(1) (2017) 014002.
\newblock \href {https://doi.org/10.1088/1361-6587/aa8977}
  {\path{doi:10.1088/1361-6587/aa8977}}.
\newline\urlprefix\url{https://dx.doi.org/10.1088/1361-6587/aa8977}

\bibitem{terzani2019envelope}
D.~Terzani, P.~Londrillo,
  \href{https://www.sciencedirect.com/science/article/pii/S0010465519301195}{A
  fast and accurate numerical implementation of the envelope model for
  laser–plasma dynamics}, Computer Physics Communications 242 (2019) 49--59.
\newblock \href {https://doi.org/https://doi.org/10.1016/j.cpc.2019.04.007}
  {\path{doi:https://doi.org/10.1016/j.cpc.2019.04.007}}.
\newline\urlprefix\url{https://www.sciencedirect.com/science/article/pii/S0010465519301195}

\bibitem{BOOK_NUMERICAL_RECIPIES}
W.~H. Press, Numerical recipes 3rd edition: The art of scientific computing,
  Cambridge university press, 2007.

\bibitem{constantinescu2002flow_cylindrical}
G.~Constantinescu, S.~Lele,
  \href{https://www.sciencedirect.com/science/article/pii/S0021999102971871}{A
  highly accurate technique for the treatment of flow equations at the polar
  axis in cylindrical coordinates using series expansions}, Journal of
  Computational Physics 183~(1) (2002) 165--186.
\newblock \href {https://doi.org/https://doi.org/10.1006/jcph.2002.7187}
  {\path{doi:https://doi.org/10.1006/jcph.2002.7187}}.
\newline\urlprefix\url{https://www.sciencedirect.com/science/article/pii/S0021999102971871}

\bibitem{godfrey2013PIC_stability}
B.~B. Godfrey, J.-L. Vay,
  \href{https://www.sciencedirect.com/science/article/pii/S0021999113002556}{Numerical
  stability of relativistic beam multidimensional {PIC} simulations employing
  the esirkepov algorithm}, Journal of Computational Physics 248 (2013) 33--46.
\newblock \href {https://doi.org/https://doi.org/10.1016/j.jcp.2013.04.006}
  {\path{doi:https://doi.org/10.1016/j.jcp.2013.04.006}}.
\newline\urlprefix\url{https://www.sciencedirect.com/science/article/pii/S0021999113002556}

\bibitem{li2016fields_tightly_focused}
J.-X. Li, Y.~I. Salamin, K.~Z. Hatsagortsyan, C.~H. Keitel,
  \href{http://opg.optica.org/josab/abstract.cfm?URI=josab-33-3-405}{Fields of
  an ultrashort tightly focused laser pulse}, J. Opt. Soc. Am. B 33~(3) (2016)
  405--411.
\newblock \href {https://doi.org/10.1364/JOSAB.33.000405}
  {\path{doi:10.1364/JOSAB.33.000405}}.
\newline\urlprefix\url{http://opg.optica.org/josab/abstract.cfm?URI=josab-33-3-405}

\bibitem{majorosi2023tightlyfocused}
S.~Majorosi, Z.~L\'{e}cz, D.~Papp, C.~Kamperidis, N.~A.~M. Hafz,
  \href{https://opg.optica.org/josab/abstract.cfm?URI=josab-40-3-551}{Numerical
  representation of tightly focused ultra-short laser pulses}, J. Opt. Soc. Am.
  B 40~(3) (2023) 551--559.
\newblock \href {https://doi.org/10.1364/JOSAB.481864}
  {\path{doi:10.1364/JOSAB.481864}}.
\newline\urlprefix\url{https://opg.optica.org/josab/abstract.cfm?URI=josab-40-3-551}

\bibitem{dijk2014tdsesource}
W.~van Dijk, F.~M. Toyama,
  \href{https://link.aps.org/doi/10.1103/PhysRevE.90.063309}{Numerical
  solutions of the {S}chrödinger equation with source terms or time-dependent
  potentials}, Phys. Rev. E 90 (2014) 063309.
\newblock \href {https://doi.org/10.1103/PhysRevE.90.063309}
  {\path{doi:10.1103/PhysRevE.90.063309}}.
\newline\urlprefix\url{https://link.aps.org/doi/10.1103/PhysRevE.90.063309}

\bibitem{blanes2009review_magnus_expansion}
S.~Blanes, F.~Casas, J.~Oteo, J.~Ros,
  \href{https://www.sciencedirect.com/science/article/pii/S0370157308004092}{The
  {M}agnus expansion and some of its applications}, Physics Reports 470~(5)
  (2009) 151--238.
\newblock \href {https://doi.org/https://doi.org/10.1016/j.physrep.2008.11.001}
  {\path{doi:https://doi.org/10.1016/j.physrep.2008.11.001}}.
\newline\urlprefix\url{https://www.sciencedirect.com/science/article/pii/S0370157308004092}

\bibitem{esarey1995tightly_focused_pulses}
E.~Esarey, P.~Sprangle, M.~Pilloff, J.~Krall,
  \href{http://opg.optica.org/josab/abstract.cfm?URI=josab-12-9-1695}{Theory
  and group velocity of ultrashort, tightly focused laser pulses}, J. Opt. Soc.
  Am. B 12~(9) (1995) 1695--1703.
\newblock \href {https://doi.org/10.1364/JOSAB.12.001695}
  {\path{doi:10.1364/JOSAB.12.001695}}.
\newline\urlprefix\url{http://opg.optica.org/josab/abstract.cfm?URI=josab-12-9-1695}

\bibitem{johnson2021PMLnotes}
S.~G. Johnson, Notes on perfectly matched layers ({PML}s), arXiv preprint
  arXiv:2108.05348 (2021).

\bibitem{barucq2007maxwell_absorbing}
H.~Barucq, M.~Fontes,
  \href{https://www.sciencedirect.com/science/article/pii/S0021782407000165}{Well-posedness
  and exponential stability of {M}axwell-like systems coupled with strongly
  absorbing layers}, Journal de Mathématiques Pures et Appliquées 87~(3)
  (2007) 253--273.
\newblock \href {https://doi.org/https://doi.org/10.1016/j.matpur.2007.01.001}
  {\path{doi:https://doi.org/10.1016/j.matpur.2007.01.001}}.
\newline\urlprefix\url{https://www.sciencedirect.com/science/article/pii/S0021782407000165}

\bibitem{higuera2017pusher}
A.~V. Higuera, J.~R. Cary,
  \href{https://doi.org/10.1063/1.4979989}{Structure-preserving second-order
  integration of relativistic charged particle trajectories in electromagnetic
  fields}, Physics of Plasmas 24~(5) (2017) 052104.
\newblock \href
  {http://arxiv.org/abs/https://pubs.aip.org/aip/pop/article-pdf/doi/10.1063/1.4979989/15988441/052104\_1\_online.pdf}
  {\path{arXiv:https://pubs.aip.org/aip/pop/article-pdf/doi/10.1063/1.4979989/15988441/052104\_1\_online.pdf}},
  \href {https://doi.org/10.1063/1.4979989} {\path{doi:10.1063/1.4979989}}.
\newline\urlprefix\url{https://doi.org/10.1063/1.4979989}

\end{thebibliography}
\end{document}